%% file: main.tex
\documentclass[fleqn,usenatbib]{mnras}

\usepackage{newtxtext,newtxmath}
\usepackage[T1]{fontenc}
\usepackage{lineno}

\usepackage{amsmath,mathrsfs,gensymb}
\usepackage{graphicx}
\usepackage{braket}

\defcitealias{Rui:2023:MagneticSuppression}{Paper I}
\defcitealias{Rui:2024:TARM}{Paper II}

\newcommand{\edit}[1]{#1}

\title[Asteroseismic magnetometry]{Extending asteroseismic magnetometry across the diverse landscape of magnetic structures}

\author[N. Z. Rui et al.]{
Nicholas Z. Rui,$^{1,2,3}$\thanks{E-mail: nrui@princeton.edu}
J. M. Joel Ong,$^4$
Armand Leclerc,$^5$
Daniel Lecoanet,$^{2,6}$
Lisa Bugnet,$^5$
\newauthor
Janosz W. Dewberry,$^7$
Bastien Liagre,$^8$
and St\'ephane Mathis$^9$
\\
$^{1}$Department of Astrophysical Sciences, Princeton University, 4 Ivy Lane, Princeton, NJ 08544, USA\\
$^{2}$Center for Interdisciplinary Exploration and Research in Astrophysics (CIERA), Northwestern University, 1800 Sherman Ave., Evanston, IL 60201, USA\\
$^{3}$NASA Hubble Fellow\\
$^{4}$Sydney Institute for Astronomy, University of Sydney, A28 Physics Road, Sydney NSW 2006, Australia\\
$^{5}$Institute of Science and Technology Austria, Am Campus 1, Klosterneuburg, 3400, Austria\\
$^{6}$Department of Engineering Sciences and Applied Mathematics, McCormick School of Engineering, Northwestern University, 2145 Sheridan Road,\\Evanston, IL 60208, USA\\
$^{7}$Department of Astronomy, University of Massachusetts Amherst, 710 N Pleasant St, Amherst, MA 01003, USA\\
$^{8}$Universit\'e Paris Cit\'e, Universit\'e Paris-Saclay, CEA, CNRS, AIM, F-91191 Gif-sur-Yvette, France\\
$^{9}$Universit\'e Paris-Saclay, Universit\'e Paris Cit\'e, CEA, CNRS, AIM, F-91191 Gif-sur-Yvette, France
}

\date{Accepted XXX. Received YYY; in original form ZZZ}
\pubyear{\the\year{}}

\begin{document}
% \linenumbers % enable for line numbers

\label{firstpage}
\pagerange{\pageref{firstpage}--\pageref{lastpage}}
\maketitle

\begin{abstract}
Magnetic fields have now been asteroseismically measured in the cores of many red giants.
However, most interpretations of these measurements assume that the magnetic field is far below the critical field strength known to be exceeded by red giants exhibiting gravity-mode suppression.
A recent method based on the traditional approximation of rotation and magnetism accurately predicts mode frequencies under fields up to this critical value by modeling gravity waves as individual magnetogravity ``polarizations'' which propagate through a waveguide-like mode cavity.
So far, this formalism has been limited to magnetic fields which are axisymmetric about the rotation axis.
In this study, we extend this approach by calculating the polarizations of magnetogravity waves under arbitrarily shaped magnetic fields under potentially rapid rotation.
We consider the special cases of a dipolar magnetic field misaligned with the rotation axis as well as a dipole-plus-quadrupole magnetic field with no rotational symmetry.
We show that non-axisymmetric field configurations can induce avoided crossings between polarizations, and that waves in such systems can convert between magnetogravity polarizations as they propagate, especially when the magnetic field strength is locally below a stratification-dependent threshold value.
This threshold is distinct from the critical field strength for gravity-mode suppression, and is instead similar to the magnetic field strength at which perturbation theory breaks down.
\end{abstract}


\input{sect_intro}

\section{Methods}
\input{sect_derive_tarm}
\input{sect_methods_numerics}
\input{sect_comparison_to_pert}

\section{Results and Discussion}
\input{sect_inclined_dipole}
\input{sect_dipole_plus_y22}
\input{sect_higher_wkb}

\input{sect_conclusion}

\section*{Acknowledgements}

We thank Lucas Barrault, Lynn Buchele, Lukas Einramhof, Jim Fuller, Jeremy Goodman, and Eliot Quataert for helpful discussions\edit{, and the anonymous referee for their valuable comments and suggestions}.
N.Z.R. acknowledges support from the NASA Hubble Fellowship grant HST-HF2-51589.001-A awarded by the Space Telescope Science Institute, which is operated by the Association of Universities for Research in Astronomy, Inc., for NASA, under contract NAS5-26555.
J.M.J.O. acknowledges support from the Australian Research Council (ARC) through grants DP210103119 and FL220100117.
L.B. and A.L. gratefully acknowledge support from the European Research Council (ERC) under the Horizon Europe programme (Calcifer; Starting Grant agreement N$^\circ$101165631).
D.L. is partially supported by NSF AAG grant AST-2405812, Sloan Foundation grant FG-2024-21548 and Simons Foundation grant SFI-MPS-T-MPS-00007353.
S.M. acknowledges support from the European Research Council (ERC) under the Horizon Europe programme (Synergy Grant agreement 101071505: 4D-STAR), from the CNES SOHO-GOLF and PLATO grants at CEA-DAp, and from PNPS (CNRS/INSU).
While partially funded by the European Union, views and opinions expressed are, however, those of the authors only and do not necessarily reflect those of the European Union or the European Research Council.
Neither the European Union nor the granting authority can be held responsible for them.

This work presents results obtained using the \texttt{Dedalus} \citep{Burns:2020:Dedalus} and \texttt{MESA} \citep{Paxton:2011:MESA,Paxton:2013:MESA,Paxton:2015:MESA,Paxton:2018:MESA,Paxton:2019:MESA,Jermyn:2023:MESA} codes.
Our analysis is conducted using the \texttt{NumPy} \citep{Harris:2020:NumPy}, \texttt{SciPy} \citep{Virtanen:2020:SciPy}, and \texttt{AstroPy} \citep{Astropy:2013:Software,Astropy:2018:Software,Astropy:2022:Software} packages.
Visualizations in this work were created using the \texttt{Matplotlib} \citep{Hunter:2007:Matplotlib} and \texttt{Manim} \citep{ManimCommunity:2026} packages.

\section*{Data Availability}

Animated figures and \edit{the MESA inlist file} used to generate our stellar model can be found in a corresponding Zenodo upload at \href{https://zenodo.org/records/20561425}{https://zenodo.org/records/20561425}.
Animated figures can also be viewed on the online version of this article.

\bibliographystyle{mnras}
\bibliography{mybib}

\appendix

\input{app_delta_mag_ell}
\input{app_quadrupole_eigenvalues}
\input{app_higher_wkb_waveguide}
\input{app_higher_wkb_derivation}
\input{app_higher_wkb_pert}

\label{lastpage}
\end{document}

%% file: sect_intro.tex
%!TEX root=./main.tex
\section{Introduction} \label{sec:intro}

Asteroseismology has recently emerged as a powerful tool to probe magnetic fields deep within stellar interiors, particularly those of red giant stars \citep{Li:2022:30to100kG}.
Techniques within asteroseismic magnetometry constrain stellar magnetism by taking advantage of the sensitivity of buoyantly restored gravity (g) modes to magnetic forces.
This area of study promises to observationally scrutinize the origin of stellar magnetic fields \citep{Cantiello:2016:EvolvingMagnetic,Einramhof:2026:MagnetoWhiteDwarfs} as well as their coupled evolution with stellar rotation \citep{Cantiello:2014:RGAMT,Fuller:2019:SlowingSpins,Skoutnev:2025:MagneticWebs}.

Gravity-mode pulsations are ubiquitous across the Hertzsprung--Russell diagram, occurring in a wide variety of stars including red giants \citep{Chaplin:2013:SolarType}, intermediate-mass main-sequence stars \citep{Balona:2011:GammaDoradus}, hot subdwarfs \citep{Heber:2009:HotSubdwarfs}, and white dwarfs \citep{Corsico:2019:PulsatingWhiteDwarfs}.
In most cases, stable stratification strongly confines the motion of the plasma in the radial direction.
Accordingly, g modes in this ``asymptotic'' regime typically have short radial wavelengths ($\lambda_r/r\ll1$), high radial orders ($\int k_r\,\mathrm{d}r\gg1$), and approximately horizontal fluid displacements ($\xi_h/\xi_r\sim N/\omega\gg1$).
Because the strength of magnetic tension primarily arises through the Alfv\'en frequency $\omega_A=\vec{k}\cdot\vec{v}_A$ (where $\vec{v}_A=\vec{B}/\sqrt{4\pi\rho}$ is the Alfv\'en velocity), the radial component of the field $B_r$ typically dominates the magnetic field's impact on g modes, especially for observable low-degree modes for which $k_h$ is small.

The degree to which a g mode is modified by magnetism is roughly set by how significantly the Lorentz force competes with buoyancy \citep{Fuller:2015:SuppressedDipole,Cantiello:2016:EvolvingMagnetic}.
A rough estimate indicates that a g mode with angular frequency $\omega$ will be strongly affected by magnetic fields whose radial components are comparable to or exceed
\begin{equation} \label{eqn:Brcrit}
    B_{r,\mathrm{crit}} \sim \sqrt{\rho}\omega^2r/N\mathrm{,}
\end{equation}
where $\rho$ is the density and $N$ is the Brunt--V\"ais\"al\"a frequency.
Magnetic red giant cores with observable g modes have typical values $B_{r,\mathrm{crit}}\sim10^5$--$10^6\,\mathrm{G}$.
Equivalently, a radial field $B_r$ strongly affects g modes with frequencies lower than roughly
\begin{equation} \label{eqn:omegaB}
    \omega_B = \sqrt{Nv_{Ar}/r}\mathrm{.}
\end{equation}
% Under these definitions, perturbation theory can be used to accurately calculate the impact of the magnetic field only when $B_r\ll B_{r,\mathrm{crit}}$, i.e., $\omega\gg\omega_B$.
Most studies then consider two categories of magnetogravity-wave behavior in radiative interiors:

\begin{enumerate}
    \item \textbf{Magnetic frequency shifts} ($B_r\ll B_{r,\mathrm{crit}}$): The magnetic tension contributes to the restoration of g modes, increasing their frequencies.
    Frequency shifts in this regime are typically calculated using first-order perturbation theory, which predicts that the magnitude of the shift is related to an overlap integral between $B_r^2$ and the unperturbed oscillation mode \citep{Gough:1990:MagneticPertTheory,Loi:2020:MGNonPerturbative,Gomes:2020:MagneticRG,Bugnet:2021:MagneticI,Bugnet:2022:MagneticII,Mathis:2021:Magnetoasteroseismology,Mathis:2023:MagneticAsymmetry,Li:2022:30to100kG,Das:2024:ComplexMagnetic,Bhattacharya:2024:MagneticSubgiant}.
    When these shifts are observed, it is often possible to directly measure the field's strength and place constraints on its geometry.
    The number of detections of magnetic frequency shifts in red giants is currently approaching one hundred \citep{Li:2022:30to100kG,Deheuvels:2023:MagneticRG,Li:2023:13Magnetic,Hatt:2024:MagneticRG,Villate:2026:MagneticOffset}.
    Magnetic frequency shifts have also been detected in a handful of main-sequence pulsators \citep{Vandersnickt:2025:BcepMagnetic,Takata:2026:GDorToroidalField,Ihallaine:2026:MagneticGammaDor}.

    \item \textbf{Magnetic mode suppression} ($B_r>B_{r,\mathrm{crit}}$): When the Lorentz force is comparable in strength to buoyancy, it is believed that gravity waves are strongly damped.
    This magnetic process is believed to cause the preferential suppression of mixed-mode oscillations in $\simeq20\%$ of red giants \citep[][although this has been the subject of some debate, see \citealt{Mosser:2017:DepressedModes}]{Garcia:2014:DepressedDipole,Fuller:2015:SuppressedDipole,Stello:2016:MagneticPrevalence,Coppee:2024:MGRadialModes}.
    Although this suppression is predicted by many global wave calculations \citep{Fuller:2015:SuppressedDipole,Loi:2017:AlfvenResonances,Lecoanet:2022:HD43317,Rui:2023:MagneticSuppression}, ray-tracing integrations \citep{Loi:2018:MGDynamicalChaos,Loi:2020:MGPackets,Mueller:2025:RayTracing}, and numerical simulations \citep{Lecoanet:2017:MagneticConversion,David:2025:Magnetogravity}, its behavior is still not fully understood.
    When $B_r>B_{r,\mathrm{crit}}$, gravity waves are usually assumed to be totally suppressed, with observation of magnetic g-mode suppression being interpreted as implying a lower bound on $B_r$.
\end{enumerate}

However, there is a intermediate regime in which $B_r<B_{r,\mathrm{crit}}$ (so that mode suppression does not occur) but in which the field is still strong enough that higher-order effects of the magnetic field are observationally important \citep[e.g.,][]{Loi:2020:MGStrongPSP}.
In these cases, perturbation theory is inapplicable, and the modified frequencies and eigenfunctions must be calculated simultaneously.
While the full three-dimensional mode problem could in principle be solved numerically, such calculations are both technically challenging and numerically expensive, limiting their application to the interpretation of observations \citep[see, e.g.,][in the context of oscillations in neutron stars]{Asai:2016:NSMagneticModes,Lee:2018:AxisymNS,Lee:2018:NSPoloidalToroidal}.

\begin{figure*}
    \centering
    \includegraphics[width=\linewidth]{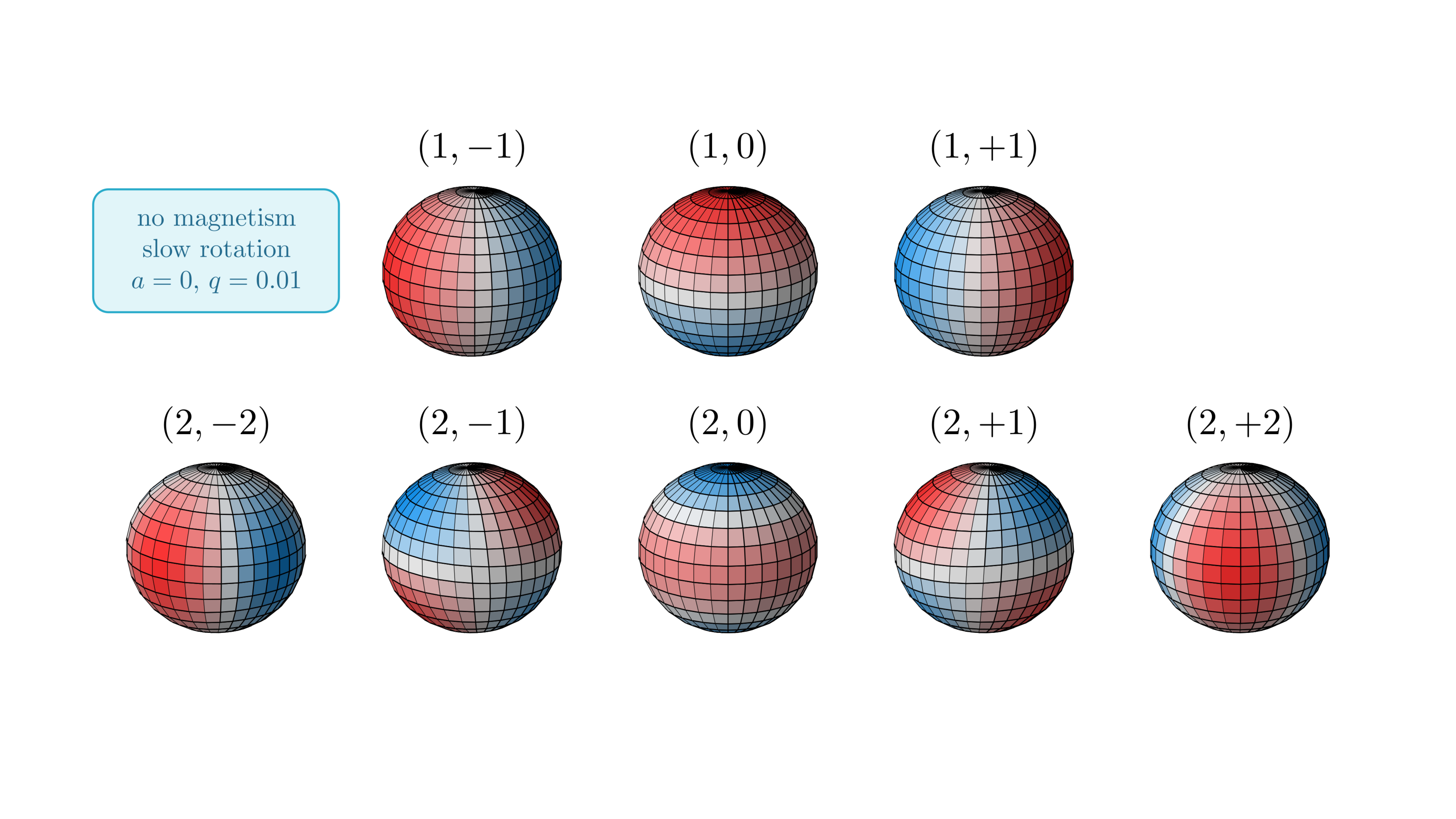}
    \caption{(Still frame of animated figure.) Dipole ($\ell=1$) and quadrupole ($\ell=2$) eigenfunctions of a non-magnetic, slowly rotating ($q=0.01$) star.
    Colors indicate the pressure perturbation, with bluer and redder regions indicating $p'>0$ (i.e., ``hotter'') and $p'<0$ (``colder''), respectively.
    The black wireframe tracks the fluid perturbations as well as the rotation of the star (the poles of the wireframe intersect with the rotation axis).
    Under these conditions, the pressure and fluid-displacement eigenfunctions are approximately scalar harmonics $Y_{\ell m}$ and spheroidal vector spherical harmonics $\bar{\nabla}_hY_{\ell m}$, respectively.
    Full animated figures from this paper can be viewed in the HTML version of the online article, or in a corresponding Zenodo upload at \href{https://zenodo.org/records/20561425}{https://zenodo.org/records/20561425}.}
    \label{fig:NoRotNoMag}
\end{figure*}

Fortunately, this calculation is dramatically simplified by g modes' short radial wavelengths and primarily horizontal fluid displacements.
\citet{Rui:2023:MagneticSuppression} \citepalias[hereafter][]{Rui:2023:MagneticSuppression} show that the effect of strong magnetism can approximately be captured by solving a lower-dimensional eigenproblem over the sphere.
While the angular structures of g modes are simply spherical harmonics in non-rotating, non-magnetic stars (Figure \ref{fig:NoRotNoMag}), \citetalias{Rui:2023:MagneticSuppression} shows that strong magnetic fields change these to modified spherical-harmonic-like angular functions.
The non-perturbative effects of strong magnetic fields is then encoded in the properties of these functions.
\citet{Rui:2024:TARM} \citepalias[hereafter][]{Rui:2024:TARM} extend these calculations to rotating stars, explicitly computing g-mode frequencies under the effect of strong magnetic fields.
This waveguide-like approach to calculating magnetogravity modes has recently led to the discovery of eight red giants with strongly magnetically distorted g-mode doublets \citep{Deheuvels:2026:NearCritical}.

Both \citetalias{Rui:2023:MagneticSuppression} and \citetalias{Rui:2024:TARM} restrict their scopes to dipolar magnetic fields, although their methods can be applied to any axisymmetric magnetic field \citep{Deheuvels:2026:NearCritical}.
While \citetalias{Rui:2024:TARM} also includes rotation, it still assumes that the rotational and magnetic axes are aligned, so that the system is still axisymmetric.
In this work, we relax any condition on the symmetry of the magnetic field.
We apply our new techniques to non-axisymmetric systems (Section \ref{sec:inclined_dipole} and \ref{sec:dipole_plus_y22}), and show that previously ignored effects (i.e., avoided crossings, mixing between polarizations) can be important in the non-axisymmetric case.

%% file: sect_derive_tarm.tex
%!TEX root=./main.tex
\subsection{Derivation of single-polarization waveguide equations} \label{sec:adiabatic_waveguide}

Calculating magnetogravity modes involves finding solutions of linearized fluid equations, which are an eigenvalue problem for a partial differential operator.
Following \citetalias{Rui:2023:MagneticSuppression} and \citetalias{Rui:2024:TARM}, we treat this \edit{eigenvalue problem} as if it \edit{can be approximately separated} into \edit{coupled} radial and horizontal equations \edit{which are individually easier to solve than the full three-dimensional problem}.
This assumption is justified by the fact that the wavevector is primarily radial ($k_r\gg k_h$).
Formally, our technique is equivalent to describing the g-mode cavity as a waveguide.
% This procedure largely extends the works of \citetalias{Rui:2023:MagneticSuppression} and \citetalias{Rui:2024:TARM}.
The basic equations of this approach are described in this Section.

We start from the linearized magneto-Boussinesq equations \citep{Proctor:1982:Magnetoconvection,Mathis:2011:MagneticRotating,Lecoanet:2017:MagneticConversion}.
The fluid displacement field $\vec{\xi}$ obeys a momentum equation:
\begin{equation} \label{eqn:momentum}
    \begin{split}
        \rho_0\partial_t^2\vec{\xi} &+ 2\rho_0\vec{\Omega}\times\partial_t\vec{\xi} \\
        &= -\nabla\left(p' + \frac{1}{4\pi}\vec{B}_0\cdot\vec{B}'\right) - \rho_0N^2\xi_r\hat{r} + \frac{1}{4\pi}\left(\vec{B}_0\cdot\nabla\right)\vec{B}'\mathrm{,}
    \end{split}
\end{equation}
where ``0'' subscripts (primes) denote equilibrium (perturbed) quantities.
Equations \ref{eqn:momentum} describe the dynamics of the perturbation in the corotating frame of the star, which has been assumed to rotate uniformly for simplicity.
We adopt the Cowling approximation, ignoring perturbations to the gravitational potential.
This is justified for g modes of high radial order \citep{Cowling:1941:Approximation}.
In Equation \ref{eqn:momentum} (and hereafter), we neglect all radial derivatives in equilibrium quantities, since for high-radial-order g modes the radial wavelength $\lambda_r=2\pi/k_r$ is much shorter than any equilibrium scale height.

Equation \ref{eqn:momentum} describes the linear dynamics of a conductive, inviscid, incompressible plasma under the effects of stratification (through the Brunt--V\"ais\"al\"a frequency $N$), the Coriolis force, and the magnetic tension.
The Coriolis force is a pseudo force in the rotating frame which causes moving fluid parcels to gyrate around the rotation vector,
\begin{equation} \label{eqn:Omega}
    \vec{\Omega} = \Omega\hat{z} = \Omega\cos\theta\hat{r}-\Omega\sin\theta\hat{\theta}\mathrm{.}
\end{equation}
Similarly, the magnetic tension restores fluid displacements which are perpendicular to the magnetic field,
\begin{equation} \label{eqn:B0}
    \vec{B}_0=B_{0r}(r)\psi(\theta,\phi;r)\hat{r}+\vec{B}_{0h}\mathrm{,}
\end{equation}
where $\psi(\theta,\phi;r)$ is a dimensionless function of $r$, $\theta$, and $\phi$ which describes the horizontal dependence of the radial component of $\vec{B}_0$.
Because this work focuses on the horizontal structure of magnetogravity waves at fixed radius $r$, we hereafter write $\psi=\psi(\theta,\phi)$ and suppress the radial argument $r$ when not relevant for the discussion.
Since fluid displacements in this problem are primarily horizontal, we will neglect the horizontal components of Equations \ref{eqn:Omega} and \ref{eqn:B0}:
\begin{subequations} \label{eqn:TARM}
    \begin{gather}
        \vec{\Omega} \simeq \Omega\cos\theta\hat{r} \\
        \vec{B}_0 \simeq B_{0r}(r)\psi(\theta,\phi;r)\hat{r}\mathrm{.}
    \end{gather}
\end{subequations}
Equations \ref{eqn:TARM} are the \textit{traditional approximation of rotation and magnetism (TARM)}, introduced by \citetalias{Rui:2024:TARM} in analogy with the traditional approximation of rotation \edit{\citep{Eckart:1960:TAR,Berthomieu:1978:TAR,Bildsten:1996:OceanGModes,Lee:1997:TAR}}.
In this work, we normalize $\psi(\theta,\phi;r)$ such that
\begin{equation} \label{eqn:psi2_norm}
    \int\psi^2\sin\theta\,\mathrm{d}\theta\,\mathrm{d}\phi=4\pi\mathrm{,}
\end{equation}
which simplifies many formulae.
However, we caution that this choice differs from the normalization conventions in \citetalias{Rui:2023:MagneticSuppression} and \citetalias{Rui:2024:TARM}.
In particular, unlike in those works, resonance of the magnetogravity wave with Alfv\'en waves \edit{occurs at $b=k_rv_{Ar}/\omega=1/\mathrm{max}(|\psi|)$ rather than $b=1$ under this definition except in the unphysical case of a monopolar field}.

As the aim of this work is to calculate the properties of standing magnetogravity waves, we assert that all perturbations have harmonic time dependence, i.e., that they are proportional to $e^{i\omega t}$, so that the identification
\begin{equation}
    \partial_t\rightarrow i\omega
\end{equation}
can be made.
In the ideal magnetohydrodynamic limit, the magnetic field perturbation $\vec{B}'$ is constrained by the linearized induction equation:
\begin{equation} \label{eqn:induction}
    \vec{B}' = \left(\vec{B}_0\cdot\nabla\right)\vec{\xi}\mathrm{.}
\end{equation}
\edit{As in Equation \ref{eqn:momentum}, Equation \ref{eqn:induction} neglects a subdominant term involving a derivative in $\vec{B}_0$.}

Under these conditions, the radial and horizontal components of Equation \ref{eqn:momentum} can be written as
\begin{subequations} \label{eqn:momentum_sep}
    \begin{equation} \label{eqn:momentum_r}
        -\rho_0\omega^2\xi_r = -\partial_rp' - \rho_0N^2\xi_r
    \end{equation}
    \begin{equation} \label{eqn:momentum_h}
        -\rho_0\omega^2\vec{\xi}_h + 2i\rho_0\omega\Omega\cos\theta\,\hat{r}\times\vec{\xi}_h = -\nabla_hp' + \frac{B_{0r}^2\psi^2}{4\pi}\partial_r^2\vec{\xi}_h\mathrm{.}
    \end{equation}
\end{subequations}
\edit{Equation \ref{eqn:momentum_r} ignores the magnetic stress, which is assumed to be much weaker than the buoyancy term $\propto N^2$.
Equation \ref{eqn:momentum_h} ignores the magnetic pressure term, which is exceptionally small compared to other terms ($\sim(\xi_r/\xi_h)^2$).}
Equation \ref{eqn:momentum_h} can be rearranged to
\begin{equation} \label{eqn:final_momentum}
    \frac{1}{\rho_0}\nabla_hp' = \omega^2\vec{\xi}_h + v_{Ar}^2\psi^2\partial_r^2\vec{\xi}_h - 2i\omega\Omega\cos\theta\,\hat{r}\times\vec{\xi}_h\mathrm{,}
\end{equation}
where $v_{Ar}\equiv B_{0r}/\sqrt{4\pi\rho_0}$ is the radial Alfv\'en speed, excluding the dimensionless $\psi$ factor.

Under the Boussinesq approximation (valid for asymptotic gravity waves), the continuity equation reads:
\begin{equation} \label{eqn:continuity}
    \nabla\cdot\vec{\xi} = 0\mathrm{,}
\end{equation}
such that the fluid motions are treated as approximately incompressible.
Using $N\gg\omega$, Equation \ref{eqn:momentum_r} relates $\xi_r$ to $p'$:
\begin{equation}
    \xi_r = -\frac{1}{\rho_0N^2}\partial_rp'\mathrm{,}
\end{equation}
so that Equation \ref{eqn:continuity} becomes
\begin{equation} \label{eqn:final_continuity}
    -\frac{1}{\rho_0N^2}\partial_r^2p' + \nabla_h\cdot\vec{\xi}_h = 0\mathrm{.}
\end{equation}

From here, we impose a Jeffreys--Wentzel--Kramers--Brillouin (JWKB) ansatz:
\begin{subequations} \label{eqn:jwkb_ansatz}
    \begin{gather}
        p' = \rho_0\omega^2r^2\sum_\alpha A_\alpha(r)\pi_\alpha(\theta,\phi;r)e^{-iS_\alpha(r)/\epsilon} \\
        \vec{\xi}_h = r\sum_\alpha A_\alpha(r)\vec{\zeta}_\alpha(\theta,\phi;r)e^{-iS_\alpha(r)/\epsilon}\mathrm{,}
    \end{gather}
\end{subequations}
where $\epsilon\sim\mathcal{O}(1/k_rr)$ is the small parameter in the JWKB expansion.
Equations \ref{eqn:jwkb_ansatz} decompose $p'$ and $\vec{\xi}_h$ into radially propagating magnetogravity ``polarizations'' $(\pi_\alpha,\vec{\zeta}_\alpha)$ labeled by $\alpha$ (so far undetermined), each with radially varying amplitudes $A_\alpha$.
At zero magnetic field and slow rotation, the polarizations $(\pi_\alpha,\vec{\zeta}_\alpha)$ become $(Y_{\ell m},\bar{\nabla}_hY_{\ell m})$ and can be indexed by the usual spherical harmonic quantum numbers $(\ell,m)$.
Our ansatz that $\pi_\alpha$ and $\vec{\zeta}_\alpha$ share an amplitude function $A_\alpha$ is also motivated by the non-rotating, non-magnetic limit, in which this assumption is exact after separation of variables with respect to these basis functions under the Cowling approximation.
The prefactors $\rho_0\omega^2r^2$ and $r$ in Equations \ref{eqn:jwkb_ansatz} are chosen to ensure that $\pi_\alpha$ and $\vec{\zeta}_\alpha$ are dimensionless.
Each polarization rapidly oscillates in the radial direction due to the factor of $e^{-iS_\alpha/\epsilon}$, where the JWKB action $S_\alpha$ is related to the radial wavenumber:
\begin{equation}
    k_r = S_\alpha'/\epsilon\mathrm{,}
\end{equation}
where hereafter primes denote differentiation by $r$.

Substituting Equations \ref{eqn:jwkb_ansatz} into Equations \ref{eqn:final_momentum} and \ref{eqn:final_continuity} gives
\begin{subequations} \label{eqn:jwkb_leading}
    \begin{equation}
        \sum_\alpha A_\alpha e^{-iS_\alpha/\epsilon}\left\lbrace\left(1 - b_\alpha^2\psi^2\right)\vec{\zeta}_\alpha - iq\mu\,\hat{r}\times\vec{\zeta}_\alpha - \bar{\nabla}_h\pi_\alpha\right\rbrace = 0
    \end{equation}
    \begin{equation}
        \sum_\alpha A_\alpha e^{-iS_\alpha/\epsilon}\left\lbrace\frac{\omega^2r^2k_{r,\alpha}^2}{N^2}\pi_\alpha + \bar{\nabla}_h\cdot\vec{\zeta}_\alpha\right\rbrace = 0\mathrm{,}
    \end{equation}
\end{subequations}
where $b_\alpha=k_{r,\alpha}v_{Ar}/\omega$, $q=2\Omega/\omega$, $\bar{\nabla}_h=r\nabla_h$, and $\mu=\cos\theta$.
\edit{We note in passing that, while this work is largely inspired by red giant cores for which $q\sim10^{-3}$ is small, our formalism should also apply to more rapidly-rotating g-mode pulsators (e.g., $\gamma$ Doradus or slowly pulsating B-type pulsators) for which $q$ can be much higher.}
Currently, Equations \ref{eqn:jwkb_leading} only retain leading-order terms in the JWKB expansion ($\sim\mathcal{O}(\epsilon^{-2})$), although we show in Section \ref{sec:higher_wkb} that higher-order terms are often important.

While the polarizations $(\pi_\alpha,\vec{\zeta}_\alpha)$ can be chosen freely, it is convenient to choose them such that Equation \ref{eqn:jwkb_leading} is trivially satisfied term by term.
Hereafter, we therefore define polarizations which solve a set of \textit{transverse equations}:
\begin{subequations} \label{eqn:transverse}
    \begin{gather}
        \left(1 - b_\alpha^2\psi^2\right)\vec{\zeta}_\alpha - iq\mu\,\hat{r}\times\vec{\zeta}_\alpha - \bar{\nabla}_h\pi_\alpha = 0 \\
        \lambda_\alpha\pi_\alpha + \bar{\nabla}_h\cdot\vec{\zeta}_\alpha = 0\mathrm{,}
    \end{gather}
\end{subequations}
where $\lambda_\alpha$, the eigenvalue of the problem, defines the dispersion relation for the polarization indexed by $\alpha$:
\begin{equation} \label{eqn:disp_rel}
    \omega^2 = \frac{\lambda_\alpha/r^2}{k_{r,\alpha}^2}N^2\mathrm{.}
\end{equation}
\edit{By writing $p'$ and $\vec{\xi}_h$ in terms of these polarizations, the radial variation of the angular structure of the oscillation mode is relegated to next-to-leading order in the JWKB expansion.}
Equations \ref{eqn:transverse} define a differential eigenvalue problem on the sphere, given the dimensionless magnetic and spin parameters $b_\alpha$ and $q$, as well as the dimensionless function $\psi^2$ which describes the angular structure of the magnetic field at a given radius.
This problem can be solved for the magnetogravity polarizations, which are described by $\lambda_\alpha$ and $(\pi_\alpha,\vec{\zeta}_\alpha)$.
Crucially, Equations \ref{eqn:transverse} can be solved separately from the radial eigenvalue problem, and therefore does not directly rely on the stellar structure.
This is a key advantage of the waveguide description to solving for the properties of magnetogravity waves.

Problematically, $b_\alpha=k_{r,\alpha}v_{Ar}/\omega$ depends on $k_{r,\alpha}$, which is not known in advance and needs to be solved post hoc using the dispersion relation (Equation \ref{eqn:disp_rel}).
In \citetalias{Rui:2023:MagneticSuppression} and \citetalias{Rui:2024:TARM}, after solving Equations \ref{eqn:transverse} \edit{as a function of $q$ and $b_\alpha$}, this problem is circumvented by retroactively considering $\lambda_\alpha$ to be a function of \edit{$q$ and}
\begin{equation} \label{eqn:a_param}
    a = \frac{b_\alpha}{\sqrt{\lambda_\alpha}} = \left(\frac{N}{\omega}\right)\left(\frac{v_{Ar}/r}{\omega}\right)\mathrm{,}
\end{equation}
which does \textit{not} depend on $\alpha$ \edit{(see Section 3.2 of \citetalias{Rui:2024:TARM} for a more detailed discussion)}.
\edit{If Equations \ref{eqn:transverse} are solved for $\lambda_\alpha$ on a rectangular grid in $q$ and $b_\alpha$ (as is done both in this work and \citetalias{Rui:2023:MagneticSuppression} and \citetalias{Rui:2024:TARM}), the reparameterization in Equation \ref{eqn:a_param} gives $\lambda_\alpha$ on a \textit{non-rectangular} grid in $q$ and $a$.}
Note that $a\sim B_r/B_{r,\mathrm{crit}}\sim\omega_B^2/\omega^2$, where $B_{r,\mathrm{crit}}$ and $\omega_B$ are defined in Equations \ref{eqn:Brcrit} and \ref{eqn:omegaB} and are related to the conditions for magnetic gravity-wave suppression \citep{Fuller:2015:SuppressedDipole}.

\edit{Although Equations \ref{eqn:jwkb_leading} are only the leading-order terms in the JWKB expansion, they exactly satisfy Equations \ref{eqn:final_momentum} and \ref{eqn:continuity} when the radial wavelength formally approaches zero} ($k_{r,\alpha}\rightarrow\infty$).
\edit{In this limit, a magnetogravity wave consisting of a single polarization $\alpha$ ($A_\beta(r_0)=\delta_{\alpha\beta}$) will remain totally within that polarization as it propagates.
This remains true even as the eigenfunctions $(\pi_\alpha,\vec{\zeta}_\alpha)$ describing that polarization change with radius with the background quantities $a=a(r)$ and $\psi=\psi(\theta,\phi;r)$.
Under such single-polarization propagation, a magnetogravity wave smoothly deforms to match the conditions within each radial shell.
This is a consequence of the ``adiabatic theorem'' in quantum mechanics \citep{Born:1928:AdiabaticTheorem}.
We caution that, throughout this work, the adjective ``adiabatic'' is used to describe wave propagation obeying the adiabatic theorem.
This is entirely distinct from the usual definition of ``adiabatic'' in stellar pulsations, in reference to the assumption that a fluid parcel does not exchange heat with its surroundings (all pulsations in this work are adiabatic in this sense).
}
% , Equations \ref{eqn:jwkb_leading} exactly represent Equations \ref{eqn:final_momentum} and \ref{eqn:final_continuity}, which are satisfied exactly by each polarization individually via Equations \ref{eqn:transverse}.
% Under these conditions, the adiabatic theorem ensures that a magnetogravity wavepacket initialized in a single polarization $\alpha$ ($A_\beta(r_0)=\delta_{\alpha\beta}$) will fully remain within polarization $\alpha$, even as its eigenfunctions $(\pi_\alpha,\vec{\zeta}_\alpha)$ smoothly deform as the local values of $a$ and $\psi(\theta,\phi;r)$ experienced by the wavepacket vary \citep{Born:1928:AdiabaticTheorem}.

\edit{Under adiabatic propagation, t}he mode frequencies can be calculated by enforcing a quantization condition in the radial direction:
\begin{equation} \label{eqn:radial_quantization}
    \pi(n+\epsilon_g) = \int_{\mathcal{R}}k_{r,\alpha}(r')\,\mathrm{d}r'\mathrm{.}
\end{equation}
The domain $\mathcal{R}$ in Equation \ref{eqn:radial_quantization} denotes the g-mode cavity, the contiguous range of radii for which $\omega<N$ and $\omega<L_\alpha=\sqrt{\lambda_\alpha}c_s/r$, where the Lamb frequency $L_\alpha$ is related to the sound speed $c_s$.
We caution that this procedure is only well defined for magnetogravity waves which do not undergo conversion to slow magnetic or Alfv\'en waves, so that the eigenvalues $\lambda$ remain real and magnetic suppression does not occur \citepalias[such as in, e.g.,][]{Rui:2023:MagneticSuppression}.

% Note that, throughout this work, we use the adjective ``adiabatic'' to describe wave propagation obeying the adiabatic theorem in the sense above.
% An adiabatically propagating magnetogravity wave follows a single polarization, even as the eigenfunctions $(\pi_\alpha,\vec{\zeta}_\alpha)$ describing that polarization change with radius.
% This is distinct from the usual definition of ``adiabatic'' in stellar pulsations, which refers to the assumption that a fluid parcel does not exchange heat with its surroundings.
% As we consider no explicit damping processes in this study, the modes we investigate in this work are all ``adiabatic'' in the latter sense.

Equations \ref{eqn:jwkb_leading} is derived by neglecting higher-order JWKB terms in Equations \ref{eqn:final_momentum} and \ref{eqn:final_continuity}.
This is the \edit{aforementioned} limit of adiabatic wave propagation, which is implicitly assumed by \citetalias{Rui:2023:MagneticSuppression} and \citetalias{Rui:2024:TARM}, as well as \citet{Lecoanet:2017:MagneticConversion}, \citet{Lecoanet:2022:HD43317}, and \citet{David:2025:Magnetogravity}, which are methodologically similar.
\edit{In Section \ref{sec:higher_wkb}, we show that non-adiabatic (multi-polarization) propagation cannot be ignored in general.}
Because non-adiabatic propagation is significantly more complicated, \edit{this study only describes magnetogravity-mode frequency prediction under single-polarization (``adiabatic'') propagation.
Theoretical predictions for mode frequencies under the more general multi-polarization case are deferred to future studies.
}

%% file: sect_methods_numerics.tex
%!TEX root=./main.tex
\subsection{Numerical formulation}

We perform a sparse solution of the two-dimensional eigenproblem defined in Equations \ref{eqn:transverse} using \texttt{Dedalus} (version 3), a general-purpose spectral code for solving partial differential equations \citep{Burns:2020:Dedalus}.
Hereafter, for conciseness, we omit the polarization subscript $\alpha$, unless the presence of multiple magnetogravity polarizations is relevant.

To reduce all unknown fields to scalar fields, we perform a Helmholtz decomposition of $\vec{\zeta}$ into curl-free and divergence-free components:
\begin{equation} \label{eqn:helmholtz}
    \vec{\zeta} = \bar{\nabla}_h\Phi + \hat{r}\times\bar{\nabla}_h\Psi\mathrm{,}
\end{equation}
where $\Phi$ and $\Psi$ are proportional to the potential and stream functions generating the flow described by $\vec{\zeta}$.
Upon substitution of Equation \ref{eqn:helmholtz}, Equations \ref{eqn:transverse} become
\begin{subequations} \label{eqn:coordinate_free_helmholtz}
    \begin{align}
        &\lambda\pi + \bar{\nabla}_h^2\Phi = 0 \\
        (1-b^2\psi^2)\bar{\nabla}_h\Phi &+ (1-b^2\psi^2)\hat{r}\times\bar{\nabla}_h\Psi \\
        - &iq\mu\hat{r}\times\bar{\nabla}_h\Phi + iq\mu\bar{\nabla}_h\Psi - \bar{\nabla}_h\pi = 0\mathrm{.} \nonumber
    \end{align}
\end{subequations}
Each distinct eigenvalue and eigenvector of Equations \ref{eqn:coordinate_free_helmholtz} corresponds to a distinct polarization $\alpha$.
In spherical-polar coordinate form, these become the equations we solve:
\begin{subequations} \label{eqn:helmholtz_numerical}
    \begin{gather}
        \lambda s^2\pi + s^2\partial_\theta^2\Phi + s\mu\partial_\theta\Phi + \partial_\phi^2\Phi = 0 \\
        s D_b\partial_\theta\Phi - D_b\partial_\phi\Psi + iq\mu\partial_\phi\Phi + iqs\mu\partial_\theta\Psi - s\partial_\theta\pi = 0 \\
        D_b\partial_\phi\Phi + s D_b\partial_\theta\Psi - iqs\mu\partial_\theta\Phi + iq\mu\partial_\phi\Psi - \partial_\phi\pi = 0\mathrm{,}
    \end{gather}
\end{subequations}
where \edit{$s=\sin\theta$ and $D_b=1-b^2\psi^2$.}

Although \texttt{Dedalus} supports decomposition into spherical harmonics for problems on the unit sphere ($S^2$), presently the code does not allow this basis to be used for differential equations whose coefficient functions (``non-constant coefficients'') depend on $\phi$.
Instead, we solve our problem over a ``Cartesian'' domain in $\theta$ and $\phi$, using a Chebyshev basis in $\theta$ and Fourier basis in $\phi$.
Specifically, we decompose our unknown fields as
\begin{subequations} \label{eqn:coord_form}
    \begin{gather}
        \pi = \sum_{nm}\bar{\pi}_{nm}T_n(2\theta/\pi-1)e^{im\phi} \\
        \Phi = \sum_{nm}\bar{\Phi}_{nm}T_n(2\theta/\pi-1)e^{im\phi} \label{eqn:phi_coord} \\
        \Psi = \sum_{nm}\bar{\Psi}_{nm}T_n(2\theta/\pi-1)e^{im\phi}\mathrm{,} \label{eqn:psi_coord}
    \end{gather}
\end{subequations}
and solve for the unknown spectral coefficients $\bar{\pi}_{nm}$, $\bar{\Phi}_{nm}$, and $\bar{\Psi}_{nm}$ together with $\lambda$.
These coefficients are then converted to spherical harmonic coefficients (indexed by $\ell$ and $m$) for ease of interpretation, i.e.,
\begin{subequations}
    \begin{gather}
        \pi = \sum_{\ell m}\pi_{\ell m}Y_{\ell m}(\theta,\phi) \\
        \Phi = \sum_{\ell m}\Phi_{\ell m}Y_{\ell m}(\theta,\phi) \\
        \Psi = \sum_{\ell m}\Psi_{\ell m}Y_{\ell m}(\theta,\phi)\mathrm{.}
    \end{gather}
\end{subequations}
The spatial resolution of the numerical solution is determined by $N_\theta$ and $N_\phi$, the number of basis functions in the $\theta$ and $\phi$ directions, respectively.
In this work, we use $(N_\theta,N_\phi)=(32,64)$ for our main numerical solutions.
The axisymmetric case from \citetalias{Rui:2023:MagneticSuppression} and \citetalias{Rui:2024:TARM} can be recovered by choosing $\psi$ to be axisymmetric.
This allows the identification $\partial_\phi=im$ for a chosen azimuthal order $m$, which reduces Equations \ref{eqn:coordinate_free_helmholtz} to an ordinary differential eigenproblem for each $m$.
For these axisymmetric solutions, we use $N_\theta=256$.

The generating fields $\Phi$ and $\Psi$ only appear under derivatives in Equations \ref{eqn:coordinate_free_helmholtz}.
They thus possess gauge freedom which, on the unit sphere, means that constant offsets to $\Phi$ and $\Psi$ are physically irrelevant.
This gauge freedom appears in two-dimensional solutions of Equations \ref{eqn:transverse}, as well as one-dimensional solutions where $m=0$.
To remove these undesired degrees of freedom, we thus enforce the gauge-fixing conditions
\begin{subequations}
    \begin{gather}
        \iint\Phi\sin\theta\,\mathrm{d}\theta\,\mathrm{d}\phi = 0 \\ \label{eqn:fix_T}
        \iint\Psi\sin\theta\,\mathrm{d}\theta\,\mathrm{d}\phi = 0 \\ \label{eqn:fix_F}
    \end{gather}
\end{subequations}
by adding ``tau terms'' to Equations \ref{eqn:phi_coord} and \ref{eqn:psi_coord} \edit{\citep[see Section II of][]{Burns:2020:Dedalus}}.

We note several subtleties in our implementation.
% First, \texttt{Dedalus} normalizes the Chebyshev polynomials $T_n$ to be orthonormal, i.e., $T_0$ contains an extra factor of $1/\sqrt{2}$ relative to $T_{n\neq0}$ when compared to the standard definition.
First, we choose the independent variable to be $\theta$ instead of the more natural-looking $\mu=\cos\theta$, because solutions with respect to the latter can diverge in slope near the poles and therefore cannot easily be accommodated by the smooth choice of basis.
For example, near the poles, spherical harmonics scale as
\begin{equation} \label{eqn:regularity}
    Y_{\ell m} \propto P_{\ell m}(\mu) \simeq (1-\mu^2)^{|m|/2} \simeq \sin^{|m|}\theta \simeq \theta^{|m|}\mathrm{.}
\end{equation}
They are therefore are differentiable near the poles when $\theta$ is used as a coordinate, but \textit{not} when $\mu$ is used instead.
The scaling of $\pi$, $\Phi$, and $\Psi$ for any magnetic field and rotation rate are also guaranteed by spherical-coordinate regularity conditions to behave as in Equation \ref{eqn:regularity} near the poles.
Second, while we do not enforce regularity conditions near the poles, the solver is found to automatically enforce these conditions due to the geometric prefactors appearing in Equations \ref{eqn:coord_form}.
Finally, at high values of $\ell$, our Cartesian grid does not represent all values of $m$ equally, so high-frequency spectral content must be interpreted with care, or discarded altogether.

%% file: sect_comparison_to_pert.tex
%!TEX root=./main.tex

\subsection{Relationship between waveguide method and perturbation theory} \label{sec:perturbation_theory}

In the literature, the standard approach to predicting the behavior of observable magnetogravity waves is to use first-order perturbation theory \citep{Gomes:2020:MagneticRG,Bugnet:2021:MagneticI,Li:2022:30to100kG,Das:2024:ComplexMagnetic}, summarized here for pure g modes (mixing parameter $\zeta=1$, though not to be confused with $\vec{\zeta}_\alpha$).
%most closely following the notation of Appendix C of \citet{Rui:2025:StochasticOblique}.
This perturbative approach hypothesizes that the magnetic field and rotation are ``weak'' in a sense which we discuss further in Section \ref{sec:higher_wkb}.

The unperturbed system has a $(2\ell+1)$-fold degeneracy at fixed $n$ and $\ell$, since modes of differing $m$ are guaranteed to have identical frequencies by the spherical symmetry of the system.
In degenerate perturbation theory, the presence of a weak perturbation selects privileged linear combinations of these degenerate eigenfunctions,
\begin{equation} \label{eqn:xi_linear_combo}
    \vec{\xi}^{(0)} = \sum^{+\ell}_{m=-\ell}a_{n\ell m}\vec{\xi}_{n\ell m}^{(0)}\mathrm{,}
\end{equation}
whose coefficients $a_{n\ell m}$, arranged into a vector $\vec{a}_{n\ell}$, must be solved for simultaneously with the frequency shift $\delta\omega$ in the following eigenvalue problem:
\begin{equation}
    (\delta\omega/\omega_{0,n\ell})\,\vec{a}_{n\ell} = (\mathbf{M}_{n\ell} + \mathbf{R}_{n\ell})\vec{a}_{n\ell}\mathrm{.}
\end{equation}
The frequency shifts are with respect to an unperturbed frequency $\omega_{0,n\ell}$.
The matrices $\mathbf{M}_{n\ell}$ and $\mathbf{R}_{n\ell}$ describe the effects of magnetism and rotation, respectively.
Hereafter, we consider the limit of large $n$, within which $\mathbf{M}_{n\ell}$ and $\mathbf{R}_{n\ell}$ become independent of $n$ so that their subscripts $n$ can be omitted.

Specializing to dipole modes ($\ell=1$) for which $\vec{a}_{\ell=1}=(a_{1,-1},a_{1,0},a_{1,+1})$, $\mathbf{R}_{\ell=1}$ is simply the diagonal matrix
\begin{equation}
    \mathbf{R}_{\ell=1} = (\delta\omega_{\mathrm{rot}}^{\ell=1}/\omega_{0,n\ell})\,\mathrm{diag}(-1, 0, +1)
\end{equation}
at first order in the rotation rate, where
\begin{equation} \label{eqn:delta_omega_rot}
    \frac{\delta\omega_{\mathrm{rot}}^{\ell=1}}{\omega_{0,n\ell}} = \frac{\Omega}{2\omega_{0,n\ell}} = \frac{1}{4}q\mathrm{.}
\end{equation}
The magnetic matrix $\mathbf{M}_{n,\ell=1}$ is given by
\begin{equation} \label{Mmatrixgeneral}
    \mathbf{M}_{\ell=1} = (\delta\omega_{\mathrm{mag}}^{\ell=1}/\omega_{0,n\ell})\int_{\mathcal{R}}\mathrm{d}r\,K(r)\iint\sin\theta\,\mathrm{d}\theta\,\mathrm{d}\phi\,\mathcal{M}_{\ell=1}\psi^2\mathrm{.}
\end{equation}
In Equation \ref{Mmatrixgeneral}, the g-mode cavity $\mathcal{R}$ is made up of contiguous shells satisfying $\omega<N,L_\ell$, where $L_\ell=\sqrt{\ell(\ell+1)}c_s/r$ is the usual Lamb frequency.
The radial integral is weighted by a function $K(r)$ given by
\begin{equation} \label{eqn:K_weight}
    K(r) \simeq
    \begin{cases}
        \frac{N^3/\rho_0r^3}{\int_{\mathcal{R}}(N^3/\rho_0r^3)\,\mathrm{d}r} & \mathrm{inside}\,\mathcal{R} \\
        0 & \mathrm{otherwise,}
    \end{cases}
\end{equation}
and the matrix weight within the angular integral is
\begin{equation} \label{matrixweight}
    \mathcal{M}_{\ell=1}(\theta,\phi) = \frac{3}{8}
    \begin{pmatrix}
        3+C_\theta & -\sqrt{2}e^{i\phi}S_\theta & e^{2i\phi}(1-C_\theta) \\
        -\sqrt{2}e^{-i\phi}S_\theta & 2-2C_\theta & \sqrt{2}e^{i\phi}S_\theta \\
        e^{-2i\phi}(1-C_\theta) & \sqrt{2}e^{-i\phi}S_\theta & 3+C_\theta \\
    \end{pmatrix}\mathrm{,}
\end{equation}
where we have abbreviated $S_\theta\equiv\sin2\theta$ and $C_\theta\equiv\cos2\theta$.
Magnetic frequency shifts are scaled by $\delta\omega_{\mathrm{mag}}^{\ell=1}$, which is given by
\begin{equation} \label{eqn:delta_omega_mag_expr}
    \frac{\delta\omega_{\mathrm{mag}}^{\ell=1}}{\omega_{0,n\ell}} = \frac{\mathscr{I}}{4\pi\omega_{0,n\ell}^4}\langle B_r^2\rangle\mathrm{,}
\end{equation}
where $\mathscr{I}$ is a stellar-structure parameter
\begin{equation} \label{eqn:curlyI}
    \mathscr{I} = \frac{\int_{\mathcal{R}}(N^3/\rho_0r^3)\,\mathrm{d}r}{\int_{\mathcal{R}}(N/r)\,\mathrm{d}r}
\end{equation}
and $\langle B_r^2\rangle$ is an averaged squared magnetic field strength weighted by $K(r)$:
\begin{equation} \label{eqn:avg_Br2}
    \langle B_r^2\rangle = \int_{\mathcal{R}}\mathrm{d}r\,K(r)B_{0r}^2(r)\mathrm{.}
\end{equation}

For general $\ell$, it is well known \citep[e.g.,][]{Ledoux:1951:Criterion} that
\begin{equation} \label{eqn:delta_omega_rot_ell}
    \frac{\delta\omega_{\mathrm{rot}}^\ell}{\omega_{0,\ell}} = \frac{\Omega}{\ell(\ell+1)\omega_{0,n\ell}} = \frac{1}{2\ell(\ell+1)}q\mathrm{.}
\end{equation}
In Appendix \ref{app:delta_mag_ell}, we also show that Equation \ref{eqn:delta_omega_mag_expr} generalizes to
\begin{equation} \label{eqn:delta_omega_mag_ell}
    \frac{\delta\omega_{\mathrm{mag}}^\ell}{\omega_{0,n\ell}} = \frac{\ell(\ell+1)\mathscr{I}}{8\pi\omega_{0,n\ell}^4}\langle B_r^2\rangle
\end{equation}
for arbitrary $\ell$.

Taken at face value, results from first-order perturbation theory are not readily comparable to those from our single-polarization waveguide description.
This is because the former predicts properties of modes in the full three-dimensional g-mode cavity, whereas the latter predicts the behavior of magnetogravity wave polarizations on a two-dimensional spherical slice through the g-mode cavity.
Although this paradigmic difference has deeper consequences (Section \ref{sec:higher_wkb}), for present purposes we can still directly compare the results of these two methods by applying perturbation theory to a thin, uniform cavity within which $N$, $B_{0r}$, $\rho_0$, and $r$ are approximately constant.
For such a cavity,
\begin{equation} \label{eqn:delta_omega_mag}
    \frac{\delta\omega_{\mathrm{mag}}^\ell}{\omega_{0,n\ell}} \approx \frac{\ell(\ell+1)\omega_B^4}{2\omega_{0,n\ell}^4} = \frac{\ell(\ell+1)}{2}a^2\mathrm{,}
\end{equation}
and the coefficients $a_{n\ell m}$ recovered by perturbation theory are the spherical harmonic coefficients of the polarization solutions of Equations \ref{eqn:transverse}.
The eigenvalue $\delta\omega/\omega_{0,n\ell}$ in perturbation theory can be compared to the polarization eigenvalue $\lambda$ by noting that the dispersion relation (Equation \ref{eqn:disp_rel}) implies that $\omega\propto\sqrt{\lambda}$ when other quantities are held fixed.
This implies for weak perturbations that \edit{$\delta\omega/\omega_{0,n\ell}\approx\delta\lambda/2\lambda_0=\delta\lambda/2\ell(\ell+1)$}, which we use to define
\begin{equation} \label{eqn:lambda_pert}
    \lambda^{\mathrm{pert}} = \ell(\ell+1) + \delta\lambda = \ell(\ell+1)\left(1 + 2\delta\omega/\omega_{0,n\ell}\right)
\end{equation}
as a perturbative estimate of $\lambda$.

%% file: sect_inclined_dipole.tex
%!TEX root=./main.tex
\subsection{Inclined dipole field} \label{sec:inclined_dipole}

\begin{figure*}
    \centering
    \includegraphics[width=\linewidth]{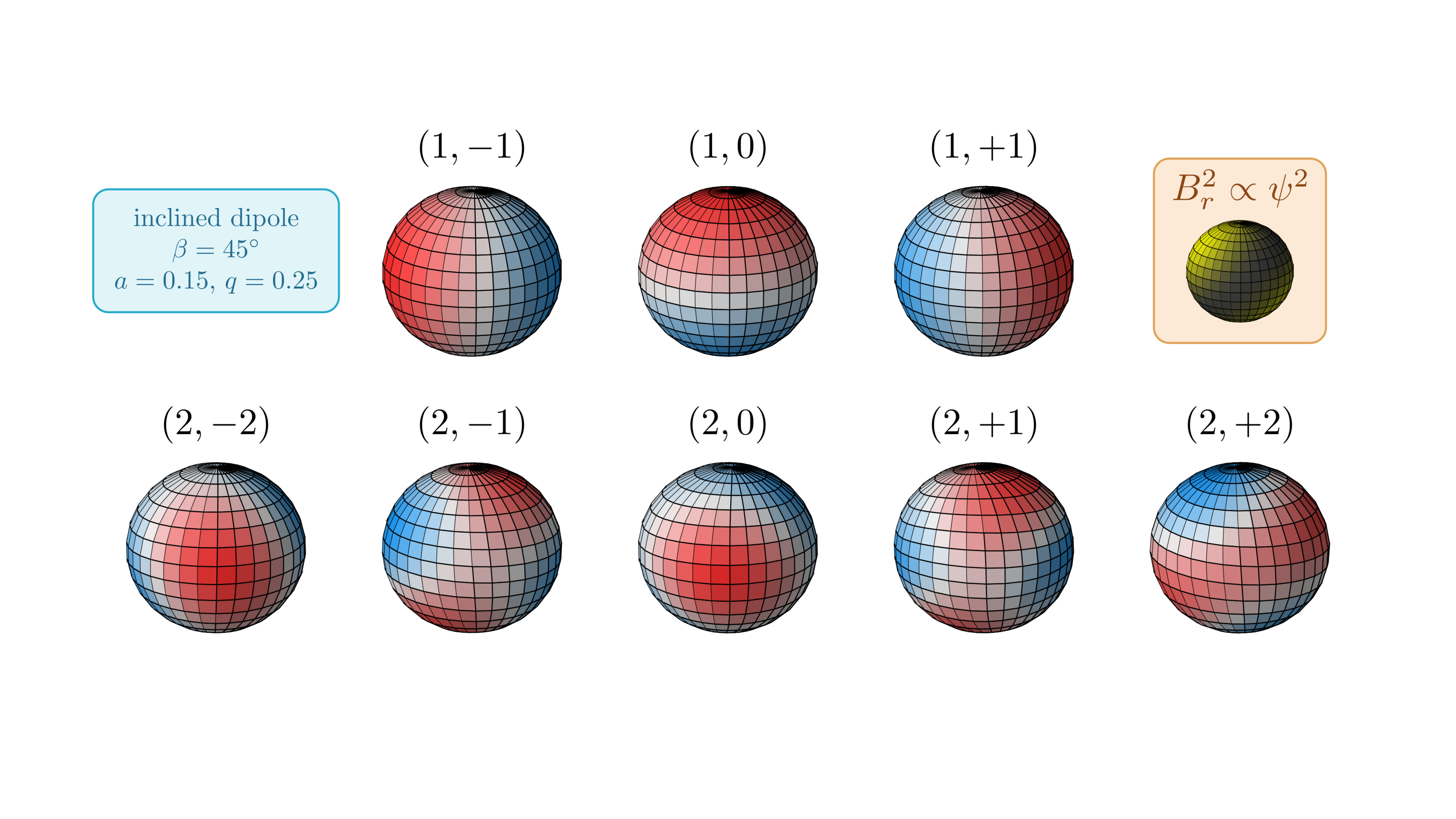}
    \caption{(Still frame of animated figure.)
    Same as Figure \ref{fig:NoRotNoMag}, but for dipole ($\ell=1$) and quadrupole ($\ell=2$) eigenfunctions under an inclined dipolar field with $\beta=45\degree$, with magnetic parameter $a=0.15$ and spin parameter $q=0.25$.
    The mode labels indicate the values of $(\ell,m)$ of the spherical harmonics approached if the eigenvalue branches approach $q\rightarrow0$ at fixed $a=0$.}
    \label{fig:IncDip45_DR_QM}
\end{figure*}

\begin{figure*}
    \centering
    \includegraphics[width=\linewidth]{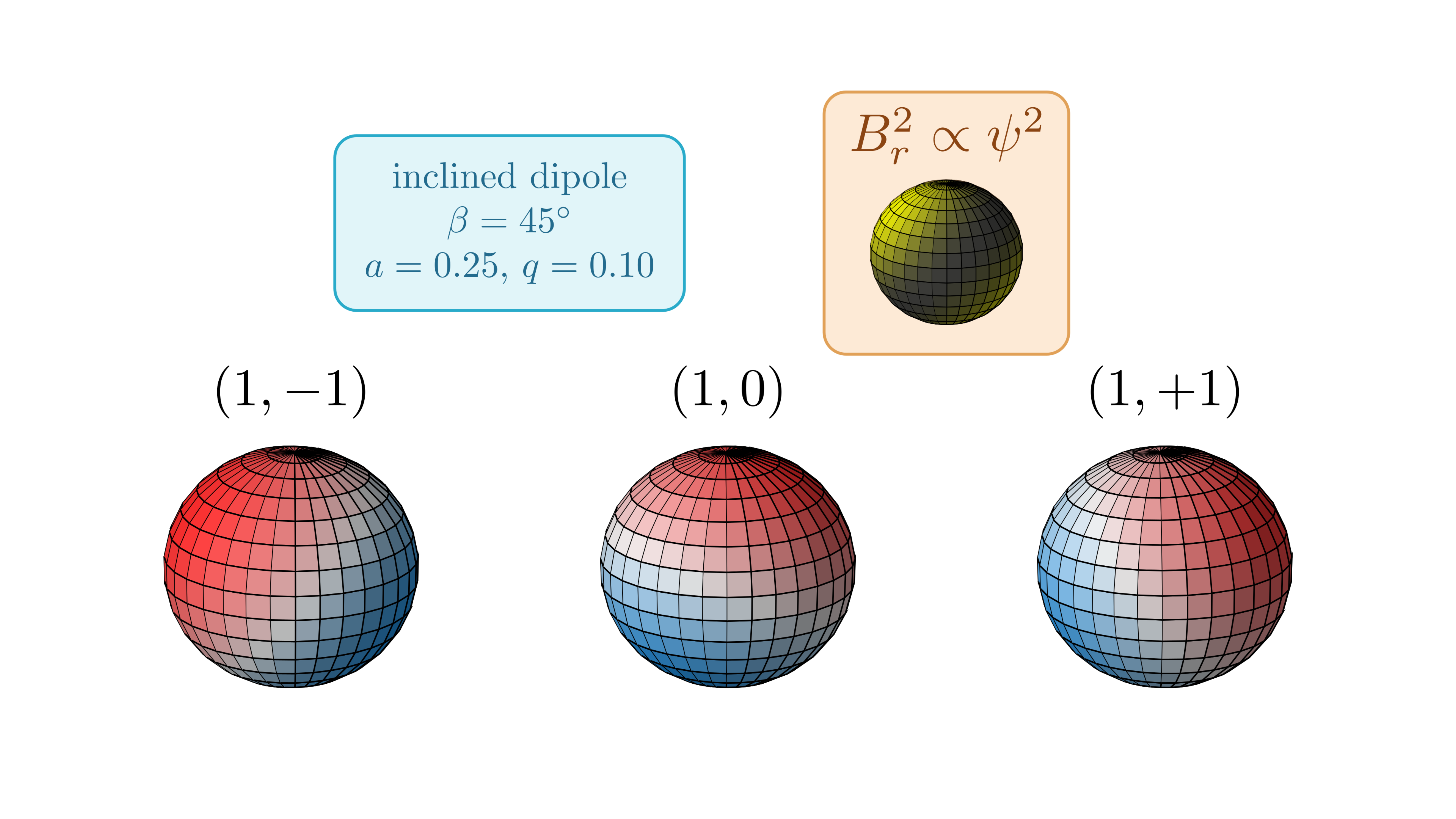}
    \caption{(Still frame of animated figure.)
    Same as Figure \ref{fig:IncDip45_DR_QM}, but for dipole ($\ell=1$) eigenfunctions with $a=0.25$ and $q=0.10$.
    Under these (near-critical) parameters, the dipole modes are aligned with the magnetic axis rather than rotation, and the quadrupole modes (not shown) are all magnetically suppressed.}
    \label{fig:IncDip45_DM}
\end{figure*}

We first consider perhaps the simplest non-axisymmetric geometry: a dipolar magnetic field inclined from the rotational axis by an angle $\beta$:
\begin{equation}
    \psi(\theta,\phi) = \sqrt{3}\left(\cos\beta\cos\theta + \sin\beta\sin\theta\cos\phi\right)\mathrm{,}
\end{equation}
with Alfv\'en resonance occurring at $b=b_{\mathrm{max}}=1/\sqrt{3}$ for all $\beta$.
This inclined dipolar field geometry is representative of the stable large-scale magnetic fields observed at the surfaces of OBA-type stars, which are exceptional laboratories for magnetism in stellar radiative regions \citep{Wade:2016:MiMeS,Shultz:2019:MagneticBStars}.
In this system, depending on the relative strengths of the Lorentz and Coriolis forces, the pulsation axis may approximately match the magnetic or rotation axes, or lie somewhere in between.
Rotationally aligned eigenfunctions conform to the approximate rotational symmetry of the system.
Their corresponding fluid perturbations are therefore all proportional to $e^{im\phi}$, and take the form of traveling waves around the rotational axis.
At low rotation rates (small $q$), the pressure and fluid-displacement eigenfunctions are approximately the spherical harmonic functions $Y_{\ell m}$ and $\bar{\nabla}_hY_{\ell m}$, respectively (Figure \ref{fig:NoRotNoMag}).
In the inertial (observer) frame, mode frequencies appear ``Doppler shifted'' by the rotation.

\begin{figure*}
    \centering
    \includegraphics[width=\textwidth]{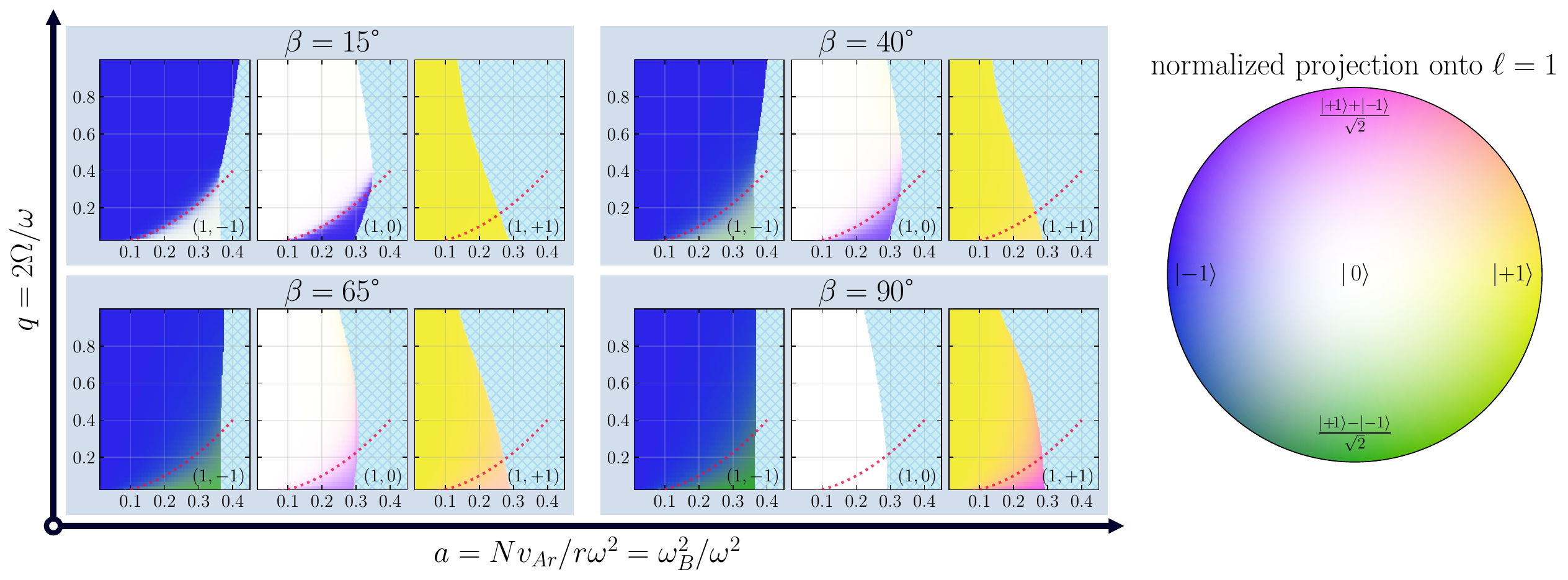}
    \caption{Dipole $(\ell=1)$ gravity wave solutions for a dipolar magnetic field inclined by an angle $\beta$ from the rotation axis, as a function of $a=Nv_{Ar}/r\omega^2$ and $q=2\Omega/\omega$.
    The ket $\ket{m}$ denotes the dipole spherical harmonic component with azimuthal order $m$.
    Each panel at fixed $\beta$ indicates a distinct dipole-mode solution, with colors representing the projection of the solution onto $\ell=1$ spherical harmonics ($\ket{0}$ and $\ket{\pm1}$ denote $m=0$ and $m=\pm1$ components, respectively).
    The pale blue hatched region denotes the conditions under which magnetic suppression is expected to occur.
    To guide the eye, the red dotted curves indicate a parabolic approximation for the threshold between rotational alignment and oblique pulsation ($q\approx2.5a^2$, cf. Equation \ref{eqn:oblique_condition}).
    Note that polarizations excluded from the color map (e.g., $\ket{+1}\pm i\ket{-1}$) do not appear in our solutions due to an antiunitary symmetry within the problem.}
    \label{fig:FIG_dipole_orientation}
\end{figure*}

In contrast, magnetically aligned eigenfunctions respect the geometry of the magnetic field, and thus rotate in and out of view with the field structure as the star rotates.
In the inertial frame, disk-integrated quantities (such as the stellar brightness) appear to beat on the rotation period, rather than modulating as a pure sinusoid.
This is the phenomenon of oblique pulsation \edit{\citep{Kurtz:1982:roAp,Bigot:2000:roAp,Loi:2021:MGTopologyObliquity,Rui:2025:StochasticOblique}}.
For example, under relatively rapid rotation with $q=0.25$ and an intermediate field strength with $a=0.15$ for $\beta=45\degree$, the dipole modes are rotationally aligned whereas the more magnetically susceptible quadrupole modes are magnetically aligned (Figure \ref{fig:IncDip45_DR_QM}).
In contrast, when the field strength is increased to $a=0.25$ and the rotation rate decreased to $q=0.10$, the dipole modes become magnetically aligned, and the quadrupole modes are magnetically suppressed \edit{(Figure \ref{fig:IncDip45_DM})}.
At weak magnetic fields and slow rotation, first-order perturbation theory provides a reasonable estimate for the threshold between rotationally and magnetically aligned pulsations.
By balancing Equations \ref{eqn:delta_omega_rot_ell} and \ref{eqn:delta_omega_mag} ($\delta\omega_{\mathrm{mag}}\simeq\delta\omega_{\mathrm{rot}}$), the threshold condition is a parabola in $a$--$q$ space:
\begin{equation} \label{eqn:oblique_condition}
    q \simeq [\ell(\ell+1)]^2a^2\mathrm{,}
\end{equation}
up to order-unity prefactors.
This condition for magnetic obliquity in Equation \ref{eqn:oblique_condition} is consistent with what is seen in our non-perturbative calculations.
Figure \ref{fig:FIG_dipole_orientation} shows the normalized projections of our dipole modes onto the $\ell=1$ spherical-harmonic manifold for various misalignment angles $\beta$.
In all cases, parabola-shaped curves demarcate qualitative transitions between rotational and magnetic mode alignment.
The red dotted curves in Figure \ref{fig:FIG_dipole_orientation} indicate the approximate location of this transition.

When $\beta=0\degree$, this system is identical to the aligned dipole field geometry considered by \citetalias{Rui:2024:TARM}.
In this axisymmetric system, mode branches at fixed $\ell$ can be uniquely indexed by $m$, with no mixing across branches as magnetogravity waves propagate radially.
However, when $\beta$ is taken to be small but nonzero, the aforementioned limiting behavior is approached in a counterintuitive fashion.
For low $\beta$ (e.g., the top-left panel of Figure \ref{fig:FIG_dipole_orientation}, for $\beta=15\degree$), there is a sharp avoided crossing at the rotational-versus-magnetic-alignment transition at which the $m=-1$ and $m=0$ branches abruptly exchange character.
In other words, even though both the rotationally and magnetically aligned limits possess zonal ($m=0$) modes which are geometrically very similar to each other, they do not formally belong to the same mode branches.
As $\beta$ is increased, the avoided crossing widens until it becomes a smooth transition in mode alignment, with the $m=0$ mode decoupling from the $m=\pm1$ entirely for $\beta=90\degree$.
Conversely, as $\beta$ approaches $0\degree$, the avoided crossing becomes infinitely sharp, in order to match the perfect eigenvalue crossing expected in the $\beta=0\degree$ case.
In other words, $\beta$ parameterizes the coupling strength of the avoided crossing.
As discussed further in Section \ref{sec:higher_wkb}, such avoided crossings are a likely site for the non-adiabatic exchange of wave amplitude between polarizations.
Indeed, this polarization mixing is required in the $\beta\rightarrow0\degree$ limit in order to match the expected behavior of the $\beta=0\degree$ case within which polarizations of different $m$ perfectly decouple.

\begin{figure}
    \centering
    \includegraphics[width=\linewidth]{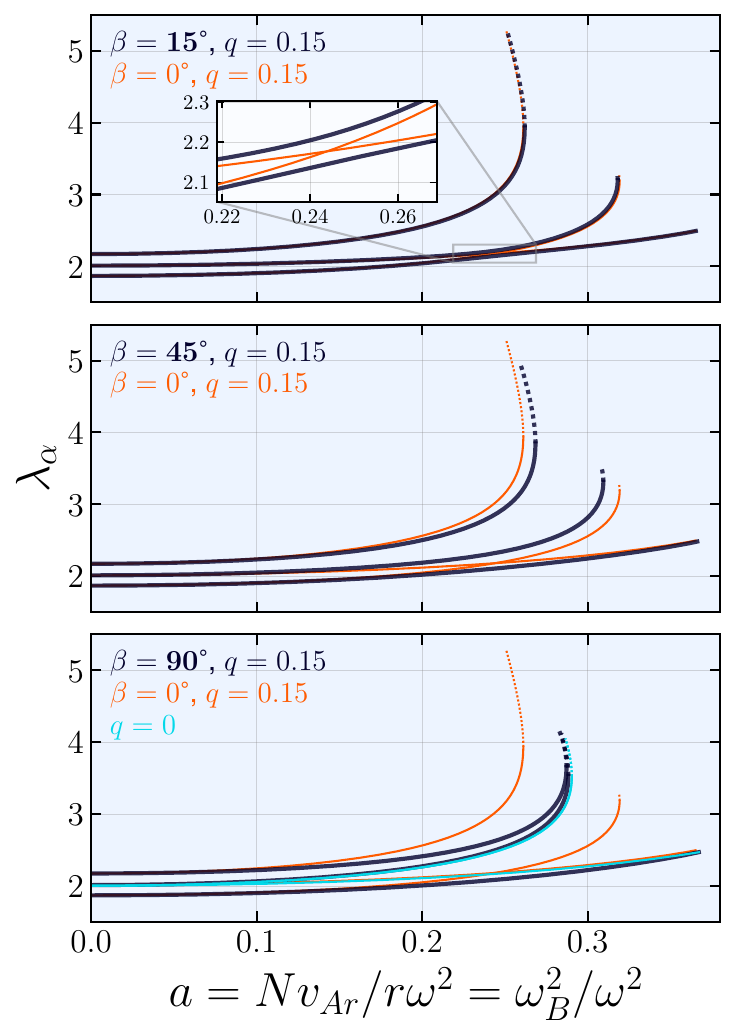}
    \caption{Eigenvalues $\lambda$ for the dipole ($\ell=1$) polarizations as a function of $a=Nv_{Ar}/r\omega^2$ for inclined dipole geometries with $\beta=15\degree$, $45\degree$, and $90\degree$, with spin parameter $q=0.15$.
    Dotted segments indicate solutions correspond to slow magnetic waves.
    The inset in the top panel zooms in on the site of an avoided crossing.
    Orange curves denote the $\beta=0\degree$ case.
    The cyan curve in the last panel shows the $q=0$ case.}
    \label{fig:FIG_dipole_eigenvalues_1}
\end{figure}

This continuum between avoided-crossing and smooth-alignment-transition behavior can also be seen in $\lambda$ (Figure \ref{fig:FIG_dipole_eigenvalues_1}).
At low but nonzero values of $\beta$ (top panel), $\lambda$ depends on $a$ and $q$ very similarly to $\beta=0\degree$.
However, at $a\approx0.24$ where an eigenvalue crossing appears in the $\beta=0\degree$ case, the eigenvalue branches repel each other in an avoided crossing, exchanging mode character.
As $\beta$ is increased (middle panel), this eigenvalue repulsion is strong enough that the eigenvalues remain well-separated throughout the rotational-to-magnetic-alignment transition.
Finally, once $\beta=90\degree$, at high $a$ (at fixed $q=0.15$), the eigenfunctions are magnetically aligned, and the eigenvalues are similar to the non-rotating ($q=0$) case (bottom panel).
Similar behavior can also be observed in the eigenvalues of the quadrupole polarizations (Appendix \ref{app:quadrupole_eigenvalues}).

\begin{figure}
    \centering
    \includegraphics[width=\linewidth]{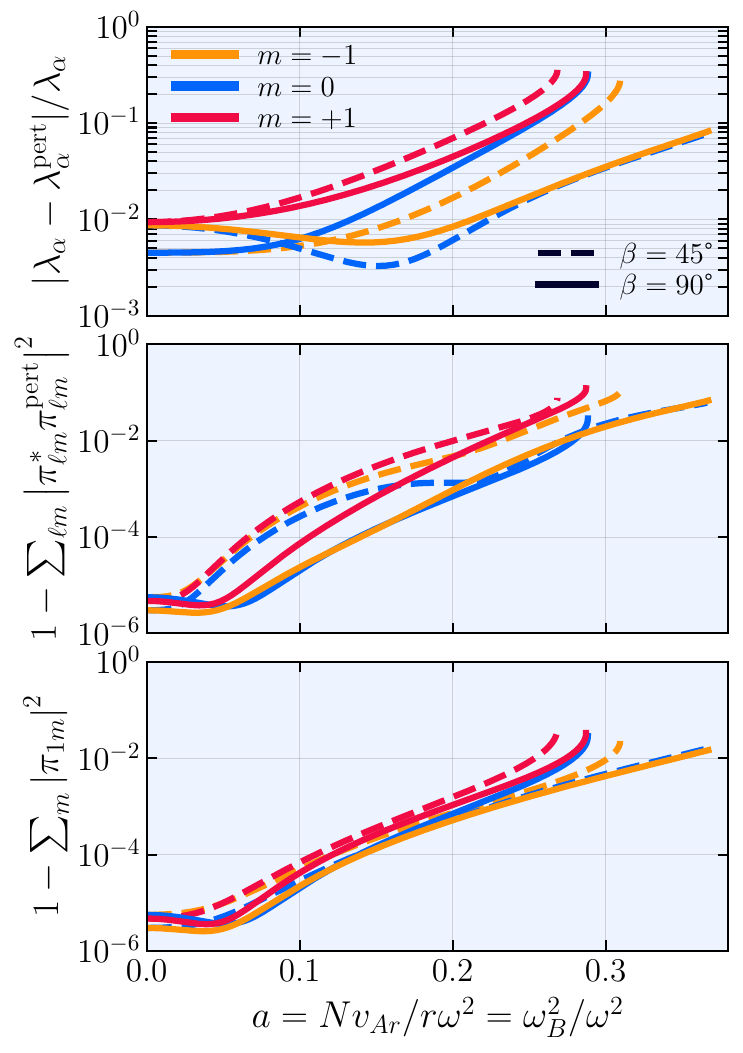}
    \caption{Comparison between perturbative and non-perturbative dipole ($\ell=1$) polarization solutions.
    Results are shown for inclined dipole fields with $\beta=45\degree$ and $\beta=90\degree$, where the spin parameter is $q=0.15$.
    \textit{Top:} Relative error in $\lambda$ if first-order perturbation theory is used.
    Because the usual perturbative formalism predicts $\delta\omega$ rather than $\lambda$, we convert $\delta\omega$ to $\lambda$ using Equation \ref{eqn:lambda_pert}.
    \textit{Center:} Deviation of non-perturbative solutions for $p'$ from perturbative solutions.
    \textit{Bottom:} Degree that non-perturbative solutions for $p'$ mix into $\ell\neq1$.
    First-order perturbation theory predicts that no such mixing across $\ell$ should occur.
    }
    \label{fig:FIG_dipole_vs_pert}
\end{figure}

A key advantage of our single-polarization waveguide description is that it incorporates both rotation and magnetism non-perturbatively.
While a perturbative approach qualitatively describes the aforementioned mode behavior correctly, a non-perturbative method is necessary to \textit{quantitatively} reproduce observations at high field strengths and/or spins.
As a benchmark of our non-perturbative single-polarization waveguide description, Figure \ref{fig:FIG_dipole_vs_pert} compares the results of our method to those of perturbation theory for inclined dipole geometries with $\beta=45\degree$ and $\beta=90\degree$ at fixed $q=0.15$.
The first panel compares $\lambda$ derived using our formalism versus its value $\lambda^{\mathrm{pert}}$ predicted by applying perturbation theory to a thin-shell g-mode cavity (Equation \ref{eqn:lambda_pert}).
For field strengths close to suppression, relative differences between $\lambda$ and $\lambda^{\mathrm{pert}}$ reach tens of percent for both $\beta=45\degree$ and $\beta=90\degree$.
The second and third panels show the residual power of the full non-perturbative solutions relative to the perturbative solutions and the $\ell=1$ spherical harmonic subspace, respectively.
In particular, the second panel shows that, at near-critical fields, the perturbative solution fails to overlap with the non-perturbative solution at the tens-of-percent level.
The third panel shows the degree of coupling between our modified dipole modes and higher-$\ell$ spherical harmonics.
This coupling is forbidden in the usual degenerate perturbation theory, but is found to rise to a few percent as the field strength approaches its critical value.
Curves in the second and third panels should be identical if the discrepancy between the non-perturbative calculation and perturbation theory is purely due to mixing in $\ell$.
The fact that these panels disagree reveals that a majority of the lack of overlap between the non-perturbative and perturbative polarizations is due to non-perturbative mixing in $m$ at fixed $\ell=1$, rather than mixing across $\ell$.

The differences between the perturbative and non-perturbative results shown in all three panels of Figure \ref{fig:FIG_dipole_vs_pert} do not converge to $0$ as $a\rightarrow0$.
This is because first-order perturbation theory omits second-order rotational effects (i.e., $\propto q^2$), which for $q>0$ are present even at $a=0$.
However, in all three cases, these differences increase with $a$, indicating that the non-perturbative magnetic effects are more significant at higher field strengths, as expected.

%% file: sect_dipole_plus_y22.tex
%!TEX root=./main.tex
\subsection{Magnetic fields with no continuous rotational symmetry} \label{sec:dipole_plus_y22}

The inclined-dipolar-field system in Section \ref{sec:inclined_dipole} is non-axisymmetric because the rotation and magnetic axes are misaligned, even though the magnetic field by itself still has an axis of symmetry.
However, it is also possible for the magnetic field to lack an axis of symmetry entirely.
In such cases, the polarizations must be solved for in a non-axisymmetric fashion even in the absence of rotation.

As a\edit{n instructive} example of such a situation, we calculate the dipole and quadrupole magnetogravity polarizations under a magnetic field whose geometry is given by
\begin{equation} \label{eqn:dipole_plus_y22}
    \psi(\theta,\phi) = \frac{1}{\sqrt{1+\mathcal{Q}^2}}\left(\sqrt{3}\cos\theta + \mathcal{Q}\sqrt{15/4}\sin^2\theta\cos(2\phi)\right)\mathrm{,}
\end{equation}
Equation \ref{eqn:dipole_plus_y22} describes an aligned dipole field superposed with a real sectoral quadrupolar field, i.e., a term proportional to $Y_{2,+2}+Y_{2,-2}$.
In general, a non-dipolar component of the field must be included in order to create a non-axisymmetric magnetic field, since real superpositions of dipole spherical harmonics are themselves dipoles \citep{Wigner:1931:RotatingSpharms}.
Hereafter, we refer to this geometry as a ``non-axisymmetric dipole-plus-quadrupole field'' \citep[although this is distinct from the dipole-plus-quadrupole field considered by][]{Das:2024:ComplexMagnetic}.
\edit{Although this field geometry lacks continuous rotational symmetry, we show that it still possesses discrete symmetries which allow g-mode frequencies to be straightforwardly calculated assuming single-polarization propagation (via Equation \ref{eqn:radial_quantization}).
Because these discrete symmetries are essential to this calculation, this dipole-plus-quadrupole field geometry (although a generalization of the axisymmetric case) is not representative of general field geometries, most of which lack these discrete symmetries.
In the general case, non-adiabatic mixing between polarizations must be directly confronted to make predictions for g-mode frequencies at strong field strengths (see Section \ref{sec:higher_wkb}).}

The dimensionless parameter $\mathcal{Q}$ sets the strength of the quadrupolar component of the field, with $\mathcal{Q}=0$ and $\mathcal{Q}\rightarrow\infty$ corresponding to pure dipolar and sectoral quadrupolar fields, respectively.
Under this field geometry, g modes resonate with Alfv\'en waves when $b=b_{\mathrm{max}}$, where $b_{\mathrm{max}}$ now depends on $\mathcal{Q}$ as
\begin{equation}
    b_{\mathrm{max}} = \frac{1}{\sqrt{\max(\psi^2)}} = 
    \begin{cases}
        \sqrt{\frac{1+\mathcal{Q}^2}{3}}&|\mathcal{Q}|\leq1/\sqrt{5}\\
        \sqrt{\frac{20\mathcal{Q}^2(1+\mathcal{Q}^2)}{3(1+5\mathcal{Q}^2)^2}}&|\mathcal{Q}|>1/\sqrt{5}
    \end{cases}
    \mathrm{.}
\end{equation}
The critical value $\mathcal{Q}=1/\sqrt{5}$ is the maximum value of $\mathcal{Q}$ such that $\psi^2$ is maximized at the poles.
As $\mathcal{Q}$ is increased beyond $1/\sqrt{5}$, the quadrupolar component dominates more significantly over the dipolar component, and the global maxima of $\psi^2$ move toward the equator.

\begin{figure*}
    \centering
    \includegraphics[width=\linewidth]{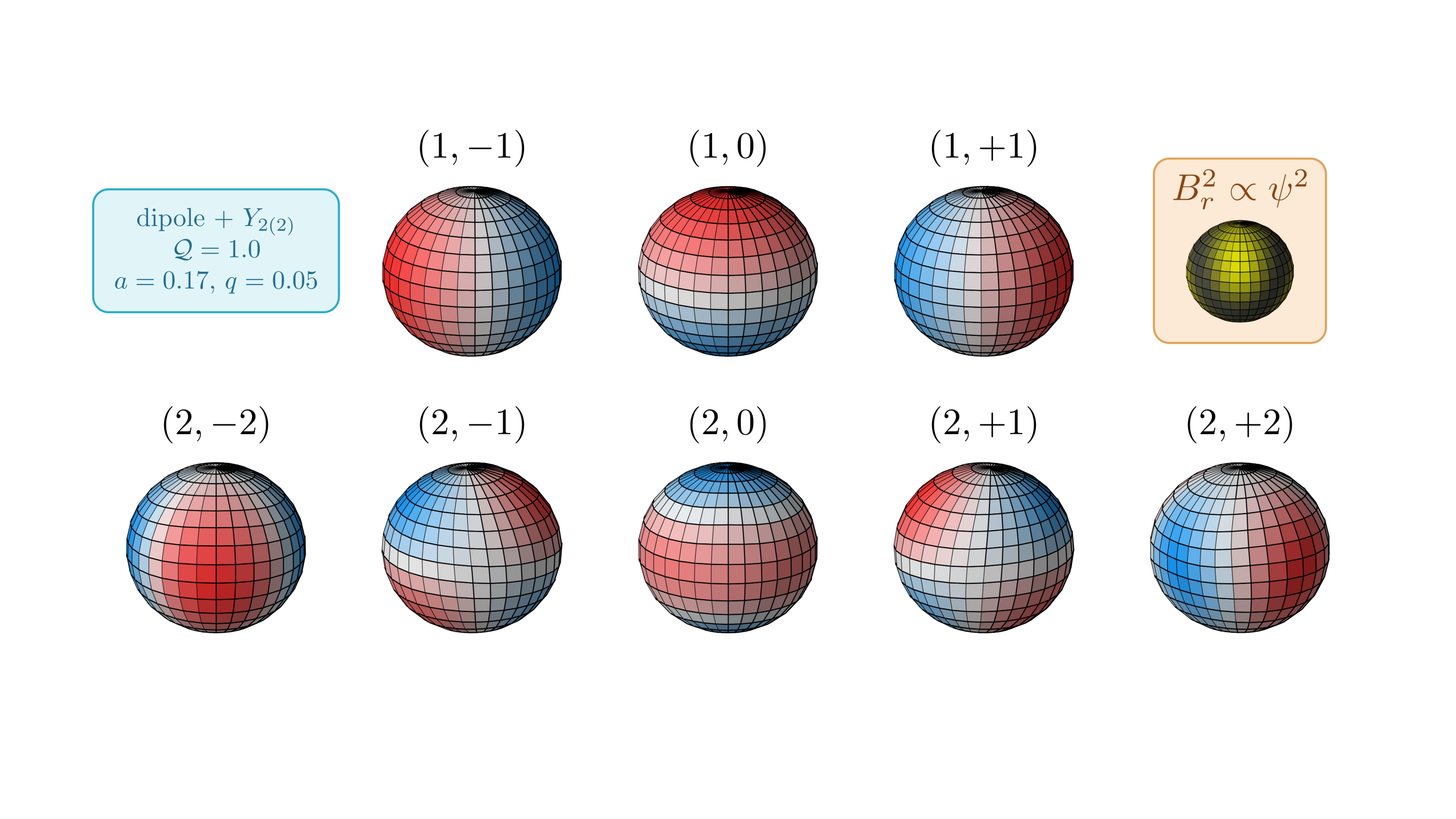}
    \caption{(Still frame of animated figure.)
    Same as Figure \ref{fig:IncDip45_DR_QM}, but for the non-axisymmetric dipole plus quadrupole geometry described in Section \ref{sec:dipole_plus_y22}, with $a=0.17$ and $q=0.05$.
    Although the quadrupole modes are near-critical at these parameters, the magnetic field only significantly mixes the $(\ell,m)=(2,\pm2)$ eigenfunctions.
    }
    \label{fig:DipY22_Q1d0}
\end{figure*}

Figure \ref{fig:DipY22_Q1d0} shows the dipole and quadrupole polarizations under this dipole-plus-quadrupole field geometry with $\mathcal{Q}=1$.
Under the adopted parameters $a=0.15$ and $q=0.05$, the quadrupole polarizations are near critical.
In spite of this, only the $(\ell,m)=(2,\pm2)$ modes are significantly mixed with each other, with the other quadrupole polarizations remaining aligned with the rotation axis (which is the same as the symmetry axis of the dipole component of the field).
This remains true even for values of $a$ under which the dipole polarizations are near critical.
Lack of mixing between $m$ across dipole spherical harmonics can easily be seen from perturbation theory.
When evaluated for the geometry in Equation \ref{eqn:dipole_plus_y22}, Equation \ref{Mmatrixgeneral} gives the dipole matrix
\begin{equation} \label{eqn:pert_dipole_plus_y22}
    \mathbf{M}_{\ell=1}=\frac{3a^2}{35(1+\mathcal{Q}^2)}
    \begin{pmatrix}
        2(7+5\mathcal{Q}^2)&0&0\\
        0&7+15\mathcal{Q}^2&0\\
        0&0&2(7+5\mathcal{Q}^2)\\
    \end{pmatrix}\mathrm{,}
\end{equation}
which is incapable of mixing across $m$ for any choice of $\mathcal{Q}$.
The vanishing off-diagonal elements in Equation \ref{eqn:pert_dipole_plus_y22} are found to be due to selection rules for the angular integral in Equation \ref{Mmatrixgeneral}. In particular, coupling is only permitted for $\Delta m = 0, \pm2, \pm4$ from angular momentum conservation, with $\Delta m = \pm 2$ further forbidden from parity symmetry. Thus, the lowest-order perturbative avoided crossing only emerges for quadrupole modes, where we have
\begin{equation} \label{eqn:pert_dipole_plus_y22_l2}
\tiny
    \mathbf{M}_{\ell=2}=\frac{a^2}{7(1+\mathcal{Q}^2)}
    \begin{pmatrix}
        5(3+5\mathcal{Q}^2)&0&0&0&-10\mathcal{Q}^2\\
        0&4(6+5\mathcal{Q}^2)&0&0&0\\
        0&0&3(9+5\mathcal{Q}^2)&0&0\\
        0&0&0&4(6+5\mathcal{Q}^2)&0\\
        -10\mathcal{Q}^2&0&0&0&5(3+5\mathcal{Q}^2)\\
    \end{pmatrix}\mathrm{.}
\end{equation}
These symmetry considerations are generic and therefore inherited by the non-perturbative calculation.
\edit{However, we reiterate that avoided crossings are expected under the more general case in which the magnetic field geometry does not possess such useful discrete symmetries.}
% Although $\psi^2$ from Equation \ref{eqn:dipole_plus_y22} appears to contain terms proportional to $e^{\pm2i\phi}$ which could na\"ively couple spherical harmonics with $\Delta m=\pm2$, these terms have the wrong parity under reflection across the equator ($\theta\mapsto\pi-\theta$) and vanish under the integral in Equation \ref{Mmatrixgeneral}.
% The only relevant coupling terms in $\psi^2$ are proportional to $e^{\pm4i\phi}$, explaining why only the $Y_{2,\pm2}$ spherical harmonics (with $\Delta m=\pm4$) are coupled to each other.

\begin{figure}
    \centering
    \includegraphics[width=\linewidth]{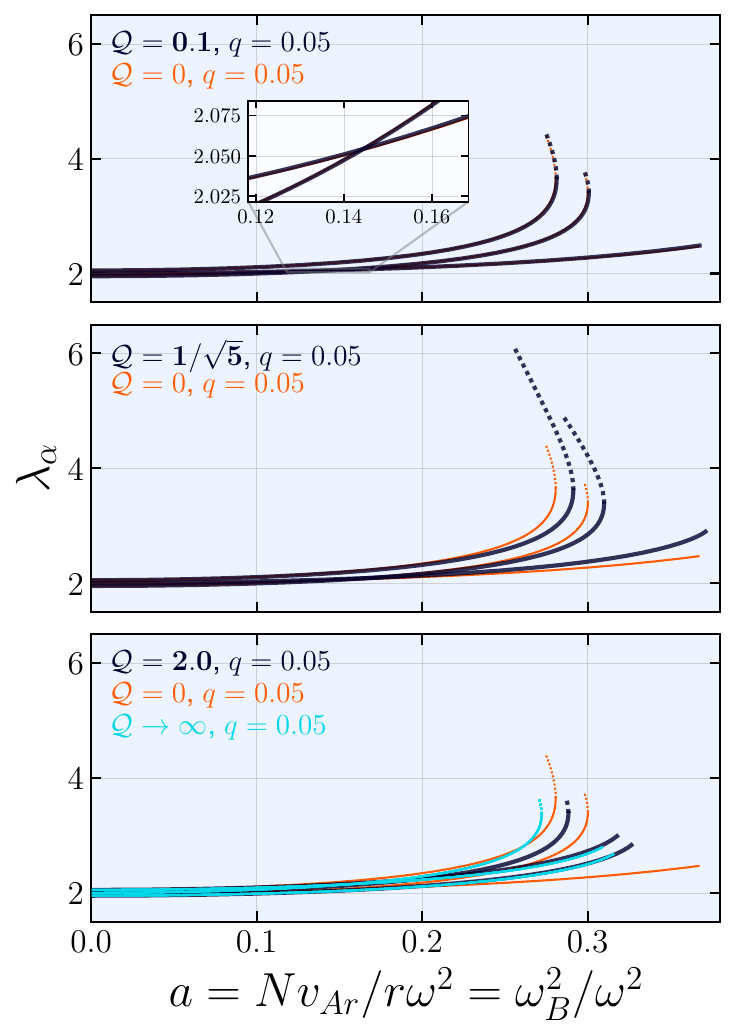}
    \caption{Same as Figure \ref{fig:FIG_dipole_eigenvalues_1}, but for the non-axisymmetric dipole plus quadrupole geometry described in Section \ref{sec:dipole_plus_y22}.
    The inset reveals that the non-axisymmetric quadrupolar contribution to $\psi$ does not cause mode repulsion.}
    \label{fig:FIG_dpy22_eigenvalues_1}
\end{figure}

As expected, under strong magnetic fields, $\lambda$ deviates significantly from $\lambda^{\mathrm{pert}}$.
Figure \ref{fig:FIG_dpy22_eigenvalues_1} shows the behavior of dipole eigenvalues $\lambda$ versus $a$ at fixed $q=0.05$ as $\mathcal{Q}$ is increased.
However, unlike in the case of a slightly inclined dipole field (Section \ref{sec:inclined_dipole}), the addition of a small non-axisymmetric quadrupolar field component (i.e., small $\mathcal{Q}$) does not produce avoided crossings.
The inset in the top panel of Figure \ref{fig:FIG_dpy22_eigenvalues_1} zooms in on a particular mode crossing for dipole polarizations with $\mathcal{Q}=0.1$, although we do not resolve avoided crossings for any value of $\mathcal{Q}$, nor do we resolve them for quadrupole polarizations (Appendix \ref{app:quadrupole_eigenvalues}).
We attribute this lack of mode repulsion to protection by discrete symmetries within $\psi^2$, which are present even though there is no longer a continuous rotational symmetry.
Curiously, we also find that the range of $\lambda$ occupied by the part of the slow magnetic branch is significantly increased at intermediate values of $\mathcal{Q}$ (middle panel of Figure \ref{fig:FIG_dpy22_eigenvalues_1}), although the impact of this on asteroseismic observations is unclear.

\begin{figure}
    \centering
    \includegraphics[width=\linewidth]{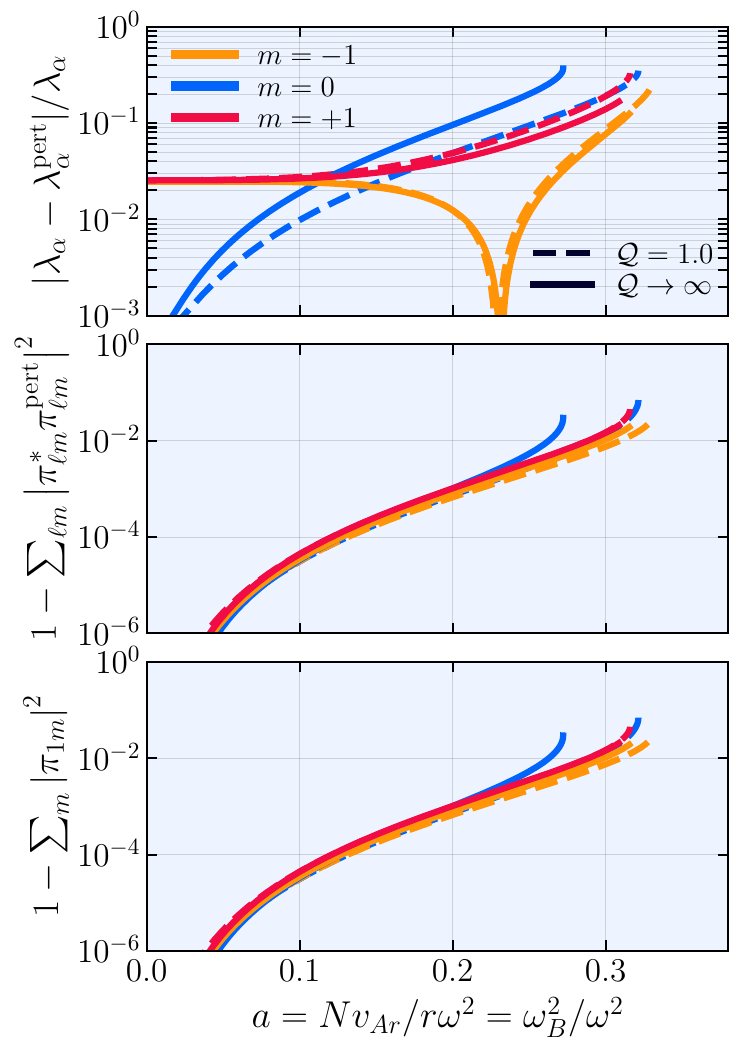}
    \caption{Same as Figure \ref{fig:FIG_dipole_vs_pert}, but for the non-axisymmetric dipole plus quadrupole geometry described in Section \ref{sec:dipole_plus_y22}, with $q=0.05$.
    }
    \label{fig:FIG_dpy22_vs_pert}
\end{figure}

Discrepancies between the single-polarization waveguide description and first-order perturbation theory for the dipole-plus-quadrupole field geometry are similar in scale to those found for the inclined dipole geometry.
We show metrics of these discrepancies for the dipole polarizations in Figure \ref{fig:FIG_dpy22_vs_pert}.
In particular, at high values of $a$, the eigenvalues predicted by these two methods differ by tens of percent, and the eigenfunctions predicted fail to overlap by a few percent in residual power.
However, unlike in the inclined dipolar field case, all of the lack of overlap between the non-perturbative and perturbative polarizations (up to machine precision) is due to mixing across $\ell$ rather than mixing across $m$ at fixed $\ell$ (i.e., the curves in the second and third panels of Figure \ref{fig:FIG_dpy22_vs_pert} are identical).
As before, this is almost certainly due to symmetry protection against mixing between certain spherical harmonics.

\begin{figure*}
    \centering
    \includegraphics[width=\linewidth]{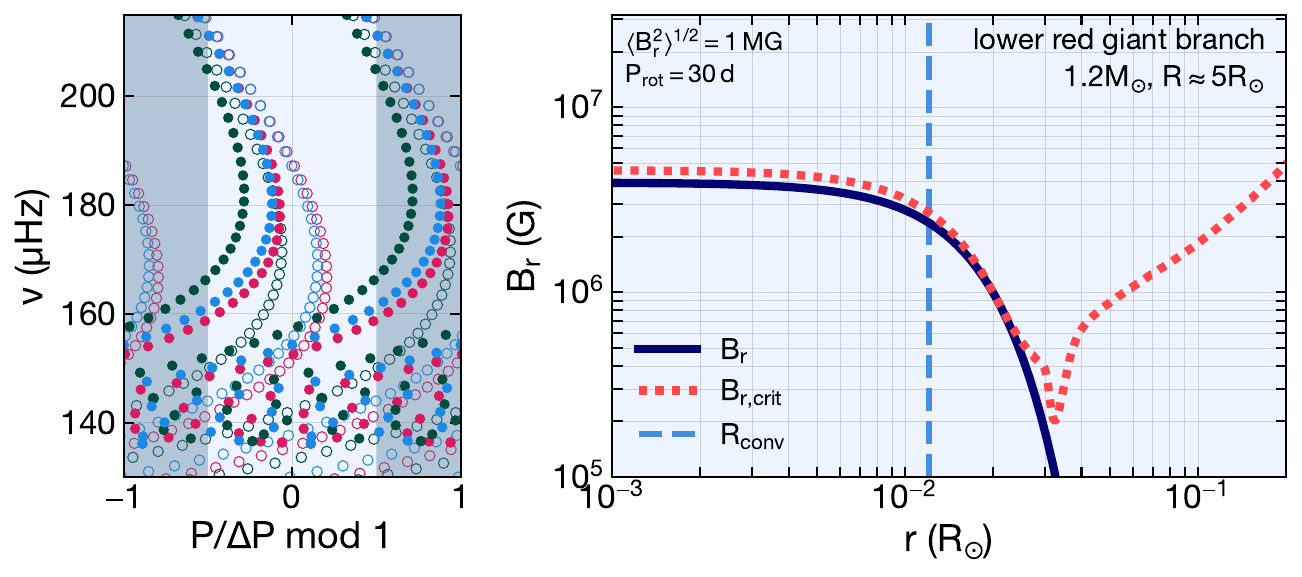}
    \caption{\textit{Left:} Mock period echelle diagram for the g modes of a lower red giant branch model with $M=1.2M_\odot$ and $R\approx5R_\odot$ ($\nu_{\mathrm{max}}\approx160\,\mu\mathrm{Hz}$) under the effects of a strong magnetic field $\langle B_r^2\rangle^{1/2}=1\,\mathrm{MG}$ \edit{($B_{0r}(r=0)\approx3.9\,\mathrm{MG}$)} and a realistic rotation rate $30\,\mathrm{d}$.
    The solid (open) circles represent predictions for g-mode frequencies obtained by our non-perturbative waveguide formalism (perturbation theory), respectively.
    \edit{Marker colors represent azimuthal order $m$, with green, red, and blue indicating $m=-1$, $0$, and $+1$, respectively.}
    Mode\edit{s} with frequencies lower than $\nu\approx136\mu\mathrm{Hz}$ are magnetically suppressed.
    For clarity, the mode periods are folded on the average period spacing of the higher-frequency part of the spectrum.
    \textit{Right:} The magnetic field profile assumed (dark purple solid), compared to the critical magnetic field $B_{r,\mathrm{crit}}$ of the lowest-frequency g mode \edit{($\nu=136\,\mu\mathrm{Hz}$)} which is not magnetically suppressed (orange dotted).
    \edit{The $B_{r,\mathrm{crit}}$ shown is derived directly from the numerical calculation rather than estimated from the \citet{Fuller:2015:SuppressedDipole} scaling.}
    The magnetic field follows the non-axisymmetric dipole-plus-quadrupole geometry described in Section \ref{sec:dipole_plus_y22} with $\mathcal{Q}=1$, with a radial dependence which follows a Gaussian whose scale length is the maximum extent of the convective core during the main sequence (light blue dashed).
    }
    \label{fig:FIG_dpy22_echelle}
\end{figure*}

Because the non-axisymmetric dipole-plus-quadrupole magnetic field geometry does not mix dipole modes, asymptotic dipole gravity waves should not undergo any magnetogravity polarization mixing with each other (cf. the inclined dipole geometry in Section \ref{sec:inclined_dipole}, under which dipole g modes can mix).
For this geometry, we can therefore predict the frequencies of g modes under strong magnetic fields using Equation \ref{eqn:radial_quantization}, which is valid in the absence of polarization mixing (i.e., ``adiabatic'' propagation).
We include the frequency shift due to the frame-change from the corotating to inertial frame and set $\epsilon_g=0$ for simplicity.
The left panel of Figure \ref{fig:FIG_dpy22_echelle} shows the mock period echelle diagram for a lower red giant branch star with a strong non-axisymmetric dipole-plus-quadrupole magnetic field with $\langle B_r^2\rangle^{1/2}=1\,\mathrm{MG}$, $P_{\mathrm{rot}}=30\,\mathrm{d}$, and $\mathcal{Q}=1$.
We create our red giant model using version \texttt{r24.08.1} of Modules for Experiments in Stellar Astrophysics \citep[\texttt{MESA};][]{Paxton:2011:MESA,Paxton:2013:MESA,Paxton:2015:MESA,Paxton:2018:MESA,Paxton:2019:MESA,Jermyn:2023:MESA}.
Our \texttt{MESA} model has a mass $1.2M_\odot$ and radius $\approx5R_\odot$, corresponding to $\nu_{\mathrm{max}}\approx160\,\mu\mathrm{Hz}$, and was evolved assuming exponential overmixing with \texttt{overshoot\_f=0.015} and \texttt{overshoot\_f0=0.005}, with a minimum mixing throughout set by \texttt{min\_D\_mix=1}.
This low mass (which is just above the Kraft break) is motivated by the preferentially low masses $\sim1.1$--$1.2M_\odot$ of the near-critical magnetic red giants discovered by \citet{Deheuvels:2026:NearCritical}.
\citet{Deheuvels:2026:NearCritical} also find that the magnetic fields in these near-critical red giants have radial extent consistent with the radius $R_{\mathrm{conv}}$ enclosing the maximum mass of the convective core during the main sequence \citep[such as would be generated by a convective dynamo;][]{Cantiello:2016:EvolvingMagnetic}.
Accordingly, following \citet{Deheuvels:2026:NearCritical}, we impose a Gaussian radial dependence on the magnetic field:
\begin{equation}
    B_r\propto \exp\left(-r^2/2R_\mathrm{conv}^2\right)\mathrm{,}
\end{equation}
where for this model $R_{\mathrm{conv}}\approx1.2\times10^{-2}R_\odot$ (below the hydrogen burning shell at $\approx3\times10^{-2}R_\odot$; right panel of Figure \ref{fig:FIG_dpy22_echelle}).
We also present perturbative estimates for the g-mode frequencies (open circles in the left panel of Figure \ref{fig:FIG_dpy22_echelle}), using the approximate expression in Equation \ref{eqn:lambda_pert} for $\lambda^{\mathrm{pert}}$ together with Equation \ref{eqn:pert_dipole_plus_y22}.
As in \citetalias{Rui:2024:TARM} for the aligned dipole geometry, we find that perturbation theory significantly underestimates the magnetic frequency shifts \edit{(by $\gtrsim60\%$ for the lowest-frequency g modes in this example)}.
Its misapplication to strongly magnetic stars would therefore preferentially overestimate the magnetic field, as in our previous work.

%% file: sect_higher_wkb.tex
%!TEX root=./main.tex

\subsection{Magnetogravity polarization mixing and oblique pulsations} \label{sec:higher_wkb}

As found in Section \ref{sec:perturbation_theory}, an \edit{immediate} apples-to-apples comparison between perturbation theory and the waveguide description is challenged by the different ``building blocks'' these two methods use to construct the global magnetogravity eigenfunctions.
Specifically, ``global'' perturbation theory \edit{forms} these eigenfunctions \edit{out of} linear combinations of unperturbed eigenfunctions (Equation \ref{eqn:xi_linear_combo}), which are each by themselves global, three-dimensional objects.
In contrast, our waveguide description assembles these eigenfunctions shell by shell as linear combinations of magnetogravity polarizations (defined by Equations \ref{eqn:transverse}), which are families of two-dimensional objects on the sphere.

Indeed, upon closer examination, first-order perturbation theory and our waveguide description (in the limit of adiabatic propagation) disagree even about what the global eigenfunctions are allowed to look like at weak field strengths.
Degenerate perturbation theory constructs the zeroth-order global eigenfunctions as linear combinations of degenerate modes in the unperturbed system, namely those with the same $n$ and $\ell$ but different $m$, i.e., $\vec{\xi}_{n\ell m}^{(0)}=f_{n\ell}(r)\nabla Y_{\ell m}$ (Equation \ref{eqn:xi_linear_combo}).
However, since at fixed $n$ and $\ell$ these basis functions all have the same radial dependence through $f_{n\ell}(r)$, the transverse structure of the zeroth-order global eigenfunction must be independent of radius.
In other words, perturbation theory asserts that eigenfunctions can approximately be written in the form $\vec{\xi}=R(r)\vec{H}(\theta,\phi)$.
Components of the global eigenfunctions which depend more generally on the coordinates, i.e., $\vec{\xi}=R(r)\vec{H}(\theta,\phi;r)$, are assumed to be first-order in the small perturbation \citep[cf.][]{Loi:2021:MGTopologyObliquity,Li:2022:30to100kG,Rui:2025:StochasticOblique}.
In contrast, Section \ref{sec:inclined_dipole} demonstrates that magnetogravity polarizations belonging to a single magnetogravity polarization branch can have very different horizontal structures depending on the relative values of $a$ and $q$, even when both $a\ll1$ and $q\ll1$ individually.
If a radially propagating wave is always assumed to follow a single polarization exactly, the transverse structure of a global eigenfunction must adjust in radius exactly as demanded by the local values of $a$ and $\psi$.

In other words, first-order perturbation theory predicts that the transverse structure of a radially propagating wave can \textit{never} \edit{significantly} adjust to its local environment as it propagates, whereas the single-polarization (``adiabatic'') waveguide description assumes that it \textit{must} do so.
While both formalisms ostensibly have validity in the limit of high $n$, weak magnetic field, and slow rotation, they are clearly in fundamental contradiction with each other.
Each formalism must therefore break down under some conditions, the details of which we explore here.

The single-polarization waveguide description breaks down when non-adiabatic propagation effects become important.
Whereas adiabatic propagation forces magnetogravity waves to conserve their polarization properties throughout the g-mode cavity, non-adiabatic effects cause waves to change polarization as they propagate.
The validity of the single-polarization waveguide approach can be derived by substituting the JWKB ansatz (Equation \ref{eqn:jwkb_ansatz}) into \ref{eqn:final_momentum} and \ref{eqn:final_continuity} and retaining terms both leading- and next-to-leading-order in $\epsilon$ (in deriving Equation \ref{eqn:jwkb_leading}, only the leading-order terms were retained).
This procedure (described in full detail in Appendix \ref{app:higher_wkb_waveguide} for weak magnetic fields and slow rotation rates) results in a rate-like transport equation describing the coupled evolution of the polarization amplitudes $A_\alpha$ with radius.
Although the total wave flux is conserved (Appendix \ref{app:higher_wkb}), the polarizations can nevertheless exchange amplitude amongst each other, particularly when they are close together in wavenumber $k_{r,\alpha}$ and when they have substantial geometric overlap.
In particular, the single-polarization waveguide description is valid when the polarization states change slowly enough in radius that the adiabatic approximation holds.
The criterion for assessing this is set by the comparison between the characteristic radial length scale $H_{\alpha\beta}$ (defined more precisely in Appendix \ref{app:higher_wkb_waveguide}) over which the polarization states $\alpha$ and $\beta$ intrinsically rotate into each other and the \textit{difference} in wavenumbers $\Delta k_{r,\alpha\beta}\equiv k_{r,\alpha}-k_{r,\beta}$ between two polarizations:
\begin{equation} \label{eqn:jwkb_adiabatic_orig_main}
    2/H_{\alpha\beta}\ll \Delta k_{r,\alpha\beta}\mathrm{.}
\end{equation}
Note that it is the (often small) difference $\Delta k_{r,\alpha\beta}$ which is relevant for this comparison, rather than $k_{r,\alpha}$ and $k_{r,\beta}$ individually (which are often much larger).
The validity criterion in Equation \ref{eqn:jwkb_adiabatic_orig_main} can be rewritten (see Appendix \ref{app:higher_wkb_waveguide} for a detailed derivation) as a condition on the magnetic field strength:
\begin{equation} \label{eqn:jwkb_validity_main}
    B_{0r} \gg B_{r,\mathrm{mix},\mathrm{wg}} \simeq \frac{4}{[\ell(\ell+1)]^{1/4}}\left(\frac{\omega}{N}\right)^{1/2}\left(\frac{r}{H_{\alpha\beta}}\right)^{1/2}B_{r,\mathrm{crit}}\mathrm{.}
\end{equation}
If the polarization states $\alpha$ and $\beta$ are protected from mixing by a global symmetry, $H_{\alpha\beta}^{-1}$ vanishes, and the criterion is automatically satisfied.
Otherwise, Equation \ref{eqn:jwkb_validity_main} is a bound on how quickly the polarization states are permitted to vary in radius (as is usual in discussions of the adiabatic approximation).
Although we have assumed $\omega/N\ll1$, the ratio $r/H_{\alpha\beta}$ can be large if $a$ is a strong function of radius, which can occur when either $B_{0r}$, $N$, or $\rho_0$ are strongly stratified.
Because the criterion in Equation \ref{eqn:jwkb_validity_main} depends on $H_{\alpha\beta}$, evaluating the validity of this condition requires assuming a background magnetic field profile $B_{0r}$ in advance (in addition to $N$, $\Omega$, etc.).

\begin{figure}
    \centering
    \includegraphics[width=\linewidth]{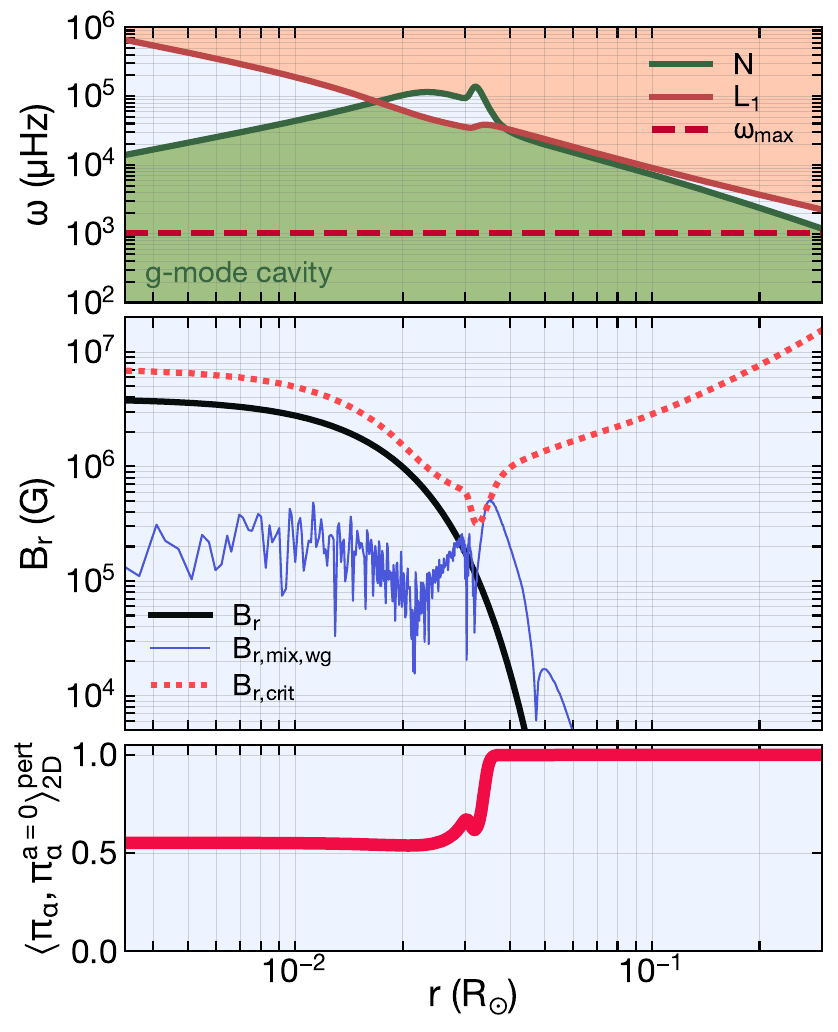}
    \caption{\textit{Top:} Propagation diagram for the lower red giant branch model described in Sections \ref{sec:dipole_plus_y22} and \ref{sec:higher_wkb}.
    \textit{Middle:} The magnetic field scale $B_{r,\mathrm{mix},\mathrm{wg}}$ at which polarization mixing is expected (Equation \ref{eqn:jwkb_validity_main}), compared to the assumed magnetic field geometry $B_r$ and critical field strength $B_{r,\mathrm{crit}}$.
    \textit{Bottom:} Overlap between the (perturbatively estimated) polarizations in a given shell with their structure at $a=0$, $\langle\pi_{m=+1},\pi^{a=0}_{m=+1}\rangle^{\mathrm{pert}}_{\mathrm{2D}}=\langle\pi_{m=-1},\pi^{a=0}_{m=-1}\rangle^{\mathrm{pert}}_{\mathrm{2D}}$.
    Change in this quantity indicates change in the structure of the polarizations at a given radius.}
    \label{fig:FIG_Brthresh}
\end{figure}

Figure \ref{fig:FIG_Brthresh} estimates $B_{r,\mathrm{mix,wg}}$ for the lower red giant branch model used in Figure \ref{fig:FIG_dpy22_echelle} evaluated at $\omega_{\mathrm{max}}=2\pi\nu_{\mathrm{max}}$.
We assume the same Gaussian radial magnetic field profile and rotation rate as used in Figure \ref{fig:FIG_dpy22_echelle}.
However, we instead assume an inclined magnetic field geometry with $\beta=90\degree$, so that only the $m=\pm1$ dipole polarizations can significantly mix with each other.
Consistent with the perturbative assumptions used to derive Equation \ref{eqn:jwkb_validity}, we estimate $H_{\alpha\beta}$ using polarization eigenfunctions calculated using perturbation theory (Section \ref{sec:perturbation_theory}) rather than solving the non-perturbative transverse equations (Equations \ref{eqn:transverse}).
Figure \ref{fig:FIG_Brthresh} shows that $B_{r,\mathrm{mix,wg}}\gg B_r$ near the hydrogen burning shell, where most of the adjustment in the polarizations occurs (bottom panel), i.e., significant polarization mixing is expected.
This polarization mixing occurs because the polarizations change too quickly in radius due to the rapid variation in the background quantities.
In our testing, adjusting the radial extent of the magnetic field does not remove the presence of a layer with significant polarization mixing \edit{(although it can change the radial coordinate of this layer)}, suggesting that these effects will usually occur unless forbidden by discrete or continuous symmetries of the system.

Correspondingly, perturbation theory breaks down under perturbations which are strong enough to significantly couple radial orders together.
Perturbation theory assumes that modes are approximately superpositions of exactly degenerate unperturbed modes of fixed $n$ and $\ell$ (cf. Equation \ref{eqn:xi_linear_combo}).
However, since g modes of nearby radial orders can be closely spaced in frequency, even superficially weak perturbations can sometimes mix them.
A validity condition for global perturbation theory can be derived by asserting that the magnetic perturbation is too weak to significantly couple unperturbed modes of different radial order (see Appendix \ref{app:higher_wkb_pert} for a detailed derivation).
This procedure again results in a condition on the magnetic field strength:
\begin{equation} \label{eqn:pert_validity_end_main}
    \begin{split}
       \langle B_r^2\rangle^{1/2} &\ll B_{r,\mathrm{mix},\mathrm{pert}} \\
       &\sim \frac{2}{[\ell(\ell+1)]^{1/4}}\left(\frac{\omega_{0,n\ell}}{X_N}\right)^{1/2}\left|\frac{\Delta n}{\tilde{\psi}_{\ell;mm';\Delta n}^2}\right|^{1/2}\langle B_{r,\mathrm{crit}}^{-2}\rangle^{-1/2}\mathrm{,}
    \end{split}
\end{equation}
where \edit{$X_N=\int_{\mathcal{R}}(N/r)\mathrm{d}r$} is the buoyancy radius of the g-mode cavity, $\Delta n$ is the difference in the radial orders being coupled, and $\langle B_{r,\mathrm{crit}}^{-2}\rangle$ is an average of $B_{r,\mathrm{crit}}$ over the g-mode cavity in buoyancy coordinate.
The shape factor $\tilde{\psi}_{\ell;mm';\Delta n}^2$ is of order unity if, roughly speaking, $B_{r,\mathrm{crit}}^{-2}$ varies significantly at the radial scale corresponding to $\Delta n$.
Like $H_{\alpha\beta}^{-1}$, it vanishes if modes are excluded from mixing by symmetry-induced selection rules (thereby rendering the approximation trivially valid).

The regimes of validity of perturbation theory (Equation \ref{eqn:pert_validity}) and the single-polarization waveguide description (Equation \ref{eqn:jwkb_validity}) can thus be seen to be complementary.
The validity condition for the single-polarization waveguide description is local in the sense of requiring the magnetic field at each radius to be strong enough to locally inhibit magnetogravity polarizations from mixing ($B_{0r}\gg B_{r,\mathrm{mix},\mathrm{wg}}$).
By contrast, the validity condition for perturbation theory is global in the sense of requiring the globally-averaged magnetic field to be too weak to significantly couple unperturbed modes ($\langle B_{0r}^2\rangle^{1/2}\ll B_{r,\mathrm{mix},\mathrm{pert}}$).
Comparing Equations \ref{eqn:jwkb_validity_main} and \ref{eqn:pert_validity_end_main}, we see that $B_{r,\mathrm{mix},\mathrm{wg}}$ and $B_{r,\mathrm{mix},\mathrm{pert}}$ are analogous quantities in the sense of scaling similarly with stellar properties.
However, their domains of applicability are not entirely mutually exclusive, overlapping in the limit of a weak magnetic field and slowly varying stellar background.
\edit{Bridging the gap between these two regimes requires the consideration of hitherto neglected terms, namely either higher-order JWKB terms within the waveguide description (cf. Appendix \ref{app:higher_wkb_waveguide}) or second-order frequency corrections due to radial-order coupling in perturbation theory.
Our forthcoming work (Liagre et al., in preparation) will explore the latter approach by considering coupling between radial orders.
Simultaneous consideration of both near-critical-field effects and polarization mixing (which appears to be generic) remains an outstanding theoretical challenge.}

% All else being equal, a strong field worsens the perturbative approximation but improves the adiabatic waveguide approximation, while sharper stratification doesn't matter for the perturbative approximation but makes the adiabatic approximation worse at the location of this sharp stratification.
% NZR: I don't think this is true -- this $\tilde{psi}^2$ quantity contains. This similar background-dependence as in r/H in the waveguide criterion is why I have refactored $\tilde{psi}^2$ onto the RHS of the pert theory criterion. Remember that this $\tilde{psi}^2$ is basically an inner product over something to do with the shape of the magnetic field, background, etc., so I think it morally contains the "FT" of the information content of r/H.

%% file: sect_conclusion.tex
%!TEX root=./main.tex

\section{Summary and outlook}

In this work, we have extended our non-perturbative formalism for predicting the frequencies of g modes which are affected by a strong, non-axisymmetric magnetorotational configurations.
This formalism describes magnetogravity waves as polarized disturbances propagating along a waveguide-like mode cavity, and generalizes the works of \citetalias{Rui:2023:MagneticSuppression} and \citetalias{Rui:2024:TARM} which require that the magnetic field be axisymmetric about the rotation axis.
The natural definitions of the ``magnetogravity polarizations'' relevant to this analysis exhibit a rich phemonenology which includes avoided crossings, mixing across angular degree $\ell$, and protection by discrete symmetries of the system even in the absence of continuous rotational symmetry.

We have focused on the case where magnetogravity waves propagate radially with a single ``magnetogravity polarization,'' which has been an unspoken assumption in previous works with similar methodologies (\citetalias{Rui:2023:MagneticSuppression} and \citetalias{Rui:2024:TARM}, as well as \citealt{Lecoanet:2017:MagneticConversion,Lecoanet:2022:HD43317,David:2025:Magnetogravity}).
However, we show that magnetogravity polarization mixing is likely to be common in realistic situations, for magnetic fields weaker than a special field strength $B_{r,\mathrm{mix},\mathrm{wg}}$ which is distinct from the critical magnetic field $B_{r,\mathrm{crit}}$ at which magnetic suppression is expected.
This polarization-mixing condition also approximately demarcates a hitherto unappreciated dividing line between the regimes of validity of global perturbation theory and the waveguide description, which are largely non-overlapping.
In this context, this work can be understood as a survey of the complexity of magnetogravity-mode behavior due to mixing in both $n$ and $\ell$, which has usually been ignored in the past.
In another study (Liagre et al., in preparation), we will provide an asymptotic description of magnetic near-degeneracy effects in order to extend the validity of the perturbative formalism to include these effects.

Although this work is primarily motivated by the study of magnetic red giants, the setup of the problem is generic and applicable to high-radial-order g-mode pulsators of all types.
Nevertheless, more work needs to be done to make asteroseismic predictions for g-mode pulsations under the full diversity of magnetorotational structures.
This study has only considered the case of pure magnetogravity modes, and has left aside the complex properties of pressure--gravity mixed modes relevant to realistic red-giant pulsations.
We have also ignored the horizontal component of the magnetic field $B_h$, whose effects may be important for predominantly toroidal fields expected to be generated by differential rotation \citep[e.g.,][]{Spruit:1999:DifferentialRotation,Spruit:2002:STDynamo,Fuller:2019:SlowingSpins}.
The recent discoveries of predominantly radial \citep[][see also \citealt{Lignieres:2024:MagnetoGravitoInertial}]{Ihallaine:2026:MagneticGammaDor} and toroidal \citep{Takata:2026:GDorToroidalField} magnetic fields in two different $\gamma$ Doradus stars illustrate the need to expand our understanding of the asteroseismic effects of magnetic fields across field strengths, magnetic geometries, and rotation rates.
Complementary approaches such as ray tracing \citep{Loi:2018:MGDynamicalChaos,Loi:2020:MGPackets,Mueller:2025:RayTracing} and traditional-approximation-like treatments of a purely toroidal field \citep{Mathis:2011:MagneticRotating,Dhouib:2022:ToroidalField} are promising for bridging this gap.

%% file: app_delta_mag_ell.tex
%!TEX root=./main.tex

\section{The average magnetic frequency shifts for any $\ell$} \label{app:delta_mag_ell}

In this Appendix, we derive Equation \ref{eqn:delta_omega_mag_ell} for the average magnetic shift $\delta\omega_{\mathrm{mag}}^\ell$ at arbitrary $\ell$.
While this result is referenced in \citet{Rui:2025:WDSeismology}, we are not aware of a published derivation.

Following \citet{Li:2022:30to100kG} and \citet{Das:2024:ComplexMagnetic}, the elements of the magnetic matrix $\mathbf{M}_\ell$ are given for high-radial-order modes by
\begin{equation} \label{eqn:M_ellmmp}
    M_{\ell;mm'} = \frac{1}{2\omega_{0,n\ell}}\langle\vec{\xi}_{h,n\ell m}^{(0)},\mathcal{L}_{\mathrm{mag}}\vec{\xi}_{h,n'\ell m'}^{(0)}\rangle_{\mathrm{3D}}\mathrm{,}
\end{equation}
where the eigenfunctions have been assumed to be normalized with respect to the inner product defined in Equation \ref{eqn:inner_product_3d}.

The inner product in Equation \ref{eqn:M_ellmmp} is given by Equation \ref{eqn:Lmag_coupling_integral}, so that Equation \ref{eqn:M_ellmmp} becomes
\begin{equation}
    M_{\ell;mm'} = \frac{1}{2\omega_{0,n\ell}}\langle\vec{\xi}_{h,n\ell m}^{(0)},\mathcal{L}_{\mathrm{mag}}\vec{\xi}_{h,n'\ell m'}^{(0)}\rangle_{\mathrm{3D}}\mathrm{,}
\end{equation}

Substituting the expression for $\vec{\xi}_{h,n\ell m}^{(0)}$ from Equation \ref{eqn:unpert_xih}, we obtain
\begin{equation}
    \begin{split}
        M_{\ell;mm'} = \frac{1}{\omega_{0,n\ell}^3X_N}&\int_{\mathcal{R}}\mathrm{d}x_N\,\omega_B^4\sin\left(\frac{n\pi}{X_N}x_N\right)^2\\
        &\times\iint\psi^2\bar{\nabla}_hY_{\ell m'}^*\cdot\bar{\nabla}_hY_{\ell m}\,\sin\theta\,\mathrm{d}\theta\,\mathrm{d}\phi\mathrm{.}
    \end{split}
\end{equation}
Under the stationary phase approximation, \edit{$\sin(n\pi x_N/X_N)^2\simeq1/2$} within the integral.
Then
\begin{equation} \label{eqn:M_with_subs}
    \begin{split}
        M_{\ell;mm'} = \frac{1}{8\pi\omega_{0,n\ell}^3X_N}&\int_{\mathcal{R}}\mathrm{d}r\,\frac{N^3B_{0r}^2}{\rho_0r^3}\\
        &\times\iint\psi^2\bar{\nabla}_hY_{\ell m'}^*\cdot\bar{\nabla}_hY_{\ell m}\,\sin\theta\,\mathrm{d}\theta\,\mathrm{d}\phi\mathrm{,}
    \end{split}
\end{equation}
where we have used the definition of $\omega_B$ in Equation \ref{eqn:omegaB} and the buoyancy coordinate \edit{$x_N$} in Equation \ref{eqn:buoyancy_coordinate}.
Using Equations \ref{eqn:K_weight} and \ref{eqn:curlyI}, Equation \ref{eqn:M_with_subs} becomes
\begin{equation}
    M_{\ell;mm'} = \frac{\mathscr{I}}{8\pi\omega_{0,n\ell}^3}\int_{\mathcal{R}}\mathrm{d}r\,K(r)B_{0r}^2\iint\psi^2\bar{\nabla}_hY_{\ell m'}^*\cdot\bar{\nabla}_hY_{\ell m}\,\sin\theta\,\mathrm{d}\theta\,\mathrm{d}\phi\mathrm{.}
\end{equation}

Since the trace of $\mathbf{M}_\ell$ is the sum of the frequency shifts in a single multiplet, the average magnetic shift $\delta\omega_{\mathrm{mag}}^\ell$ is
\begin{equation} \label{eqn:delta_omega_mag_sum}
    \begin{split}
        \delta\omega_{\mathrm{mag}}^\ell &= \frac{\mathrm{Tr}(\mathrm{M}_\ell)}{2\ell+1} \\
        &= \frac{\mathscr{I}}{8\pi(2\ell+1)\omega_{0,n\ell}^3}\int_{\mathcal{R}}\mathrm{d}r\,K(r)B_{0r}^2\\
        &\quad\times\iint\psi^2\left(\sum^{+\ell}_{m=-\ell}\left\lvert\bar{\nabla}_hY_{\ell m}\right\rvert^2\right)\sin\theta\,\mathrm{d}\theta\,\mathrm{d}\phi\mathrm{.}
    \end{split}
\end{equation}

To make progress, we notice that spherical harmonics obey Uns\"old's theorem \citep{Unsoeld:1927:SpharmTheorem}:
\begin{equation} \label{eqn:spharm_sum}
    \sum^{+\ell}_{m=-\ell}|Y_{\ell m}(\theta,\phi)|^2 = \frac{2\ell+1}{4\pi}\mathrm{.}
\end{equation}
Applying the two-dimensional Laplacian $\bar{\nabla}_h^2$ to Equation \ref{eqn:spharm_sum} gives
\begin{equation} \label{eqn:spharm_sum_2}
    \sum^{+\ell}_{m=-\ell}\left(\left(\bar{\nabla}_h^2Y_{\ell m}^*\right)Y_{\ell m} + Y_{\ell m}^*\left(\bar{\nabla}_h^2Y_{\ell m}\right) + 2\left\lvert\bar{\nabla}_hY_{\ell m}\right\rvert^2\right) = 0\mathrm{.}
\end{equation}
We can use $\bar{\nabla}_h^2Y_{\ell m}=-\ell(\ell+1)Y_{\ell m}$ to write \ref{eqn:spharm_sum_2} as
\begin{equation} \label{eqn:spharm_identity}
    \sum^{+\ell}_{m=-\ell}\left\lvert\bar{\nabla}_hY_{\ell m}\right\rvert^2 = \frac{\ell(\ell+1)(2\ell+1)}{4\pi}\mathrm{.}
\end{equation}

However, we recognize the sum in Equation \ref{eqn:spharm_identity} as appearing in Equation \ref{eqn:delta_omega_mag_sum}.
Our expression for $\delta\omega_{\mathrm{mag}}^\ell$ then becomes
\begin{equation} \label{eqn:delta_omega_mag_ell_final}
    \delta\omega_{\mathrm{mag}}^\ell = \frac{\ell(\ell+1)\mathscr{I}}{8\pi\omega_{0,n\ell}^3}\int_{\mathcal{R}}\mathrm{d}r\,K(r)B_{0r}^2\mathrm{,}
\end{equation}
where we have used the normalization convention of $\psi^2$ (Equation \ref{eqn:psi2_norm}).
We recognize Equation \ref{eqn:delta_omega_mag_ell_final} as our desired result (Equation \ref{eqn:delta_omega_mag_ell}).

%% file: app_quadrupole_eigenvalues.tex
%!TEX root=./main.tex
\section{Quadrupole eigenvalues and avoided crossings for inclined dipole fields} \label{app:quadrupole_eigenvalues}

\begin{figure*}
    \centering
    \includegraphics[width=\linewidth]{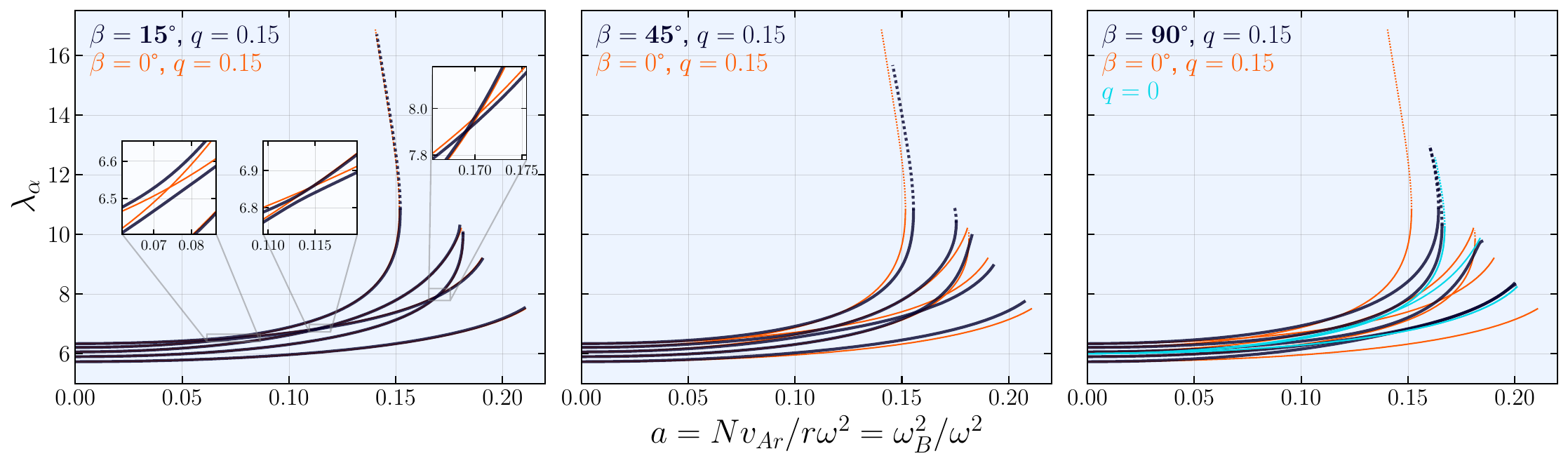}
    \caption{Same as Figure \ref{fig:FIG_dipole_eigenvalues_1}, but for quadrupole ($\ell=2$) polarizations under inclined dipolar magnetic fields.
    Insets on the left panel zoom in on three sites of avoided crossings which occur in the quintuplet.}
    \label{fig:FIG_dipole_eigenvalues_2}
\end{figure*}

\begin{figure*}
    \centering
    \includegraphics[width=\linewidth]{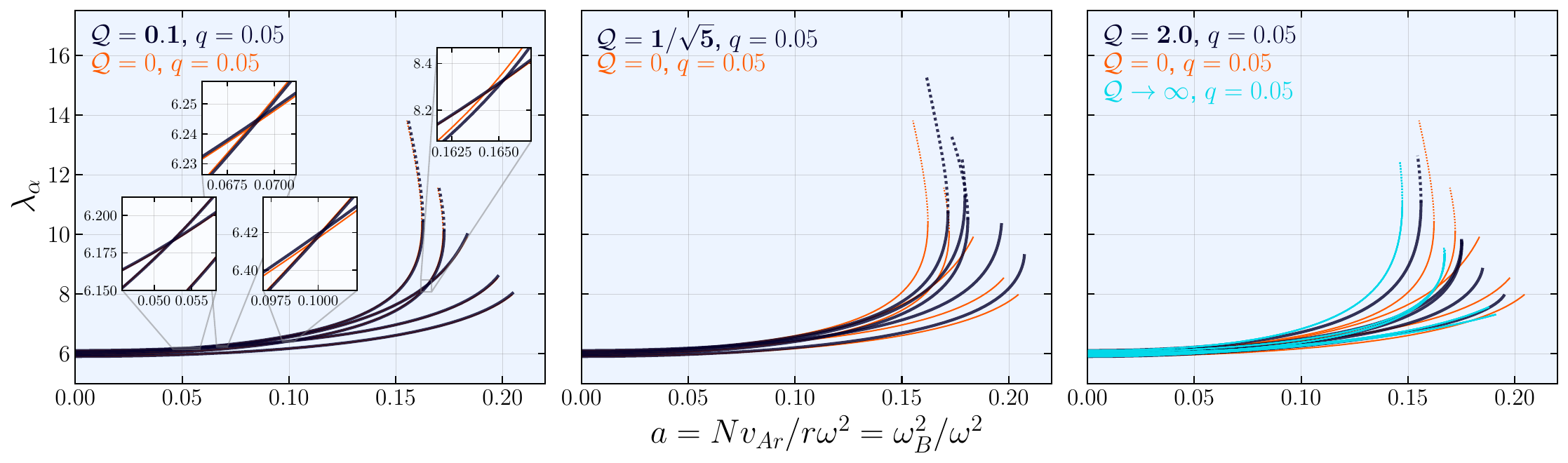}
    \caption{Same as Figure \ref{fig:FIG_dipole_eigenvalues_2}, but for the non-axisymmetric dipole plus quadrupole geometry described in Section \ref{sec:dipole_plus_y22}.
    As in Figure \ref{fig:FIG_dpy22_eigenvalues_1} for the dipole polarizations, the insets show that the non-axisymmetric quadrupolar component of $\psi$ does not cause mode repulsion.}
    \label{fig:FIG_dpy22_eigenvalues_2}
\end{figure*}

This Appendix presents the eigenvalues $\lambda$ versus $a$ for quadrupole polarizations under the inclined-dipole (Figure \ref{fig:FIG_dipole_eigenvalues_2}) and non-axisymmetric (Figure \ref{fig:FIG_dpy22_eigenvalues_2}) field geometries described in Sections \ref{sec:inclined_dipole} and \ref{sec:dipole_plus_y22}, respectively.
Eigenvalues for the quadrupole polarizations behave similarly to their dipole counterparts (Figures \ref{fig:FIG_dipole_eigenvalues_1} and \ref{fig:FIG_dpy22_eigenvalues_1}).

%% file: app_higher_wkb_waveguide.tex
%!TEX root=./main.tex
\section{Validity condition for single-polarization waveguide description} \label{app:higher_wkb_waveguide}

In this Appendix, we derive a validity condition for the single-polarization waveguide description, which involves finding the conditions under which different magnetogravity polarizations can couple with each other.

While Equations \ref{eqn:jwkb_leading} include only leading-order JWKB terms, if next-to-leading-order terms are included and Equations \ref{eqn:transverse} are asserted, Equations \ref{eqn:final_momentum} and \ref{eqn:final_continuity} become
\begin{subequations} \label{eqn:jwkb_next_order}
    \begin{equation} \label{eqn:jwkb_next_order_momentum}
        0 = \sum_\alpha e^{-iS_\alpha/\epsilon}\left\lbrace2k_{r,\alpha}A_\alpha\vec{\zeta}_\alpha' + 2k_{r,\alpha}A_\alpha'\vec{\zeta}_\alpha + k_{r,\alpha}'A_\alpha\vec{\zeta}_\alpha\right\rbrace
    \end{equation} \\
    \begin{equation} \label{eqn:jwkb_next_order_continuity}
        0 = \sum_\alpha e^{-iS_\alpha/\epsilon}\left\lbrace2k_{r,\alpha}A_\alpha\pi_\alpha' + 2k_{r,\alpha}A_\alpha'\pi_\alpha + k_{r,\alpha}'A_\alpha\pi_\alpha\right\rbrace\mathrm{,}
    \end{equation}
\end{subequations}
where for simplicity we have still neglected derivatives of equilibrium quantities.

Equations \ref{eqn:transverse} can be written into a single second-order differential eigenvalue problem by eliminating $\vec{\zeta}_\alpha$:
\begin{equation} \label{eqn:single_operator}
    \lambda_\alpha\pi_\alpha + \bar{\nabla}_h\cdot\left[\frac{(1-b_\alpha^2\psi^2)\bar{\nabla}_h\pi_\alpha + iq\mu\hat{r}\times\bar{\nabla}_h\pi_\alpha}{(1-b_\alpha^2\psi^2)^2 - q^2\mu^2}\right] = 0\mathrm{.}
\end{equation}
The differential operator in Equation \ref{eqn:single_operator} is manifestly Hermitian as long as \edit{the denominator $(1-b_\alpha^2\psi^2)^2-q^2\mu^2$ does not change sign over the sphere.
This condition is satisfied as long as} $b_\alpha$ and $q$ are not \edit{too} large.
\edit{Under these conditions, t}he normalized pressure perturbations are orthogonal under the usual inner product over the unit sphere:
\begin{equation} \label{eqn:inner_product}
    \langle\pi_\alpha,\pi_\beta\rangle_{\mathrm{2D}} = \delta_{\alpha\beta}\mathrm{,}
\end{equation}
where
\begin{equation}
    \langle f,g\rangle_{\mathrm{2D}} \equiv \int f^*g\,\sin\theta\,\mathrm{d}\theta\,\mathrm{d}\phi\mathrm{.}
\end{equation}
The fluid displacements $\vec{\zeta}_\alpha$ can then be recovered using Equations \ref{eqn:transverse}.

The usual next step in a JWKB analysis would be to use this inner product to project Equations \ref{eqn:jwkb_next_order} onto each magnetogravity polarization to derive a system of transport equations.
The most serious issue with this is that, while the operator in Equation \ref{eqn:single_operator} is Hermitian at fixed $b_\alpha$, each magnetogravity polarization does not have the same $b_\alpha$ at a fixed radius but rather a fixed $a=b_\alpha/\sqrt{\lambda_\alpha}$.
Therefore, at a given radius, the magnetogravity polarizations are eigenfunctions of \textit{different} operators and are therefore not orthogonal to each other under the inner product in Equation \ref{eqn:inner_product}.
This makes the usual projection step difficult to perform.

While this issue should be explored in the future, here for illustrative purposes we specialize to slow rotation ($q\ll1$) and weak magnetic fields ($b_\alpha\approx a\sqrt{\ell(\ell+1)}\ll1$).
In this limit, Equation \ref{eqn:jwkb_next_order_momentum} is approximately satisfied since all of its terms (which were already next-to-leading JWKB order) are multiplied by $b_\alpha\ll1$.
Moreover, since the polarizations are now given by perturbation theory (Section \ref{sec:perturbation_theory}), \edit{each of them is a linear combination of spherical harmonics of a single fixed $\ell$}.
Also, in this limit, polarizations of the same $\ell$ correspond to the same $b_\alpha=a/\sqrt{\ell(\ell+1)}$ at fixed $a$, and can therefore be chosen to be orthonormal with respect to each other.
Therefore, the non-trivial inner product $\langle\pi_\alpha,\pi_\beta'\rangle_{\mathrm{2D}}$ which appears when projecting Equation \ref{eqn:jwkb_next_order_continuity} is only nonzero if the polarizations match in $\ell$.
Performing this projection and simplifying yields the \textit{transport equations}:
\begin{equation} \label{eqn:transport}
    k_{r,\alpha}A_\alpha' = -\frac{1}{2}k_{r,\alpha}'A_\alpha + \sum_\beta ik_{r,\beta}\Gamma_{\alpha\beta}A_\beta e^{i(S_\alpha-S_\beta)/\epsilon}\mathrm{,}
\end{equation}
where we have defined the matrix elements
\begin{equation} \label{eqn:berry_connection}
    \Gamma_{\alpha\beta} = i\langle\pi_\alpha,\pi_\beta'\rangle_{\mathrm{2D}}\mathrm{.}
\end{equation}
\edit{The connection matrix $\mathbf{\Gamma}$ stores information about how a magnetogravity wave's polarization changes as it propagates radially \citep[this is formally a non-Abelian Berry connection;][]{Wilczek:1984:GaugeStructure,Shapere:1989:GeometricPhases}.}
$\mathbf{\Gamma}$ can be seen to be self-adjoint by noticing that
\begin{equation}
    \frac{\mathrm d}{\mathrm d r}\langle\pi_\alpha,\pi_\beta\rangle_{\mathrm{2D}} = \langle\pi_\alpha,\pi_\beta'\rangle_{\mathrm{2D}} + \langle\pi_\beta,\pi_\alpha'\rangle_{\mathrm{2D}}^* = \frac{\mathrm d\delta_{\alpha\beta}}{\mathrm d r} = 0
\end{equation}
so that
\begin{equation}
    \Gamma_{\alpha\beta} = \Gamma_{\beta\alpha}^*\mathrm{.}
\end{equation}
The diagonal elements of $\mathbf{\Gamma}$ \edit{(formally the classical Berry connections)} describe the geometric phase accumulated by a single polarization as it propagates in radius.
Although they can be chosen to vanish at a single part of parameter space using cleverly chosen overall complex phases, they cannot be made to vanish in general \edit{(if there is nonzero Berry curvature)}.
The off-diagonal elements of $\mathbf{\Gamma}$ store information about how polarizations ``mix'' as a magnetogravity wave propagates.
These elements encode geometrical information about how the polarizations overlap with each other as they change due to the changing background and, as such, will vanish if mixing between two polarization states is forbidden by symmetry-enforced selection rules.
% The diagonal elements of $\mathbf{\Gamma}$ are nonzero in general, but are gauge-dependent.
% In the absence of Berry curvature, these diagonal elements can be chosen to vanish with a cleverly chosen overall complex phase.
% In the illustrative case which follows, we assume these diagonal elements can be ignored.

While the first term on the right-hand side of Equation \ref{eqn:transport} simply imposes an adiabatic relationship between $k_{r,\alpha}$ and $A_\alpha$, the second term explicitly allows for exchange between polarizations.
This exchange is controlled by the \edit{difference} in the actions $\Delta S_{\alpha\beta}\equiv S_\alpha-S_\beta$ and the degree to which the changing structure of a polarization overlaps with other polarizations via $\mathbf{\Gamma}$.
\edit{We reiterate that the adjective ``adiabatic'' here refers to the preservation of polarization state during magnetogravity-wave propagation, rather than anything to do with heat exchange between fluid parcels.}

We make the substitution
\begin{equation}
    A_\alpha=\frac{1}{\sqrt{k_{r,\alpha}}}e^{i\int\Gamma_{\alpha\alpha}\mathrm{d}r}\bar{A}_\alpha\mathrm{,}
\end{equation}
explicitly factoring out adiabatic evolution of the amplitude as well as the contribution from the classical Berry phase.
This yields the rate equations
\begin{equation} \label{eqn:general_rate}
    \bar{A}_\alpha' = \sum_{\beta\neq\alpha}ik_{r,\beta}\Gamma_{\alpha\beta}\sqrt{\frac{k_{r,\alpha}}{k_{r,\beta}}}\bar{A}_\beta e^{i(S_\alpha-S_\beta)/\epsilon-i\int(\Gamma_{\alpha\alpha}-\Gamma_{\beta\beta})\mathrm{d}r}\mathrm{.}
\end{equation}
To gain intuition, we schematically consider the case of only two nearly-degenerate polarizations $\alpha$ and $\beta$, under which Equation \ref{eqn:general_rate} becomes
\begin{subequations} \label{eqn:two_modes}
    \begin{align}
        \bar{A}_\alpha' &= i\Gamma_{\alpha\beta}\bar{A}_\beta e^{+i(S_\alpha-S_\beta)/\epsilon} \\
        \bar{A}_\beta' &= i\Gamma_{\alpha\beta}^*\bar{A}_\alpha e^{-i(S_\alpha-S_\beta)/\epsilon}\mathrm{,}
    \end{align}
\end{subequations}
where we have ignored small differences in the wavenumber except when they are exponentiated, and ignored the small contribution from the classical Berry phase.
Equations \ref{eqn:two_modes} very closely resemble the equations describing Rabi oscillations in the quantum mechanical two-level system \citep{Griffiths:2016:QM}.
This analogy motivates the following definition:
\begin{subequations}
    \begin{gather}
        C_\alpha \equiv \bar{A}_\alpha e^{-iS_\alpha/\epsilon} \\
        C_\beta \equiv \bar{A}_\beta e^{-iS_\beta/\epsilon}\mathrm{,}
    \end{gather}
\end{subequations}
which is mathematically equivalent to going into the interaction picture in quantum mechanics.
The Equations \ref{eqn:two_modes} then take the form
\begin{subequations} \label{eqn:two_modes_heisenberg}
    \begin{align}
        C_\alpha' &= -ik_{r,\alpha}C_\alpha + i\Gamma_{\alpha\beta}C_\beta \\
        C_\beta' &= +i\Gamma_{\alpha\beta}^*C_\alpha - ik_{r,\beta}C_\beta\mathrm{,}
    \end{align}
\end{subequations}
which can be solved exactly.

Equations \ref{eqn:two_modes_heisenberg} can be written as the matrix equation
\begin{equation}
    \vec{C}' = -i\mathbf{K}\vec{C}
\end{equation}
where $\vec{C}=(C_\alpha,C_\beta)$, and
\begin{subequations}
    \begin{gather}
        \mathbf{K} =
        \begin{pmatrix}
            k_{r,\alpha} & -\Gamma_{\alpha\beta} \\
            -\Gamma_{\alpha\beta}^* & k_{r,\beta}
        \end{pmatrix}\mathrm{.}
    \end{gather}
\end{subequations}
The eigenvalues of $\mathbf{K}$ should be interpreted as wavenumbers,
\begin{equation} \label{eqn:K_evals}
    k_{r,\pm} = \bar{k}_{r,\alpha\beta} \pm \frac{1}{2}\sqrt{\Delta k_{r,\alpha\beta}^2 + 4|\Gamma_{\alpha\beta}|^2}\mathrm{,}
\end{equation}
where $\bar{k}_{r,\alpha\beta}=(k_{r,\alpha}+k_{r,\beta})/2$ and $\Delta k_{r,\alpha\beta}=k_{r,\alpha}-k_{r,\beta}$ are the carrier (averaged) and beat wavenumbers between polarizations $\alpha$ and $\beta$, respectively.
These eigenvalues correspond to the eigenvectors
\begin{subequations} \label{eqn:K_evecs}
    \begin{gather}
        \vec{C}_+ =
        \begin{pmatrix}
            \frac{1}{2}\Delta k_{r,\alpha\beta} + \frac{1}{2}\sqrt{\Delta k_{r,\alpha\beta}^2 + 4|\Gamma_{\alpha\beta}|^2}\\ -\Gamma_{\alpha\beta}^*
        \end{pmatrix}\\
        \vec{C}_- =
        \begin{pmatrix}
            +\Gamma_{\alpha\beta}\\ \frac{1}{2}\Delta k_{r,\alpha\beta} + \frac{1}{2}\sqrt{\Delta k_{r,\alpha\beta}^2 + 4|\Gamma_{\alpha\beta}|^2}
        \end{pmatrix}\mathrm{.}
    \end{gather}
\end{subequations}

The behavior of the system then depends sensitively on the hierarchy between $|\Delta k_{r,\alpha\beta}|$ and $|\Gamma_{\alpha\beta}|$.
In particular, taking $k_{r,\alpha}>k_{r,\beta}$ without loss of generality, when $|\Delta k_{r,\alpha\beta}|\gg2|\Gamma_{\alpha\beta}|$,
\begin{subequations}
    \begin{gather}
        k_{r,+} \rightarrow k_{r,\alpha} \;\;\mathrm{with}\;\; \vec{C}_+ \rightarrow \begin{pmatrix}1\\0\end{pmatrix}\\
        k_{r,-} \rightarrow k_{r,\beta} \;\;\mathrm{with}\;\; \vec{C}_- \rightarrow \begin{pmatrix}0\\1\end{pmatrix}\mathrm{.}
    \end{gather}
\end{subequations}
This regime corresponds to the adiabatic limit, within which $k_{r,\alpha}$ and $k_{r,\beta}$ are well-separated enough that the two polarizations decouple.
In the complementary regime, when $|\Delta k_{r,\alpha\beta}|\ll2|\Gamma_{\alpha\beta}|$,
\begin{subequations}
    \begin{gather}
        k_{r,+} \rightarrow \bar{k}_{r,\alpha\beta}+|\Gamma_{\alpha\beta}| \;\;\mathrm{with}\;\; \vec{C}_+ \rightarrow \begin{pmatrix}e^{i\Theta_{\alpha\beta}}\\-1\end{pmatrix}\\
        k_{r,-} \rightarrow \bar{k}_{r,\alpha\beta}-|\Gamma_{\alpha\beta}| \;\;\mathrm{with}\;\; \vec{C}_- \rightarrow \begin{pmatrix}e^{i\Theta_{\alpha\beta}}\\+1\end{pmatrix}\mathrm{,}
    \end{gather}
\end{subequations}
where $\Theta_{\alpha\beta}=\arg(\Gamma_{\alpha\beta})$.
Since the eigenvectors $\vec{C}_\pm$ have equal support in each polarization in this limit, the two polarizations exchange polarization on the beat wavenumber $2(k_{r,+}-k_{r,-})=2|\Gamma_{\alpha\beta}|$.
The definition of $\Gamma_{\alpha\beta}$ (Equation \ref{eqn:berry_connection}) supplies the interpretation that this polarization-exchange wavelength corresponds to the length scale $H_{\alpha\beta}\equiv|\Gamma_{\alpha\beta}|^{-1}$ over which $\pi_\alpha$ and $\pi_\beta$ ``rotate'' into each other.
This analysis defines an approximate condition for adiabaticity:
\begin{equation} \label{eqn:jwkb_adiabatic_orig}
    2|\Gamma_{\alpha\beta}| = 2/H_{\alpha\beta} \ll \Delta k_{r,\alpha\beta}\mathrm{,}
\end{equation}
cf. Equation \ref{eqn:jwkb_adiabatic_orig_main}.
Although the presence or absence of non-adiabatic dynamics clearly depends sensitively balance between $|\Delta k_{\alpha\beta}|$ and $|\Gamma_{\alpha\beta}|$, these quantities can still both remain very small relative to, e.g., the carrier wavenumber $\bar{k}_{r,\alpha\beta}$.

At weak fields and slow rotation, the typical difference between wavenumbers of the same $\ell$ are of the form
\begin{equation} \label{eqn:estimate_dkr}
    \Delta k_{r,\alpha\beta} \sim \frac{N}{\omega r}\left(\kappa_aa^2 + \kappa_qq\right)\mathrm{,}
\end{equation}
where $\kappa_a$ and $\kappa_q$ are dimensionless numbers which are typically of order unity.
At transitions between rotational and magnetic alignment, the terms in Equation \ref{eqn:estimate_dkr} are of similar size.
Even without rotation, polarization mixing can still happen if the horizontal structure of the magnetic field varies with radius, i.e., due to the radial dependence of $\psi(\theta,\phi;r)$.
We reiterate that adiabaticity can still break even if the magnetic field profile changes much more slowly than the gravity wave's radial wavelength, since Equation \ref{eqn:jwkb_adiabatic_orig} is only a condition on $\Delta k_{r,\alpha\beta}$ and does not depend on $\bar{k}_{r,\alpha\beta}$ at all.
In either case, it is roughly justified to drop the rotational term in Equation \ref{eqn:estimate_dkr} and set $\kappa_a=[\ell(\ell+1)]^{3/2}/2$ (implied by Equations \ref{eqn:delta_omega_mag} and \ref{eqn:lambda_pert}).

Recalling the definition of $a$ (Equation \ref{eqn:a_param}), Equation \ref{eqn:jwkb_adiabatic_orig} can be rewritten as a validity condition for the adiabatic approximation underpinning the single-polarization waveguide description:
\begin{equation}
    \frac{1}{H_{\alpha\beta}} \ll \frac{\left[\ell(\ell + 1)\right]^{3/2}N^3B_{0r}^2}{16\pi\rho_0\omega^5r^3}.
\end{equation}
This can in turn be rewritten a bound on the magnetic field $B_{0r}$ relative to the critical field strength $B_{r,\text{crit}}$,
\begin{equation} \label{eqn:jwkb_validity}
    B_{0r} \gg \frac{4}{[\ell(\ell+1)]^{1/4}}\left(\frac{\omega}{N}\right)^{1/2}\left(\frac{r}{H_{\alpha\beta}}\right)^{1/2}B_{r,\mathrm{crit}}\mathrm{,}
\end{equation}
where for the prefactor in Equation \ref{eqn:Brcrit} for $B_{r,\mathrm{crit}}$ we adopt $\sqrt{\pi/\ell(\ell+1)}$ \citep[cf.][]{Fuller:2015:SuppressedDipole}.
Equation \ref{eqn:jwkb_validity} is the desired result which we have quoted in Equation \ref{eqn:jwkb_validity_main} in the main text.

%% file: app_higher_wkb_derivation.tex
%!TEX root=./main.tex
\section{Conservation of magnetogravity-wave flux} \label{app:higher_wkb}

In Section \ref{sec:higher_wkb}, we show that, under certain conditions, magnetogravity waves can non-adiabatically exchange amplitude between magnetogravity polarizations.
In this Appendix, we show for weakly perturbed polarizations that such exchanges are constrained by a conservation-like equation for the total wave flux.

Upon multiplying Equation \ref{eqn:transport} by $A_\alpha^*$ and taking the real part, we obtain
\begin{equation} \label{eqn:multiplied_transport}
    k_{r,\alpha}(|A_\alpha|^2)' = -k_{r,\alpha}'|A_\alpha|^2 - \sum_{\beta\neq\alpha}2k_{r,\beta}\mathrm{Im}\left[\Gamma_{\alpha\beta}A_\alpha^*A_\beta e^{i(S_\alpha-S_\beta)/\epsilon}\right]\mathrm{.}
\end{equation}

The first two terms in Equation \ref{eqn:multiplied_transport} can be combined into the derivative of a single wave flux, such that
\begin{equation} \label{eqn:transport_flux}
    F_\alpha' + \sum_{\beta\neq\alpha}2k_{r,\beta}\mathrm{Im}\left[\Gamma_{\alpha\beta}A_\alpha^*A_\beta e^{i(S_\alpha-S_\beta)/\epsilon}\right] = 0\mathrm{,}
\end{equation}
where
\begin{equation}
    F_\alpha \equiv k_{r,\alpha}|A_\alpha|^2\mathrm{.}
\end{equation}

We can multiply Equation \ref{eqn:transport_flux} by $k_{r,\alpha}$ and sum over polarization index $\alpha$ to obtain
\begin{equation} \label{eqn:transport_flux_summed}
    \sum_\alpha k_{r,\alpha}F_\alpha' + \sum_\alpha\sum_{\beta\neq\alpha}2k_{r,\alpha}k_{r,\beta}\mathrm{Im}\left[\Gamma_{\alpha\beta}A_\alpha^*A_\beta e^{i(S_\alpha-S_\beta)/\epsilon}\right] = 0\mathrm{.}
\end{equation}

The nested sum in the second term of Equation \ref{eqn:transport_flux_summed} vanishes.
To see this, we can partition the summand into two identical terms, and exchange $\alpha\leftrightarrow\beta$:
\begin{equation} \label{eqn:nested_sum_cancel}
    \begin{split}
        \sum_\alpha\sum_{\beta\neq\alpha}&2k_{r,\alpha}k_{r,\beta}\mathrm{Im}\left[\Gamma_{\alpha\beta}A_\alpha^*A_\beta e^{i(S_\alpha-S_\beta)/\epsilon}\right] \\
        &= \sum_\alpha\sum_{\beta\neq\alpha}2k_{r,\alpha}k_{r,\beta}\Bigl\lbrace\mathrm{Im}\left[\Gamma_{\alpha\beta}A_\alpha^*A_\beta e^{i(S_\alpha-S_\beta)/\epsilon}\right] \\
        &\quad+ \mathrm{Im}\left[\Gamma_{\beta\alpha}A_\beta^*A_\alpha e^{i(S_\beta-S_\alpha)/\epsilon}\right]\Bigr\rbrace\mathrm{.}
    \end{split}
\end{equation}

By noticing $\Gamma_{\alpha\beta}=\Gamma_{\beta\alpha}^*$, it can be seen that the arguments of the $\mathrm{Im}(\circ)$ operations on the right-hand side are complex conjugates of each other.
However, because complex conjugation flips the imaginary part, the two terms on the right-hand side of Equation \ref{eqn:nested_sum_cancel} cancel out.

We are left with a simple form for a conservation-like equation for the wave flux:
\begin{equation}
    \sum_\alpha k_{r,\alpha}F_\alpha' = 0\mathrm{.}
\end{equation}

%% file: app_higher_wkb_pert.tex
%!TEX root=./main.tex
\section{Validity condition for global perturbation theory} \label{app:higher_wkb_pert}

In this Appendix, we derive a validity condition for global perturbation theory, which involves finding the conditions under which the magnetic perturbation can couple unperturbed modes of different radial order.

Equation \ref{eqn:xi_linear_combo} gives the zeroth-order g-mode eigenfunction in perturbation theory as a linear combination of initially degenerate modes.
The first-order perturbation to the eigenfunction is given by a sum over states outside of the degenerate subspace:
\begin{equation} \label{eqn:perturbed_xih}
    \delta\vec{\xi}_h = \sum_{(n',\ell')\neq(n,\ell),m'}c_{n'\ell'm'}\vec{\xi}^{(0)}_{h,n'\ell'm'}\mathrm{,}
\end{equation}
where, for asymptotic g modes, Section 16 of \citet{Unno:1979:NonradialOsc} gives
\begin{equation} \label{eqn:asymptotic_g_mode}
    \vec{\xi}^{(0)}_{h,n\ell m} \approx A\rho^{-1/2}r^{-3/2}N^{1/2}\sin\left(\int^r_{r_{\mathrm{in}}}k_r(r')\,\mathrm{d}r'\right)\bar{\nabla}_hY_{\ell m}\mathrm{.}
\end{equation}
In Equation \ref{eqn:asymptotic_g_mode}, $r_{\mathrm{in}}$ is the radius of the inner boundary of $\mathcal{R}$, and we ignore constant phase offsets, which are small.
The natural independent variable is the buoyancy coordinate \edit{$x_N$}, defined as
\begin{equation} \label{eqn:buoyancy_coordinate}
    x_N(r) = \int^r_{r_{\mathrm{in}}}\frac{N(r')}{r'}\mathrm{d}r'\mathrm{.}
\end{equation}
The g-mode cavity has a natural buoyancy radius \edit{$X_N=x_N(r_{\mathrm{out}})$}, where $r_{\mathrm{out}}$ is the radius at the outer boundary of the g-mode cavity.
Normalization of the unperturbed eigenfunctions with respect to the inner product
\begin{equation} \label{eqn:inner_product_3d}
    \langle\vec{f},\vec{g}\rangle_{\mathrm{3D}} = \iiint\vec{f}^*\cdot\vec{g}\,\rho r^2\,\sin\theta\,\mathrm{d}r\,\mathrm{d}\theta\,\mathrm{d}\phi
\end{equation}
requires \edit{$A=\sqrt{2/\ell(\ell+1)X_N}$}. The subscript ``3D'' distinguishes the inner product in Equation \ref{eqn:inner_product_3d} from the inner product in Equation \ref{eqn:inner_product} which appeared in the discussion related to JWKB theory.

Using also the fact that g modes are evenly spaced in period as
\begin{equation}
    \omega_{0,n\ell} = \frac{\sqrt{\ell(\ell+1)}}{n\pi}X_N\mathrm{,}
\end{equation}
the unperturbed eigenfunctions become
\begin{equation} \label{eqn:unpert_xih}
    \vec{\xi}^{(0)}_{h,n\ell m} \approx \sqrt{\frac{2}{\ell(\ell+1)X_N}}\rho^{-1/2}r^{-3/2}N^{1/2}\sin\left(\frac{n\pi}{X_N}x_N\right)\bar{\nabla}_hY_{\ell m}\mathrm{.}
\end{equation}

The coefficients $c_{n'\ell'm'}$ which appear in Equation \ref{eqn:perturbed_xih} are given by \edit{the usual perturbative expression for the first-order correction to the eigenfunctions:}
\begin{equation} \label{eqn:perturbed_cnlm}
    c_{n'\ell'm'}=\sum_m\frac{\langle\vec{\xi}_{h,n'\ell'm'}^{(0)},\mathcal{L}_{\mathrm{mag}}\vec{\xi}_{h,n\ell m}^{(0)}\rangle_{\mathrm{3D}}}{\omega_{0,n\ell}^2-\omega_{0,n'\ell'}^2}c_{n\ell m}\mathrm{,}
\end{equation}
where the Lorentz operator is approximately
\begin{equation}
    \mathcal{L}_{\mathrm{mag}}[\vec{\xi}_h] = -v_{Ar}^2\psi^2\partial_r^2\vec{\xi}_h\mathrm{.}
\end{equation}
\edit{Intuitively, as can be seen in Equation \ref{eqn:perturbed_cnlm}, the value of $c_{n'\ell'm'}$ is most strongly influenced by pairs of eigenfunctions which are both nearby in unperturbed frequency and efficiently coupled by the perturbation $\mathcal{L}_{\mathrm{mag}}$.}
Although perturbation theory assumes that $|c_{n'\ell'm'}|\ll1$, Equation \ref{eqn:perturbed_cnlm} shows that $|c_{n'\ell'm'}|$ can be large if the unperturbed frequency separation is smaller than the coupling introduced by $\mathcal{L}_{\mathrm{mag}}$.
That is to say, the condition for the validity of perturbation theory is that
\begin{equation}
    \langle\vec{\xi}_{h,n'\ell'm'}^{(0)},\mathcal{L}_{\mathrm{mag}}\vec{\xi}_{h,n\ell m}^{(0)}\rangle_{\mathrm{3D}} \ll \omega_{0,n\ell}^2-\omega_{0,n'\ell'}^2\mathrm{.}
\end{equation}
Note the algebraic similarity with Equation \ref{eqn:jwkb_adiabatic_orig} for the adiabatic approximation: in both cases, the size of an off-diagonal matrix element, which may be zero under symmetry-determined selection rules, has to be small relative to the difference between two on-diagonal elements in order for the respective approximations to hold good.

Hereafter, for illustrative purposes, we consider only coupling between different radial orders ($n'\neq n$), fixing $\ell=\ell'$.
The matrix element which appears in Equation \ref{eqn:perturbed_cnlm} is
\begin{equation} \label{eqn:Lmag_coupling_integral}
    \begin{split}
        \langle&\vec{\xi}_{h,n'\ell m'}^{(0)},\mathcal{L}_{\mathrm{mag}}\vec{\xi}_{h,n\ell m}^{(0)}\rangle_{\mathrm{3D}} \\
        &= \frac{1}{\omega_{0,n\ell}^2}\frac{2}{X_N}\int_{\mathcal{R}}\mathrm{d}x_N\,\omega_B^4\sin\left(\frac{n'\pi}{X_N}x_N\right)\sin\left(\frac{n\pi}{X_N}x_N\right)\\
        &\times\iint\psi^2\bar{\nabla}_hY_{\ell m'}^*\cdot\bar{\nabla}_hY_{\ell m}\,\sin\theta\,\mathrm{d}\theta\,\mathrm{d}\phi\mathrm{.}
    \end{split}
\end{equation}
Although the angular integral in Equation \ref{eqn:Lmag_coupling_integral} can be written exactly in terms of the spherical harmonic coefficients of $\psi^2$, for now it suffices to define
\begin{equation} \label{eqn:tilde_psi2}
    4\pi\ell(\ell+1)\tilde{\psi}_{\ell;mm'}^2(r) = \iint\psi^2\bar{\nabla}_hY_{\ell m'}^*\cdot\bar{\nabla}_hY_{\ell m}\,\sin\theta\,\mathrm{d}\theta\,\mathrm{d}\phi\mathrm{.}
\end{equation}
This definition normalizes $|\tilde{\psi}_{\ell;mm'}^2|$ to order unity, although it also encodes geometrical information about the angular integral in Equation \ref{eqn:tilde_psi2}.
If symmetry considerations in $\psi$ should forbid any coupling between $m$ and $m'$, then corresponding selection rules will cause this integral to vanish.
Under this definition,
\begin{equation}
    \begin{split}
        \langle&\vec{\xi}_{h,n'\ell m'}^{(0)},\mathcal{L}_{\mathrm{mag}}\vec{\xi}_{h,n\ell m}^{(0)}\rangle_{\mathrm{3D}} \\
        &= \frac{4\pi\ell(\ell+1)}{\omega_{0,n\ell}^2}\frac{2}{X_N}\int_{\mathcal{R}}\omega_B^4\tilde{\psi}_{\ell;mm'}^2\sin\left(\frac{n'\pi}{X_N}x_N\right)\sin\left(\frac{n\pi}{X_N}x_N\right)\mathrm{d}x_N\mathrm{.}
    \end{split}
\end{equation}

We observe the trigonometric identity
\begin{equation} \label{eqn:trig_identity}
    \begin{split}
        \sin&\left(\frac{n'\pi}{X_N}x_N\right)\sin\left(\frac{n\pi}{X_N}x_N\right) \\
        &= \frac{1}{2}\left[\cos\left(\frac{(n'-n)\pi}{X_N}x_N\right) - \cos\left(\frac{(n'+n)\pi}{X_N}x_N\right)\right]\mathrm{,}
    \end{split}
\end{equation}
where the second term varies very quickly in radius and approximately averages to zero in the g-mode cavity, such that we can ignore it under a stationary phase approximation.
Then
\begin{equation}
    \langle\vec{\xi}_{h,n'\ell m'}^{(0)},\mathcal{L}_{\mathrm{mag}}\vec{\xi}_{h,n\ell m}^{(0)}\rangle_{\mathrm{3D}} \approx \frac{2\pi\ell(\ell+1)}{\omega_{0,n\ell}^2}\langle\omega_B^4\tilde{\psi}_{\ell;mm'}^2\rangle_{\Delta n}\mathrm{,}
\end{equation}
where $\Delta n\equiv n'-n$, and we have defined the coefficient of the cosine series of $\omega_B^4\tilde{\psi}_{\ell;mm'}^2$:
\begin{equation} \label{eqn:cosine_series_coeff}
    \langle\omega_B^4\tilde{\psi}_{\ell;mm'}^2\rangle_{\Delta n} = \frac{2}{X_N}\int_{\mathcal{R}}\omega_B^4\tilde{\psi}_{\ell;mm'}^2\cos\left(\frac{\Delta n\pi}{X_N}x_N\right)\mathrm{d}x_N\mathrm{.}
\end{equation}
Equation \ref{eqn:cosine_series_coeff} indicates that smaller differences in radial order $\Delta n$ are coupled by larger-scale features in $\omega_B^4\tilde{\psi}_{\ell;mm'}^2$.

Finally, $c_{n'\ell m'}$ becomes
\begin{equation} \label{eqn:cnlm_final}
    c_{n'\ell m'}=\frac{2\pi\ell(\ell+1)}{\omega_{0,n\ell}^2}\sum_m\frac{\langle\omega_B^4\tilde{\psi}_{\ell;mm'}^2\rangle_{\Delta n}}{\omega_{0,n\ell}^2-\omega_{0,n'\ell}^2}c_{n\ell m}\mathrm{.}
\end{equation}

To evaluate the size of $c_{n'\ell m'}$ in Equation \ref{eqn:cnlm_final}, we next define
\begin{equation}
    \begin{split}
        \langle\omega_B^4\tilde{\psi}_{\ell;mm'}^2\rangle_{\Delta n} &\equiv \frac{\tilde{\psi}_{\ell;mm';\Delta n}^2}{X_N}\int_{\mathcal{R}}\omega_B^4\,\mathrm{d}x_N \\
        &= \frac{\langle B_{0r}^2\rangle}{4\pi X_N}\tilde{\psi}_{\ell;mm';\Delta n}^2\int_{\mathcal{R}}\frac{N^3}{\rho_0r^3}\,\mathrm{d}r\mathrm{,}
    \end{split}
\end{equation}
where $\tilde{\psi}_{\ell;mm';\Delta n}^2$ is of order unity if $\omega_B^4\tilde{\psi}_{\ell;mm'}^2$ has significant support at the radial scale corresponding to a difference in radial orders $\Delta n$.
We assume that $\omega_{n\ell}$ and $\omega_{n'\ell}$ are close, such that
\begin{equation}
    \omega_{0,n\ell}^2-\omega_{0,n'\ell}^2 \simeq 2\omega_{0,n\ell}^2(\Delta n/n)\mathrm{.}
\end{equation}
Further, we ignore the sum over $m$ in Equation \ref{eqn:cnlm_final} against the unperturbed coefficients $c_{n\ell m}$, under the rough justification that the whole sum scales like the summand.

Under these assumptions, the validity condition for perturbation theory takes the form
\begin{equation}
    \frac{\ell(\ell+1)}{4\omega_{0,n\ell}^4X_N}\frac{\langle B_{0r}^2\rangle}{|\Delta n|/n}\tilde{\psi}_{\ell;mm';\Delta n}^2\int_{\mathcal{R}}\frac{N^3}{\rho_0r^3}\,\mathrm{d}r \ll 1
\end{equation}
or, as a condition on the magnetic field strength,
\begin{equation} \label{eqn:pert_validity}
    \langle B_{0r}^2\rangle^{1/2} \ll \frac{\sqrt{4\pi}\omega_{0,n\ell}^{5/2}}{[\ell(\ell+1)]^{3/4}\sqrt{\tilde{\psi}_{\ell;mm';\Delta n}^2\int_{\mathcal{R}}(N^3/\rho_0r^3)\,\mathrm{d}r}}|\Delta n|^{1/2}\mathrm{.}
\end{equation}

We can define buoyant average
\begin{equation}
    \langle B_{r,\mathrm{crit}}^{-2}\rangle \equiv \frac{1}{X_N}\int_{\mathcal{R}}B_{r,\mathrm{crit}}^{-2}\mathrm{d}x_N\mathrm{,}
\end{equation}
whereupon Equation \ref{eqn:pert_validity} becomes
\begin{equation} \label{eqn:pert_validity_end}
    \langle B_{0r}^2\rangle^{1/2} \ll \frac{2}{[\ell(\ell+1)]^{1/4}}\left(\frac{\omega_{0,n\ell}}{X_N}\right)^{1/2}\left|\frac{\Delta n}{\tilde{\psi}_{\ell;mm';\Delta n}^2}\right|^{1/2}\langle B_{r,\mathrm{crit}}^{-2}\rangle^{-1/2}\mathrm{.}
\end{equation}
Equation \ref{eqn:pert_validity_end} is the desired criterion which we have quoted in the main text (Equation \ref{eqn:pert_validity_end_main}).

%% file: mybib.bib
@ARTICLE{David:2025:Magnetogravity,
       author = {{David}, Cy S. and {Lecoanet}, Daniel and {Garaud}, Pascale},
        title = "{Conversion and Damping of Non-axisymmetric Internal Gravity Waves in Magnetized Stellar Cores}",
      journal = {arXiv e-prints},
         year = 2025,
        month = oct,
          eid = {arXiv:2510.14026},
        pages = {arXiv:2510.14026},
          doi = {10.48550/arXiv.2510.14026},
archivePrefix = {arXiv},
       eprint = {2510.14026},
 primaryClass = {astro-ph.SR},
       adsurl = {https://ui.adsabs.harvard.edu/abs/2025arXiv251014026D}
}

@ARTICLE{Mosser:2017:DepressedModes,
       author = {{Mosser}, B. and {Belkacem}, K. and {Pin{\c{c}}on}, C. and {Takata}, M. and {Vrard}, M. and {Barban}, C. and {Goupil}, M.-J. and {Kallinger}, T. and {Samadi}, R.},
        title = "{Dipole modes with depressed amplitudes in red giants are mixed modes}",
      journal = {\aap},
         year = 2017,
        month = feb,
       volume = {598},
          eid = {A62},
        pages = {A62},
          doi = {10.1051/0004-6361/201629494},
archivePrefix = {arXiv},
       eprint = {1610.03872},
 primaryClass = {astro-ph.SR},
       adsurl = {https://ui.adsabs.harvard.edu/abs/2017A&A...598A..62M}
}

@ARTICLE{Mueller:2025:RayTracing,
       author = {{M{\"u}ller}, Jonas and {Copp{\'e}e}, Quentin and {Hekker}, Saskia},
        title = "{Oscillations of red giant stars with magnetic damping in the core: I. Dissipation of mode energy in dipole-like magnetic fields}",
      journal = {\aap},
         year = 2025,
        month = apr,
       volume = {696},
          eid = {A134},
        pages = {A134},
          doi = {10.1051/0004-6361/202553888},
archivePrefix = {arXiv},
       eprint = {2503.11451},
 primaryClass = {astro-ph.SR},
       adsurl = {https://ui.adsabs.harvard.edu/abs/2025A&A...696A.134M}
}

@ARTICLE{Skoutnev:2025:MagneticWebs,
       author = {{Skoutnev}, Valentin A. and {Beloborodov}, Andrei M.},
        title = "{Magnetic Webs in Stellar Radiative Zones}",
      journal = {\apjl},
         year = 2025,
        month = aug,
       volume = {989},
       number = {1},
          eid = {L4},
        pages = {L4},
          doi = {10.3847/2041-8213/adefda},
archivePrefix = {arXiv},
       eprint = {2504.07223},
 primaryClass = {astro-ph.SR},
       adsurl = {https://ui.adsabs.harvard.edu/abs/2025ApJ...989L...4S}
}

@ARTICLE{Spruit:2002:STDynamo,
       author = {{Spruit}, H.~C.},
        title = "{Dynamo action by differential rotation in a stably stratified stellar interior}",
      journal = {\aap},
         year = 2002,
        month = jan,
       volume = {381},
        pages = {923-932},
          doi = {10.1051/0004-6361:20011465},
archivePrefix = {arXiv},
       eprint = {astro-ph/0108207},
 primaryClass = {astro-ph},
       adsurl = {https://ui.adsabs.harvard.edu/abs/2002A&A...381..923S}
}

@ARTICLE{Proctor:1982:Magnetoconvection,
       author = {{Proctor}, M.~R.~E. and {Weiss}, N.~O.},
        title = "{REVIEW ARTICLE: Magnetoconvection}",
      journal = {Reports on Progress in Physics},
         year = 1982,
        month = nov,
       volume = {45},
       number = {11},
        pages = {1317-1379},
          doi = {10.1088/0034-4885/45/11/003},
       adsurl = {https://ui.adsabs.harvard.edu/abs/1982RPPh...45.1317P}
}

@ARTICLE{Fuller:2015:SuppressedDipole,
       author = {{Fuller}, Jim and {Cantiello}, Matteo and {Stello}, Dennis and {Garcia}, Rafael A. and {Bildsten}, Lars},
        title = "{Asteroseismology can reveal strong internal magnetic fields in red giant stars}",
      journal = {Science},
         year = 2015,
        month = oct,
       volume = {350},
       number = {6259},
        pages = {423-426},
          doi = {10.1126/science.aac6933},
archivePrefix = {arXiv},
       eprint = {1510.06960},
 primaryClass = {astro-ph.SR},
       adsurl = {https://ui.adsabs.harvard.edu/abs/2015Sci...350..423F}
}

@ARTICLE{Cantiello:2014:RGAMT,
       author = {{Cantiello}, Matteo and {Mankovich}, Christopher and {Bildsten}, Lars and {Christensen-Dalsgaard}, J{\o}rgen and {Paxton}, Bill},
        title = "{Angular Momentum Transport within Evolved Low-mass Stars}",
      journal = {\apj},
         year = 2014,
        month = jun,
       volume = {788},
       number = {1},
          eid = {93},
        pages = {93},
          doi = {10.1088/0004-637X/788/1/93},
archivePrefix = {arXiv},
       eprint = {1405.1419},
 primaryClass = {astro-ph.SR},
       adsurl = {https://ui.adsabs.harvard.edu/abs/2014ApJ...788...93C}
}

@ARTICLE{Rui:2025:StochasticOblique,
       author = {{Rui}, Nicholas Z. and {Fuller}, Jim and {Ong}, J.~M. Joel},
        title = "{It's Not Just a Phase: Oblique Pulsations in Magnetic Red Giants and Other Stochastic Oscillators}",
      journal = {\apjl},
         year = 2025,
        month = jun,
       volume = {985},
       number = {2},
          eid = {L39},
        pages = {L39},
          doi = {10.3847/2041-8213/add5e2},
archivePrefix = {arXiv},
       eprint = {2505.03169},
 primaryClass = {astro-ph.SR},
       adsurl = {https://ui.adsabs.harvard.edu/abs/2025ApJ...985L..39R}
}

@ARTICLE{Rui:2025:WDSeismology,
       author = {{Rui}, Nicholas Z. and {Fuller}, Jim and {Hermes}, J.~J.},
        title = "{Supersensitive Seismic Magnetometry of White Dwarfs}",
      journal = {\apj},
         year = 2025,
        month = mar,
       volume = {981},
       number = {1},
          eid = {72},
        pages = {72},
          doi = {10.3847/1538-4357/adaf9e},
archivePrefix = {arXiv},
       eprint = {2410.20557},
 primaryClass = {astro-ph.SR},
       adsurl = {https://ui.adsabs.harvard.edu/abs/2025ApJ...981...72R}
}

@ARTICLE{Rui:2024:TARM,
       author = {{Rui}, Nicholas Z. and {Ong}, J.~M. Joel and {Mathis}, St{\'e}phane},
        title = "{Asteroseismic g-mode period spacings in strongly magnetic rotating stars}",
      journal = {\mnras},
         year = 2024,
        month = jan,
       volume = {527},
       number = {3},
        pages = {6346-6362},
          doi = {10.1093/mnras/stad3461},
archivePrefix = {arXiv},
       eprint = {2310.19873},
 primaryClass = {astro-ph.SR},
       adsurl = {https://ui.adsabs.harvard.edu/abs/2024MNRAS.527.6346R}
}

@ARTICLE{Rui:2023:MagneticSuppression,
       author = {{Rui}, Nicholas Z. and {Fuller}, Jim},
        title = "{Gravity waves in strong magnetic fields}",
      journal = {\mnras},
         year = 2023,
        month = jul,
       volume = {523},
       number = {1},
        pages = {582-602},
          doi = {10.1093/mnras/stad1424},
archivePrefix = {arXiv},
       eprint = {2303.08147},
 primaryClass = {astro-ph.SR},
       adsurl = {https://ui.adsabs.harvard.edu/abs/2023MNRAS.523..582R}
}

@ARTICLE{Lecoanet:2022:HD43317,
       author = {{Lecoanet}, Daniel and {Bowman}, Dominic M. and {Van Reeth}, Timothy},
        title = "{Asteroseismic inference of the near-core magnetic field strength in the main-sequence B star HD 43317}",
      journal = {\mnras},
         year = 2022,
        month = may,
       volume = {512},
       number = {1},
        pages = {L16-L20},
          doi = {10.1093/mnrasl/slac013},
archivePrefix = {arXiv},
       eprint = {2202.03440},
 primaryClass = {astro-ph.SR},
       adsurl = {https://ui.adsabs.harvard.edu/abs/2022MNRAS.512L..16L}
}

@ARTICLE{Lecoanet:2017:MagneticConversion,
       author = {{Lecoanet}, D. and {Vasil}, G.~M. and {Fuller}, J. and {Cantiello}, M. and {Burns}, K.~J.},
        title = "{Conversion of internal gravity waves into magnetic waves}",
      journal = {\mnras},
         year = 2017,
        month = apr,
       volume = {466},
       number = {2},
        pages = {2181-2193},
          doi = {10.1093/mnras/stw3273},
archivePrefix = {arXiv},
       eprint = {1610.08506},
 primaryClass = {astro-ph.SR},
       adsurl = {https://ui.adsabs.harvard.edu/abs/2017MNRAS.466.2181L}
}

@ARTICLE{Bugnet:2022:MagneticII,
       author = {{Bugnet}, L.},
        title = "{Magnetic signatures on mixed-mode frequencies. II. Period spacings as a probe of the internal magnetism of red giants}",
      journal = {\aap},
         year = 2022,
        month = nov,
       volume = {667},
          eid = {A68},
        pages = {A68},
          doi = {10.1051/0004-6361/202243167},
archivePrefix = {arXiv},
       eprint = {2208.14954},
 primaryClass = {astro-ph.SR},
       adsurl = {https://ui.adsabs.harvard.edu/abs/2022A&A...667A..68B}
}

@ARTICLE{Bugnet:2021:MagneticI,
       author = {{Bugnet}, L. and {Prat}, V. and {Mathis}, S. and {Astoul}, A. and {Augustson}, K. and {Garc{\'\i}a}, R.~A. and {Mathur}, S. and {Amard}, L. and {Neiner}, C.},
        title = "{Magnetic signatures on mixed-mode frequencies. I. An axisymmetric fossil field inside the core of red giants}",
      journal = {\aap},
         year = 2021,
        month = jun,
       volume = {650},
          eid = {A53},
        pages = {A53},
          doi = {10.1051/0004-6361/202039159},
archivePrefix = {arXiv},
       eprint = {2102.01216},
 primaryClass = {astro-ph.SR},
       adsurl = {https://ui.adsabs.harvard.edu/abs/2021A&A...650A..53B}
}

@ARTICLE{Cantiello:2016:EvolvingMagnetic,
       author = {{Cantiello}, Matteo and {Fuller}, Jim and {Bildsten}, Lars},
        title = "{Asteroseismic Signatures of Evolving Internal Stellar Magnetic Fields}",
      journal = {\apj},
         year = 2016,
        month = jun,
       volume = {824},
       number = {1},
          eid = {14},
        pages = {14},
          doi = {10.3847/0004-637X/824/1/14},
archivePrefix = {arXiv},
       eprint = {1602.03056},
 primaryClass = {astro-ph.SR},
       adsurl = {https://ui.adsabs.harvard.edu/abs/2016ApJ...824...14C}
}

@ARTICLE{Deheuvels:2023:MagneticRG,
       author = {{Deheuvels}, S. and {Li}, G. and {Ballot}, J. and {Ligni{\`e}res}, F.},
        title = "{Strong magnetic fields detected in the cores of 11 red giant stars using gravity-mode period spacings}",
      journal = {\aap},
         year = 2023,
        month = feb,
       volume = {670},
          eid = {L16},
        pages = {L16},
          doi = {10.1051/0004-6361/202245282},
archivePrefix = {arXiv},
       eprint = {2301.01308},
 primaryClass = {astro-ph.SR},
       adsurl = {https://ui.adsabs.harvard.edu/abs/2023A&A...670L..16D}
}

@ARTICLE{Li:2022:30to100kG,
       author = {{Li}, Gang and {Deheuvels}, S{\'e}bastien and {Ballot}, J{\'e}r{\^o}me and {Ligni{\`e}res}, Fran{\c{c}}ois},
        title = "{Magnetic fields of 30 to 100 kG in the cores of red giant stars}",
      journal = {\nat},
         year = 2022,
        month = oct,
       volume = {610},
       number = {7930},
        pages = {43-46},
          doi = {10.1038/s41586-022-05176-0},
archivePrefix = {arXiv},
       eprint = {2208.09487},
 primaryClass = {astro-ph.SR},
       adsurl = {https://ui.adsabs.harvard.edu/abs/2022Natur.610...43L}
}

@ARTICLE{Takata:2026:GDorToroidalField,
       author = {{Takata}, Masao and {Murphy}, Simon J. and {Kurtz}, Donald W. and {Saio}, Hideyuki and {Shibahashi}, Hiromoto},
        title = "{Asteroseismic detection of a predominantly toroidal magnetic field in the deep interior of the main-sequence F star KIC 9244992}",
      journal = {\mnras},
         year = 2026,
        month = jan,
       volume = {545},
       number = {3},
          eid = {staf2153},
        pages = {staf2153},
          doi = {10.1093/mnras/staf2153},
archivePrefix = {arXiv},
       eprint = {2512.00786},
 primaryClass = {astro-ph.SR},
       adsurl = {https://ui.adsabs.harvard.edu/abs/2026MNRAS.545f2153T}
}

@ARTICLE{Stello:2016:MagneticPrevalence,
       author = {{Stello}, Dennis and {Cantiello}, Matteo and {Fuller}, Jim and {Huber}, Daniel and {Garc{\'\i}a}, Rafael A. and {Bedding}, Timothy R. and {Bildsten}, Lars and {Silva Aguirre}, Victor},
        title = "{A prevalence of dynamo-generated magnetic fields in the cores of intermediate-mass stars}",
      journal = {\nat},
         year = 2016,
        month = jan,
       volume = {529},
       number = {7586},
        pages = {364-367},
          doi = {10.1038/nature16171},
archivePrefix = {arXiv},
       eprint = {1601.00004},
 primaryClass = {astro-ph.SR},
       adsurl = {https://ui.adsabs.harvard.edu/abs/2016Natur.529..364S}
}

@ARTICLE{Chaplin:2013:SolarType,
       author = {{Chaplin}, William J. and {Miglio}, Andrea},
        title = "{Asteroseismology of Solar-Type and Red-Giant Stars}",
      journal = {\araa},
         year = 2013,
        month = aug,
       volume = {51},
       number = {1},
        pages = {353-392},
          doi = {10.1146/annurev-astro-082812-140938},
archivePrefix = {arXiv},
       eprint = {1303.1957},
 primaryClass = {astro-ph.SR},
       adsurl = {https://ui.adsabs.harvard.edu/abs/2013ARA&A..51..353C}
}

@ARTICLE{Balona:2011:GammaDoradus,
       author = {{Balona}, L.~A. and {Guzik}, J.~A. and {Uytterhoeven}, K. and {Smith}, J.~C. and {Tenenbaum}, P. and {Twicken}, J.~D.},
        title = "{The Kepler view of {\ensuremath{\gamma}} Doradus stars}",
      journal = {\mnras},
         year = 2011,
        month = aug,
       volume = {415},
       number = {4},
        pages = {3531-3538},
          doi = {10.1111/j.1365-2966.2011.18973.x},
       adsurl = {https://ui.adsabs.harvard.edu/abs/2011MNRAS.415.3531B}
}

@ARTICLE{Heber:2009:HotSubdwarfs,
       author = {{Heber}, Ulrich},
        title = "{Hot Subdwarf Stars}",
      journal = {\araa},
         year = 2009,
        month = sep,
       volume = {47},
       number = {1},
        pages = {211-251},
          doi = {10.1146/annurev-astro-082708-101836},
       adsurl = {https://ui.adsabs.harvard.edu/abs/2009ARA&A..47..211H}
}

@ARTICLE{Corsico:2019:PulsatingWhiteDwarfs,
       author = {{C{\'o}rsico}, Alejandro H. and {Althaus}, Leandro G. and {Miller Bertolami}, Marcelo M. and {Kepler}, S.~O.},
        title = "{Pulsating white dwarfs: new insights}",
      journal = {\aapr},
         year = 2019,
        month = sep,
       volume = {27},
       number = {1},
          eid = {7},
        pages = {7},
          doi = {10.1007/s00159-019-0118-4},
archivePrefix = {arXiv},
       eprint = {1907.00115},
 primaryClass = {astro-ph.SR},
       adsurl = {https://ui.adsabs.harvard.edu/abs/2019A&ARv..27....7C}
}

@ARTICLE{Garcia:2014:DepressedDipole,
       author = {{Garc{\'\i}a}, R.~A. and {P{\'e}rez Hern{\'a}ndez}, F. and {Benomar}, O. and {Silva Aguirre}, V. and {Ballot}, J. and {Davies}, G.~R. and {Do{\u{g}}an}, G. and {Stello}, D. and {Christensen-Dalsgaard}, J. and {Houdek}, G. and {Ligni{\`e}res}, F. and {Mathur}, S. and {Takata}, M. and {Ceillier}, T. and {Chaplin}, W.~J. and {Mathis}, S. and {Mosser}, B. and {Ouazzani}, R.~M. and {Pinsonneault}, M.~H. and {Reese}, D.~R. and {R{\'e}gulo}, C. and {Salabert}, D. and {Thompson}, M.~J. and {van Saders}, J.~L. and {Neiner}, C. and {De Ridder}, J.},
        title = "{Study of KIC 8561221 observed by Kepler: an early red giant showing depressed dipolar modes}",
      journal = {\aap},
         year = 2014,
        month = mar,
       volume = {563},
          eid = {A84},
        pages = {A84},
          doi = {10.1051/0004-6361/201322823},
archivePrefix = {arXiv},
       eprint = {1311.6990},
 primaryClass = {astro-ph.SR},
       adsurl = {https://ui.adsabs.harvard.edu/abs/2014A&A...563A..84G}
}

@ARTICLE{Hatt:2024:MagneticRG,
       author = {{Hatt}, Emily J. and {Ong}, J.~M. Joel and {Nielsen}, Martin B. and {Chaplin}, William J. and {Davies}, Guy R. and {Deheuvels}, S{\'e}bastien and {Ballot}, J{\'e}r{\^o}me and {Li}, Gang and {Bugnet}, Lisa},
        title = "{Asteroseismic signatures of core magnetism and rotation in hundreds of low-luminosity red giants}",
      journal = {\mnras},
         year = 2024,
        month = oct,
       volume = {534},
       number = {2},
        pages = {1060-1076},
          doi = {10.1093/mnras/stae2053},
archivePrefix = {arXiv},
       eprint = {2409.01157},
 primaryClass = {astro-ph.SR},
       adsurl = {https://ui.adsabs.harvard.edu/abs/2024MNRAS.534.1060H}
}

@ARTICLE{Mathis:2021:Magnetoasteroseismology,
       author = {{Mathis}, S. and {Bugnet}, L. and {Prat}, V. and {Augustson}, K. and {Mathur}, S. and {Garcia}, R.~A.},
        title = "{Probing the internal magnetism of stars using asymptotic magneto-asteroseismology}",
      journal = {\aap},
         year = 2021,
        month = mar,
       volume = {647},
          eid = {A122},
        pages = {A122},
          doi = {10.1051/0004-6361/202039180},
archivePrefix = {arXiv},
       eprint = {2012.11050},
 primaryClass = {astro-ph.SR},
       adsurl = {https://ui.adsabs.harvard.edu/abs/2021A&A...647A.122M}
}

@ARTICLE{Dhouib:2022:ToroidalField,
       author = {{Dhouib}, H. and {Mathis}, S. and {Bugnet}, L. and {Van Reeth}, T. and {Aerts}, C.},
        title = "{Detecting deep axisymmetric toroidal magnetic fields in stars. The traditional approximation of rotation for differentially rotating deep spherical shells with a general azimuthal magnetic field}",
      journal = {\aap},
         year = 2022,
        month = may,
       volume = {661},
          eid = {A133},
        pages = {A133},
          doi = {10.1051/0004-6361/202142956},
archivePrefix = {arXiv},
       eprint = {2202.10026},
 primaryClass = {astro-ph.SR},
       adsurl = {https://ui.adsabs.harvard.edu/abs/2022A&A...661A.133D}
}

@ARTICLE{Gomes:2020:MagneticRG,
       author = {{Gomes}, Pedro and {Lopes}, Il{\'\i}dio},
        title = "{Core magnetic field imprint in the non-radial oscillations of red giant stars}",
      journal = {\mnras},
         year = 2020,
        month = jul,
       volume = {496},
       number = {1},
        pages = {620-628},
          doi = {10.1093/mnras/staa1585},
archivePrefix = {arXiv},
       eprint = {2007.09632},
 primaryClass = {astro-ph.SR},
       adsurl = {https://ui.adsabs.harvard.edu/abs/2020MNRAS.496..620G}
}

@ARTICLE{Das:2024:ComplexMagnetic,
       author = {{Das}, S.~B. and {Einramhof}, L. and {Bugnet}, L.},
        title = "{Unveiling complex magnetic field configurations in red giant stars}",
      journal = {\aap},
         year = 2024,
        month = oct,
       volume = {690},
          eid = {A217},
        pages = {A217},
          doi = {10.1051/0004-6361/202450918},
archivePrefix = {arXiv},
       eprint = {2405.20133},
 primaryClass = {astro-ph.SR},
       adsurl = {https://ui.adsabs.harvard.edu/abs/2024A&A...690A.217D}
}

@ARTICLE{Mathis:2023:MagneticAsymmetry,
       author = {{Mathis}, S. and {Bugnet}, L.},
        title = "{Asymmetries of frequency splittings of dipolar mixed modes: A window on the topology of deep magnetic fields}",
      journal = {\aap},
         year = 2023,
        month = aug,
       volume = {676},
          eid = {L9},
        pages = {L9},
          doi = {10.1051/0004-6361/202346832},
archivePrefix = {arXiv},
       eprint = {2306.11587},
 primaryClass = {astro-ph.SR},
       adsurl = {https://ui.adsabs.harvard.edu/abs/2023A&A...676L...9M}
}

@ARTICLE{Li:2023:13Magnetic,
       author = {{Li}, Gang and {Deheuvels}, S{\'e}bastien and {Li}, Tanda and {Ballot}, J{\'e}r{\^o}me and {Ligni{\`e}res}, Fran{\c{c}}ois},
        title = "{Internal magnetic fields in 13 red giants detected by asteroseismology}",
      journal = {\aap},
         year = 2023,
        month = dec,
       volume = {680},
          eid = {A26},
        pages = {A26},
          doi = {10.1051/0004-6361/202347260},
archivePrefix = {arXiv},
       eprint = {2309.13756},
 primaryClass = {astro-ph.SR},
       adsurl = {https://ui.adsabs.harvard.edu/abs/2023A&A...680A..26L}
}

@ARTICLE{Spruit:1999:DifferentialRotation,
       author = {{Spruit}, H.~C.},
        title = "{Differential rotation and magnetic fields in stellar interiors}",
      journal = {\aap},
         year = 1999,
        month = sep,
       volume = {349},
        pages = {189-202},
          doi = {10.48550/arXiv.astro-ph/9907138},
archivePrefix = {arXiv},
       eprint = {astro-ph/9907138},
 primaryClass = {astro-ph},
       adsurl = {https://ui.adsabs.harvard.edu/abs/1999A&A...349..189S}
}

@ARTICLE{Bhattacharya:2024:MagneticSubgiant,
       author = {{Bhattacharya}, Shatanik and {Das}, Srijan Bharati and {Bugnet}, Lisa and {Panda}, Subrata and {Hanasoge}, Shravan M.},
        title = "{Detectability of Axisymmetric Magnetic Fields from the Core to the Surface of Oscillating Post-main-sequence Stars}",
      journal = {\apj},
         year = 2024,
        month = jul,
       volume = {970},
       number = {1},
          eid = {42},
        pages = {42},
          doi = {10.3847/1538-4357/ad4708},
archivePrefix = {arXiv},
       eprint = {2404.17167},
 primaryClass = {astro-ph.SR},
       adsurl = {https://ui.adsabs.harvard.edu/abs/2024ApJ...970...42B}
}

@ARTICLE{Bildsten:1996:OceanGModes,
       author = {{Bildsten}, Lars and {Ushomirsky}, Greg and {Cutler}, Curt},
        title = "{Ocean g-Modes on Rotating Neutron Stars}",
      journal = {\apj},
         year = 1996,
        month = apr,
       volume = {460},
        pages = {827},
          doi = {10.1086/177012},
       adsurl = {https://ui.adsabs.harvard.edu/abs/1996ApJ...460..827B}
}

@ARTICLE{Lee:1997:TAR,
       author = {{Lee}, Umin and {Saio}, Hideyuki},
        title = "{Low-Frequency Nonradial Oscillations in Rotating Stars. I. Angular Dependence}",
      journal = {\apj},
         year = 1997,
        month = dec,
       volume = {491},
       number = {2},
        pages = {839-845},
          doi = {10.1086/304980},
       adsurl = {https://ui.adsabs.harvard.edu/abs/1997ApJ...491..839L}
}

@ARTICLE{Fuller:2019:SlowingSpins,
       author = {{Fuller}, Jim and {Piro}, Anthony L. and {Jermyn}, Adam S.},
        title = "{Slowing the spins of stellar cores}",
      journal = {\mnras},
         year = 2019,
        month = may,
       volume = {485},
       number = {3},
        pages = {3661-3680},
          doi = {10.1093/mnras/stz514},
archivePrefix = {arXiv},
       eprint = {1902.08227},
 primaryClass = {astro-ph.SR},
       adsurl = {https://ui.adsabs.harvard.edu/abs/2019MNRAS.485.3661F}
}

@ARTICLE{Paxton:2011:MESA,
  author = {{Paxton}, B. and {Bildsten}, L. and {Dotter}, A. and {Herwig}, F. and {Lesaffre}, P. and {Timmes}, F.},
  title = {{Modules for Experiments in Stellar Astrophysics (MESA)}},
  journal = {\apjs},
  archivePrefix = {arXiv},
  eprint = {1009.1622},
  primaryClass = {astro-ph.SR},
  year = {2011},
  month = {jan},
  volume = {192},
  eid = {3},
  pages = {3},
  doi = {10.1088/0067-0049/192/1/3},
  adsurl = {https://ui.adsabs.harvard.edu/abs/2011ApJS..192....3P},
}

@ARTICLE{Paxton:2013:MESA,
  author = {{Paxton}, B. and {Cantiello}, M. and {Arras}, P. and {Bildsten}, L. and {Brown}, E.~F. and {Dotter}, A. and {Mankovich}, C. and {Montgomery}, M.~H. and {Stello}, D. and {Timmes}, F.~X. and {Townsend}, R.},
  title = {{Modules for Experiments in Stellar Astrophysics (MESA): Planets, Oscillations, Rotation, and Massive Stars}},
  journal = {\apjs},
  archivePrefix = {arXiv},
  eprint = {1301.0319},
  primaryClass = {astro-ph.SR},
  year = {2013},
  month = {sep},
  volume = {208},
  eid = {4},
  pages = {4},
  doi = {10.1088/0067-0049/208/1/4},
  adsurl = {https://ui.adsabs.harvard.edu/abs/2013ApJS..208....4P},
}

@ARTICLE{Paxton:2015:MESA,
  author = {{Paxton}, B. and {Marchant}, P. and {Schwab}, J. and {Bauer}, E.~B. and {Bildsten}, L. and {Cantiello}, M. and {Dessart}, L. and {Farmer}, R. and {Hu}, H. and {Langer}, N. and {Townsend}, R.~H.~D. and {Townsley}, D.~M. and {Timmes}, F.~X.},
  title = {{Modules for Experiments in Stellar Astrophysics (MESA): Binaries, Pulsations, and Explosions}},
  journal = {\apjs},
  archivePrefix = {arXiv},
  eprint = {1506.03146},
  primaryClass = {astro-ph.SR},
  year = {2015},
  month = {sep},
  volume = {220},
  eid = {15},
  pages = {15},
  doi = {10.1088/0067-0049/220/1/15},
  adsurl = {https://ui.adsabs.harvard.edu/abs/2015ApJS..220...15P},
}

@ARTICLE{Paxton:2018:MESA,
  author = {{Paxton}, B. and {Schwab}, J. and {Bauer}, E.~B. and {Bildsten}, L. and {Blinnikov}, S. and {Duffell}, P. and {Farmer}, R. and {Goldberg}, J.~A. and {Marchant}, P. and {Sorokina}, E. and {Thoul}, A. and {Townsend}, R.~H.~D. and {Timmes}, F.~X.},
  title = {{Modules for Experiments in Stellar Astrophysics (MESA): Convective Boundaries, Element Diffusion, and Massive Star Explosions}},
  journal = {\apjs},
  archivePrefix = {arXiv},
  eprint = {1710.08424},
  primaryClass = {astro-ph.SR},
  year = {2018},
  month = {feb},
  volume = {234},
  eid = {34},
  pages = {34},
  doi = {10.3847/1538-4365/aaa5a8},
  adsurl = {https://ui.adsabs.harvard.edu/abs/2018ApJS..234...34P},
}

@ARTICLE{Paxton:2019:MESA,
       author = {{Paxton}, Bill and {Smolec}, R. and {Schwab}, Josiah and {Gautschy}, A. and
         {Bildsten}, Lars and {Cantiello}, Matteo and {Dotter}, Aaron and
         {Farmer}, R. and {Goldberg}, Jared A. and {Jermyn}, Adam S. and
         {Kanbur}, S.~M. and {Marchant}, Pablo and {Thoul}, Anne and
         {Townsend}, Richard H.~D. and {Wolf}, William M. and {Zhang}, Michael and
         {Timmes}, F.~X.},
        title = "{Modules for Experiments in Stellar Astrophysics (MESA): Pulsating Variable Stars, Rotation, Convective Boundaries, and Energy Conservation}",
      journal = {\apjs},
         year = "2019",
        month = "Jul",
       volume = {243},
       number = {1},
          eid = {10},
        pages = {10},
          doi = {10.3847/1538-4365/ab2241},
archivePrefix = {arXiv},
       eprint = {1903.01426},
 primaryClass = {astro-ph.SR},
       adsurl = {https://ui.adsabs.harvard.edu/abs/2019ApJS..243...10P}
}

@ARTICLE{Jermyn:2023:MESA,
       author = {{Jermyn}, Adam S. and {Bauer}, Evan B. and {Schwab}, Josiah and {Farmer}, R. and {Ball}, Warrick H. and {Bellinger}, Earl P. and {Dotter}, Aaron and {Joyce}, Meridith and {Marchant}, Pablo and {Mombarg}, Joey S.~G. and {Wolf}, William M. and {Sunny Wong}, Tin Long and {Cinquegrana}, Giulia C. and {Farrell}, Eoin and {Smolec}, R. and {Thoul}, Anne and {Cantiello}, Matteo and {Herwig}, Falk and {Toloza}, Odette and {Bildsten}, Lars and {Townsend}, Richard H.~D. and {Timmes}, F.~X.},
        title = "{Modules for Experiments in Stellar Astrophysics (MESA): Time-dependent Convection, Energy Conservation, Automatic Differentiation, and Infrastructure}",
      journal = {\apjs},
         year = 2023,
        month = mar,
       volume = {265},
       number = {1},
          eid = {15},
        pages = {15},
          doi = {10.3847/1538-4365/acae8d},
archivePrefix = {arXiv},
       eprint = {2208.03651},
 primaryClass = {astro-ph.SR},
       adsurl = {https://ui.adsabs.harvard.edu/abs/2023ApJS..265...15J}
}

@ARTICLE{Burns:2020:Dedalus,
       author = {{Burns}, Keaton J. and {Vasil}, Geoffrey M. and {Oishi}, Jeffrey S. and {Lecoanet}, Daniel and {Brown}, Benjamin P.},
        title = "{Dedalus: A flexible framework for numerical simulations with spectral methods}",
      journal = {Physical Review Research},
         year = 2020,
        month = apr,
       volume = {2},
       number = {2},
          eid = {023068},
        pages = {023068},
          doi = {10.1103/PhysRevResearch.2.023068},
archivePrefix = {arXiv},
       eprint = {1905.10388},
 primaryClass = {astro-ph.IM},
       adsurl = {https://ui.adsabs.harvard.edu/abs/2020PhRvR...2b3068B}
}

@ARTICLE{Villate:2026:MagneticOffset,
       author = {{Villate}, Matisse and {Deheuvels}, S{\'e}bastien and {Ballot}, J{\'e}r{\^o}me},
        title = "{Seismic detection of core magnetic fields in red giants using the gravity offset}",
      journal = {arXiv e-prints},
         year = 2026,
        month = feb,
          eid = {arXiv:2602.14570},
        pages = {arXiv:2602.14570},
          doi = {10.48550/arXiv.2602.14570},
archivePrefix = {arXiv},
       eprint = {2602.14570},
 primaryClass = {astro-ph.SR},
       adsurl = {https://ui.adsabs.harvard.edu/abs/2026arXiv260214570V}
}

@ARTICLE{Einramhof:2026:MagnetoWhiteDwarfs,
       author = {{Einramhof}, Lukas and {Bugnet}, Lisa and {Magdalena Calcaferro}, Leila and {Barrault}, Lucas and {Bharati Das}, Srijan},
        title = "{Magneto-Archeology of White Dwarfs. Revisiting the fossil field scenario with observational constraints during the red giant branch}",
      journal = {arXiv e-prints},
         year = 2026,
        month = jan,
          eid = {arXiv:2601.15203},
        pages = {arXiv:2601.15203},
          doi = {10.48550/arXiv.2601.15203},
archivePrefix = {arXiv},
       eprint = {2601.15203},
 primaryClass = {astro-ph.SR},
       adsurl = {https://ui.adsabs.harvard.edu/abs/2026arXiv260115203E}
}

@ARTICLE{Loi:2020:MGPackets,
       author = {{Loi}, Shyeh Tjing},
        title = "{Magneto-gravity wave packet dynamics in strongly magnetized cores of evolved stars}",
      journal = {\mnras},
         year = 2020,
        month = apr,
       volume = {493},
       number = {4},
        pages = {5726-5742},
          doi = {10.1093/mnras/staa581},
archivePrefix = {arXiv},
       eprint = {2002.11130},
 primaryClass = {astro-ph.SR},
       adsurl = {https://ui.adsabs.harvard.edu/abs/2020MNRAS.493.5726L}
}

@ARTICLE{Loi:2018:MGDynamicalChaos,
       author = {{Loi}, Shyeh Tjing and {Papaloizou}, John C.~B.},
        title = "{Effects of a strong magnetic field on internal gravity waves: trapping, phase mixing, reflection, and dynamical chaos}",
      journal = {\mnras},
         year = 2018,
        month = jul,
       volume = {477},
       number = {4},
        pages = {5338-5357},
          doi = {10.1093/mnras/sty917},
archivePrefix = {arXiv},
       eprint = {1804.03664},
 primaryClass = {astro-ph.SR},
       adsurl = {https://ui.adsabs.harvard.edu/abs/2018MNRAS.477.5338L}
}

@ARTICLE{Loi:2020:MGNonPerturbative,
       author = {{Loi}, Shyeh Tjing and {Papaloizou}, John C.~B.},
        title = "{Low-degree mixed modes in red giant stars with moderate core magnetic fields}",
      journal = {\mnras},
         year = 2020,
        month = jan,
       volume = {491},
       number = {1},
        pages = {708-724},
          doi = {10.1093/mnras/stz2987},
archivePrefix = {arXiv},
       eprint = {1910.09940},
 primaryClass = {astro-ph.SR},
       adsurl = {https://ui.adsabs.harvard.edu/abs/2020MNRAS.491..708L}
}

@ARTICLE{Eckart:1960:TAR,
       author = {{Eckart}, Carl},
        title = "{Variation Principles of Hydrodynamics}",
      journal = {Physics of Fluids},
         year = 1960,
        month = may,
       volume = {3},
       number = {3},
        pages = {421-427},
          doi = {10.1063/1.1706053},
       adsurl = {https://ui.adsabs.harvard.edu/abs/1960PhFl....3..421E}
}

@ARTICLE{Berthomieu:1978:TAR,
       author = {{Berthomieu}, G. and {Gonczi}, G. and {Graff}, Ph. and {Provost}, J. and {Rocca}, A.},
        title = "{Low-frequency Gravity Modes of a Rotating Star}",
      journal = {\aap},
         year = 1978,
        month = nov,
       volume = {70},
        pages = {597},
       adsurl = {https://ui.adsabs.harvard.edu/abs/1978A&A....70..597B}
}

@ARTICLE{Asai:2016:NSMagneticModes,
       author = {{Asai}, Hidetaka and {Lee}, Umin and {Yoshida}, Shijun},
        title = "{Non-axisymmetric magnetic modes of neutron stars with purely poloidal magnetic fields}",
      journal = {\mnras},
         year = 2016,
        month = jan,
       volume = {455},
       number = {2},
        pages = {2228-2241},
          doi = {10.1093/mnras/stv2368},
archivePrefix = {arXiv},
       eprint = {1507.00314},
 primaryClass = {astro-ph.HE},
       adsurl = {https://ui.adsabs.harvard.edu/abs/2016MNRAS.455.2228A}
}

@ARTICLE{Wade:2016:MiMeS,
       author = {{Wade}, G.~A. and {Neiner}, C. and {Alecian}, E. and {Grunhut}, J.~H. and {Petit}, V. and {Batz}, B. de and {Bohlender}, D.~A. and {Cohen}, D.~H. and {Henrichs}, H.~F. and {Kochukhov}, O. and {Landstreet}, J.~D. and {Manset}, N. and {Martins}, F. and {Mathis}, S. and {Oksala}, M.~E. and {Owocki}, S.~P. and {Rivinius}, Th. and {Shultz}, M.~E. and {Sundqvist}, J.~O. and {Townsend}, R.~H.~D. and {ud-Doula}, A. and {Bouret}, J.-C. and {Braithwaite}, J. and {Briquet}, M. and {Carciofi}, A.~C. and {David-Uraz}, A. and {Folsom}, C.~P. and {Fullerton}, A.~W. and {Leroy}, B. and {Marcolino}, W.~L.~F. and {Moffat}, A.~F.~J. and {Naz{\'e}}, Y. and {Louis}, N. St and {Auri{\`e}re}, M. and {Bagnulo}, S. and {Bailey}, J.~D. and {Barb{\'a}}, R.~H. and {Blaz{\`e}re}, A. and {B{\"o}hm}, T. and {Catala}, C. and {Donati}, J.-F. and {Ferrario}, L. and {Harrington}, D. and {Howarth}, I.~D. and {Ignace}, R. and {Kaper}, L. and {L{\"u}ftinger}, T. and {Prinja}, R. and {Vink}, J.~S. and {Weiss}, W.~W. and {Yakunin}, I.},
        title = "{The MiMeS survey of magnetism in massive stars: introduction and overview}",
      journal = {\mnras},
         year = 2016,
        month = feb,
       volume = {456},
       number = {1},
        pages = {2-22},
          doi = {10.1093/mnras/stv2568},
archivePrefix = {arXiv},
       eprint = {1511.08425},
 primaryClass = {astro-ph.SR},
       adsurl = {https://ui.adsabs.harvard.edu/abs/2016MNRAS.456....2W}
}

@ARTICLE{Shultz:2019:MagneticBStars,
       author = {{Shultz}, M.~E. and {Wade}, G.~A. and {Rivinius}, Th and {Alecian}, E. and {Neiner}, C. and {Petit}, V. and {Owocki}, S. and {ud-Doula}, A. and {Kochukhov}, O. and {Bohlender}, D. and {Keszthelyi}, Z. and {MiMeS Collaboration} and {BinaMIcS Collaboration}},
        title = "{The magnetic early B-type stars - III. A main-sequence magnetic, rotational, and magnetospheric biography}",
      journal = {\mnras},
         year = 2019,
        month = nov,
       volume = {490},
       number = {1},
        pages = {274-295},
          doi = {10.1093/mnras/stz2551},
archivePrefix = {arXiv},
       eprint = {1909.02530},
 primaryClass = {astro-ph.SR},
       adsurl = {https://ui.adsabs.harvard.edu/abs/2019MNRAS.490..274S}
}

@ARTICLE{Lee:2018:AxisymNS,
       author = {{Lee}, Umin},
        title = "{Axisymmetric spheroidal modes of neutron stars magnetized with poloidal magnetic fields}",
      journal = {\mnras},
         year = 2018,
        month = jan,
       volume = {473},
       number = {3},
        pages = {3661-3670},
          doi = {10.1093/mnras/stx2558},
archivePrefix = {arXiv},
       eprint = {1709.10102},
 primaryClass = {astro-ph.HE},
       adsurl = {https://ui.adsabs.harvard.edu/abs/2018MNRAS.473.3661L}
}

@ARTICLE{Lee:2018:NSPoloidalToroidal,
       author = {{Lee}, Umin},
        title = "{Axisymmetric magnetic modes of neutron stars having mixed poloidal and toroidal magnetic fields}",
      journal = {\mnras},
         year = 2018,
        month = may,
       volume = {476},
       number = {3},
        pages = {3399-3414},
          doi = {10.1093/mnras/sty406},
archivePrefix = {arXiv},
       eprint = {1802.04957},
 primaryClass = {astro-ph.HE},
       adsurl = {https://ui.adsabs.harvard.edu/abs/2018MNRAS.476.3399L}
}

@ARTICLE{Mathis:2011:MagneticRotating,
       author = {{Mathis}, S. and {de Brye}, N.},
        title = "{Low-frequency internal waves in magnetized rotating stellar radiation zones. I. Wave structure modification by a toroidal field}",
      journal = {\aap},
         year = 2011,
        month = feb,
       volume = {526},
          eid = {A65},
        pages = {A65},
          doi = {10.1051/0004-6361/201015571},
       adsurl = {https://ui.adsabs.harvard.edu/abs/2011A&A...526A..65M}
}

@ARTICLE{Loi:2017:AlfvenResonances,
       author = {{Loi}, Shyeh Tjing and {Papaloizou}, John C.~B.},
        title = "{Torsional Alfv{\'e}n resonances as an efficient damping mechanism for non-radial oscillations in red giant stars}",
      journal = {\mnras},
         year = 2017,
        month = may,
       volume = {467},
       number = {3},
        pages = {3212-3225},
          doi = {10.1093/mnras/stx281},
archivePrefix = {arXiv},
       eprint = {1701.08771},
 primaryClass = {astro-ph.SR},
       adsurl = {https://ui.adsabs.harvard.edu/abs/2017MNRAS.467.3212L}
}

@ARTICLE{Loi:2020:MGStrongPSP,
       author = {{Loi}, Shyeh Tjing},
        title = "{Effect of a strong magnetic field on gravity-mode period spacings in red giant stars}",
      journal = {\mnras},
         year = 2020,
        month = aug,
       volume = {496},
       number = {3},
        pages = {3829-3840},
          doi = {10.1093/mnras/staa1823},
archivePrefix = {arXiv},
       eprint = {2006.08635},
 primaryClass = {astro-ph.SR},
       adsurl = {https://ui.adsabs.harvard.edu/abs/2020MNRAS.496.3829L}
}

@ARTICLE{Loi:2021:MGTopologyObliquity,
       author = {{Loi}, Shyeh Tjing},
        title = "{Topology and obliquity of core magnetic fields in shaping seismic properties of slowly rotating evolved stars}",
      journal = {\mnras},
         year = 2021,
        month = jul,
       volume = {504},
       number = {3},
        pages = {3711-3729},
          doi = {10.1093/mnras/stab991},
archivePrefix = {arXiv},
       eprint = {2104.03112},
 primaryClass = {astro-ph.SR},
       adsurl = {https://ui.adsabs.harvard.edu/abs/2021MNRAS.504.3711L}
}

@ARTICLE{Coppee:2024:MGRadialModes,
       author = {{Copp{\'e}e}, Q. and {M{\"u}ller}, J. and {Bazot}, M. and {Hekker}, S.},
        title = "{The radial modes of stars with suppressed dipole modes}",
      journal = {\aap},
         year = 2024,
        month = oct,
       volume = {690},
          eid = {A324},
        pages = {A324},
          doi = {10.1051/0004-6361/202450037},
archivePrefix = {arXiv},
       eprint = {2409.12692},
 primaryClass = {astro-ph.SR},
       adsurl = {https://ui.adsabs.harvard.edu/abs/2024A&A...690A.324C}
}

@software{ManimCommunity:2026,
    author = {{The Manim Community Developers}},
    license = {MIT},
    month = feb,
    title = {{Manim – Mathematical Animation Framework}},
    url = {https://www.manim.community/},
    version = {v0.20.1},
    year = {2026}
}

@article{Virtanen:2020:SciPy,
  author  = {Virtanen, Pauli and Gommers, Ralf and Oliphant, Travis E. and
            Haberland, Matt and Reddy, Tyler and Cournapeau, David and
            Burovski, Evgeni and Peterson, Pearu and Weckesser, Warren and
            Bright, Jonathan and {van der Walt}, St{\'e}fan J. and
            Brett, Matthew and Wilson, Joshua and Millman, K. Jarrod and
            Mayorov, Nikolay and Nelson, Andrew R. J. and Jones, Eric and
            Kern, Robert and Larson, Eric and Carey, C J and
            Polat, {\.I}lhan and Feng, Yu and Moore, Eric W. and
            {VanderPlas}, Jake and Laxalde, Denis and Perktold, Josef and
            Cimrman, Robert and Henriksen, Ian and Quintero, E. A. and
            Harris, Charles R. and Archibald, Anne M. and
            Ribeiro, Ant{\^o}nio H. and Pedregosa, Fabian and
            {van Mulbregt}, Paul and {SciPy 1.0 Contributors}},
  title   = {{{SciPy} 1.0: Fundamental Algorithms for Scientific
            Computing in Python}},
  journal = {Nature Methods},
  year    = {2020},
  volume  = {17},
  pages   = {261--272},
  adsurl  = {https://rdcu.be/b08Wh},
  doi     = {10.1038/s41592-019-0686-2},
}

@ARTICLE{Unsoeld:1927:SpharmTheorem,
       author = {{Uns{\"o}ld}, Albrecht},
        title = "{Beitr{\"a}ge zur Quantenmechanik der Atome}",
      journal = {Annalen der Physik},
         year = 1927,
        month = jan,
       volume = {387},
       number = {3},
        pages = {355-393},
          doi = {10.1002/andp.19273870304},
       adsurl = {https://ui.adsabs.harvard.edu/abs/1927AnP...387..355U}
}

@ARTICLE{Astropy:2022:Software,
       author = {{Astropy Collaboration} and {Price-Whelan}, Adrian M. and {Lim}, Pey Lian and {Earl}, Nicholas and {Starkman}, Nathaniel and {Bradley}, Larry and {Shupe}, David L. and {Patil}, Aarya A. and {Corrales}, Lia and {Brasseur}, C.~E. and {N{\"o}the}, Maximilian and {Donath}, Axel and {Tollerud}, Erik and {Morris}, Brett M. and {Ginsburg}, Adam and {Vaher}, Eero and {Weaver}, Benjamin A. and {Tocknell}, James and {Jamieson}, William and {van Kerkwijk}, Marten H. and {Robitaille}, Thomas P. and {Merry}, Bruce and {Bachetti}, Matteo and {G{\"u}nther}, H. Moritz and {Aldcroft}, Thomas L. and {Alvarado-Montes}, Jaime A. and {Archibald}, Anne M. and {B{\'o}di}, Attila and {Bapat}, Shreyas and {Barentsen}, Geert and {Baz{\'a}n}, Juanjo and {Biswas}, Manish and {Boquien}, M{\'e}d{\'e}ric and {Burke}, D.~J. and {Cara}, Daria and {Cara}, Mihai and {Conroy}, Kyle E. and {Conseil}, Simon and {Craig}, Matthew W. and {Cross}, Robert M. and {Cruz}, Kelle L. and {D'Eugenio}, Francesco and {Dencheva}, Nadia and {Devillepoix}, Hadrien A.~R. and {Dietrich}, J{\"o}rg P. and {Eigenbrot}, Arthur Davis and {Erben}, Thomas and {Ferreira}, Leonardo and {Foreman-Mackey}, Daniel and {Fox}, Ryan and {Freij}, Nabil and {Garg}, Suyog and {Geda}, Robel and {Glattly}, Lauren and {Gondhalekar}, Yash and {Gordon}, Karl D. and {Grant}, David and {Greenfield}, Perry and {Groener}, Austen M. and {Guest}, Steve and {Gurovich}, Sebastian and {Handberg}, Rasmus and {Hart}, Akeem and {Hatfield-Dodds}, Zac and {Homeier}, Derek and {Hosseinzadeh}, Griffin and {Jenness}, Tim and {Jones}, Craig K. and {Joseph}, Prajwel and {Kalmbach}, J. Bryce and {Karamehmetoglu}, Emir and {Ka{\l}uszy{\'n}ski}, Miko{\l}aj and {Kelley}, Michael S.~P. and {Kern}, Nicholas and {Kerzendorf}, Wolfgang E. and {Koch}, Eric W. and {Kulumani}, Shankar and {Lee}, Antony and {Ly}, Chun and {Ma}, Zhiyuan and {MacBride}, Conor and {Maljaars}, Jakob M. and {Muna}, Demitri and {Murphy}, N.~A. and {Norman}, Henrik and {O'Steen}, Richard and {Oman}, Kyle A. and {Pacifici}, Camilla and {Pascual}, Sergio and {Pascual-Granado}, J. and {Patil}, Rohit R. and {Perren}, Gabriel I. and {Pickering}, Timothy E. and {Rastogi}, Tanuj and {Roulston}, Benjamin R. and {Ryan}, Daniel F. and {Rykoff}, Eli S. and {Sabater}, Jose and {Sakurikar}, Parikshit and {Salgado}, Jes{\'u}s and {Sanghi}, Aniket and {Saunders}, Nicholas and {Savchenko}, Volodymyr and {Schwardt}, Ludwig and {Seifert-Eckert}, Michael and {Shih}, Albert Y. and {Jain}, Anany Shrey and {Shukla}, Gyanendra and {Sick}, Jonathan and {Simpson}, Chris and {Singanamalla}, Sudheesh and {Singer}, Leo P. and {Singhal}, Jaladh and {Sinha}, Manodeep and {Sip{\H{o}}cz}, Brigitta M. and {Spitler}, Lee R. and {Stansby}, David and {Streicher}, Ole and {{\v{S}}umak}, Jani and {Swinbank}, John D. and {Taranu}, Dan S. and {Tewary}, Nikita and {Tremblay}, Grant R. and {de Val-Borro}, Miguel and {Van Kooten}, Samuel J. and {Vasovi{\'c}}, Zlatan and {Verma}, Shresth and {de Miranda Cardoso}, Jos{\'e} Vin{\'\i}cius and {Williams}, Peter K.~G. and {Wilson}, Tom J. and {Winkel}, Benjamin and {Wood-Vasey}, W.~M. and {Xue}, Rui and {Yoachim}, Peter and {Zhang}, Chen and {Zonca}, Andrea and {Astropy Project Contributors}},
        title = "{The Astropy Project: Sustaining and Growing a Community-oriented Open-source Project and the Latest Major Release (v5.0) of the Core Package}",
      journal = {\apj},
         year = 2022,
        month = aug,
       volume = {935},
       number = {2},
          eid = {167},
        pages = {167},
          doi = {10.3847/1538-4357/ac7c74},
archivePrefix = {arXiv},
       eprint = {2206.14220},
 primaryClass = {astro-ph.IM},
       adsurl = {https://ui.adsabs.harvard.edu/abs/2022ApJ...935..167A}
}

@ARTICLE{Bigot:2000:roAp,
       author = {{Bigot}, L. and {Provost}, J. and {Berthomieu}, G. and {Dziembowski}, W.~A. and {Goode}, P.~R.},
        title = "{Non-axisymmetric oscillations of roAp stars}",
      journal = {\aap},
         year = 2000,
        month = apr,
       volume = {356},
        pages = {218-233},
       adsurl = {https://ui.adsabs.harvard.edu/abs/2000A&A...356..218B}
}

@ARTICLE{Astropy:2018:Software,
       author = {{Astropy Collaboration} and {Price-Whelan}, A.~M. and {Sip{\H{o}}cz}, B.~M. and {G{\"u}nther}, H.~M. and {Lim}, P.~L. and {Crawford}, S.~M. and {Conseil}, S. and {Shupe}, D.~L. and {Craig}, M.~W. and {Dencheva}, N. and {Ginsburg}, A. and {VanderPlas}, J.~T. and {Bradley}, L.~D. and {P{\'e}rez-Su{\'a}rez}, D. and {de Val-Borro}, M. and {Aldcroft}, T.~L. and {Cruz}, K.~L. and {Robitaille}, T.~P. and {Tollerud}, E.~J. and {Ardelean}, C. and {Babej}, T. and {Bach}, Y.~P. and {Bachetti}, M. and {Bakanov}, A.~V. and {Bamford}, S.~P. and {Barentsen}, G. and {Barmby}, P. and {Baumbach}, A. and {Berry}, K.~L. and {Biscani}, F. and {Boquien}, M. and {Bostroem}, K.~A. and {Bouma}, L.~G. and {Brammer}, G.~B. and {Bray}, E.~M. and {Breytenbach}, H. and {Buddelmeijer}, H. and {Burke}, D.~J. and {Calderone}, G. and {Cano Rodr{\'\i}guez}, J.~L. and {Cara}, M. and {Cardoso}, J.~V.~M. and {Cheedella}, S. and {Copin}, Y. and {Corrales}, L. and {Crichton}, D. and {D'Avella}, D. and {Deil}, C. and {Depagne}, {\'E}. and {Dietrich}, J.~P. and {Donath}, A. and {Droettboom}, M. and {Earl}, N. and {Erben}, T. and {Fabbro}, S. and {Ferreira}, L.~A. and {Finethy}, T. and {Fox}, R.~T. and {Garrison}, L.~H. and {Gibbons}, S.~L.~J. and {Goldstein}, D.~A. and {Gommers}, R. and {Greco}, J.~P. and {Greenfield}, P. and {Groener}, A.~M. and {Grollier}, F. and {Hagen}, A. and {Hirst}, P. and {Homeier}, D. and {Horton}, A.~J. and {Hosseinzadeh}, G. and {Hu}, L. and {Hunkeler}, J.~S. and {Ivezi{\'c}}, {\v{Z}}. and {Jain}, A. and {Jenness}, T. and {Kanarek}, G. and {Kendrew}, S. and {Kern}, N.~S. and {Kerzendorf}, W.~E. and {Khvalko}, A. and {King}, J. and {Kirkby}, D. and {Kulkarni}, A.~M. and {Kumar}, A. and {Lee}, A. and {Lenz}, D. and {Littlefair}, S.~P. and {Ma}, Z. and {Macleod}, D.~M. and {Mastropietro}, M. and {McCully}, C. and {Montagnac}, S. and {Morris}, B.~M. and {Mueller}, M. and {Mumford}, S.~J. and {Muna}, D. and {Murphy}, N.~A. and {Nelson}, S. and {Nguyen}, G.~H. and {Ninan}, J.~P. and {N{\"o}the}, M. and {Ogaz}, S. and {Oh}, S. and {Parejko}, J.~K. and {Parley}, N. and {Pascual}, S. and {Patil}, R. and {Patil}, A.~A. and {Plunkett}, A.~L. and {Prochaska}, J.~X. and {Rastogi}, T. and {Reddy Janga}, V. and {Sabater}, J. and {Sakurikar}, P. and {Seifert}, M. and {Sherbert}, L.~E. and {Sherwood-Taylor}, H. and {Shih}, A.~Y. and {Sick}, J. and {Silbiger}, M.~T. and {Singanamalla}, S. and {Singer}, L.~P. and {Sladen}, P.~H. and {Sooley}, K.~A. and {Sornarajah}, S. and {Streicher}, O. and {Teuben}, P. and {Thomas}, S.~W. and {Tremblay}, G.~R. and {Turner}, J.~E.~H. and {Terr{\'o}n}, V. and {van Kerkwijk}, M.~H. and {de la Vega}, A. and {Watkins}, L.~L. and {Weaver}, B.~A. and {Whitmore}, J.~B. and {Woillez}, J. and {Zabalza}, V. and {Astropy Contributors}},
        title = "{The Astropy Project: Building an Open-science Project and Status of the v2.0 Core Package}",
      journal = {\aj},
         year = 2018,
        month = sep,
       volume = {156},
       number = {3},
          eid = {123},
        pages = {123},
          doi = {10.3847/1538-3881/aabc4f},
archivePrefix = {arXiv},
       eprint = {1801.02634},
 primaryClass = {astro-ph.IM},
       adsurl = {https://ui.adsabs.harvard.edu/abs/2018AJ....156..123A}
}

@ARTICLE{Astropy:2013:Software,
       author = {{Astropy Collaboration} and {Robitaille}, Thomas P. and {Tollerud}, Erik J. and {Greenfield}, Perry and {Droettboom}, Michael and {Bray}, Erik and {Aldcroft}, Tom and {Davis}, Matt and {Ginsburg}, Adam and {Price-Whelan}, Adrian M. and {Kerzendorf}, Wolfgang E. and {Conley}, Alexander and {Crighton}, Neil and {Barbary}, Kyle and {Muna}, Demitri and {Ferguson}, Henry and {Grollier}, Fr{\'e}d{\'e}ric and {Parikh}, Madhura M. and {Nair}, Prasanth H. and {Unther}, Hans M. and {Deil}, Christoph and {Woillez}, Julien and {Conseil}, Simon and {Kramer}, Roban and {Turner}, James E.~H. and {Singer}, Leo and {Fox}, Ryan and {Weaver}, Benjamin A. and {Zabalza}, Victor and {Edwards}, Zachary I. and {Azalee Bostroem}, K. and {Burke}, D.~J. and {Casey}, Andrew R. and {Crawford}, Steven M. and {Dencheva}, Nadia and {Ely}, Justin and {Jenness}, Tim and {Labrie}, Kathleen and {Lim}, Pey Lian and {Pierfederici}, Francesco and {Pontzen}, Andrew and {Ptak}, Andy and {Refsdal}, Brian and {Servillat}, Mathieu and {Streicher}, Ole},
        title = "{Astropy: A community Python package for astronomy}",
      journal = {\aap},
         year = 2013,
        month = oct,
       volume = {558},
          eid = {A33},
        pages = {A33},
          doi = {10.1051/0004-6361/201322068},
archivePrefix = {arXiv},
       eprint = {1307.6212},
 primaryClass = {astro-ph.IM},
       adsurl = {https://ui.adsabs.harvard.edu/abs/2013A&A...558A..33A}
}

@article{Hunter:2007:Matplotlib,
  Author    = {Hunter, J. D.},
  Title     = {Matplotlib: A 2D graphics environment},
  Journal   = {Computing in Science \& Engineering},
  Volume    = {9},
  Number    = {3},
  Pages     = {90--95},
  publisher = {IEEE COMPUTER SOC},
  doi       = {10.1109/MCSE.2007.55},
  year      = 2007
}

@article{Harris:2020:NumPy,
 title         = {Array programming with {NumPy}},
 author        = {Charles R. Harris and K. Jarrod Millman and St{\'{e}}fan J.
                 van der Walt and Ralf Gommers and Pauli Virtanen and David
                 Cournapeau and Eric Wieser and Julian Taylor and Sebastian
                 Berg and Nathaniel J. Smith and Robert Kern and Matti Picus
                 and Stephan Hoyer and Marten H. van Kerkwijk and Matthew
                 Brett and Allan Haldane and Jaime Fern{\'{a}}ndez del
                 R{\'{i}}o and Mark Wiebe and Pearu Peterson and Pierre
                 G{\'{e}}rard-Marchant and Kevin Sheppard and Tyler Reddy and
                 Warren Weckesser and Hameer Abbasi and Christoph Gohlke and
                 Travis E. Oliphant},
 year          = {2020},
 month         = sep,
 journal       = {Nature},
 volume        = {585},
 number        = {7825},
 pages         = {357--362},
 doi           = {10.1038/s41586-020-2649-2},
 publisher     = {Springer Science and Business Media {LLC}},
 url           = {https://doi.org/10.1038/s41586-020-2649-2}
}

@book{Wigner:1931:RotatingSpharms,
  title={Gruppentheorie und ihre Anwendung auf die Quantenmechanik der Atomspektren},
  author={Wigner, Eugen},
  year={1931},
  publisher={Springer}
}

@ARTICLE{Vandersnickt:2025:BcepMagnetic,
       author = {{Vandersnickt}, Jelle and {Vanlaer}, Vincent and {Vanrespaille}, Mathijs and {Aerts}, Conny},
        title = "{Asteroseismic detection of an internal magnetic field in the B0.5V pulsator HD 192575}",
      journal = {\aap},
         year = 2025,
        month = dec,
       volume = {704},
          eid = {L13},
        pages = {L13},
          doi = {10.1051/0004-6361/202556850},
archivePrefix = {arXiv},
       eprint = {2511.21812},
 primaryClass = {astro-ph.SR},
       adsurl = {https://ui.adsabs.harvard.edu/abs/2025A&A...704L..13V}
}

@BOOK{Griffiths:2016:QM,
       author = {{Griffiths}, David J.},
        title = "{Introduction to Quantum Mechanics}",
         year = 2016,
       adsurl = {https://ui.adsabs.harvard.edu/abs/2016iqm..book.....G}
}

@BOOK{Unno:1979:NonradialOsc,
       author = {{Unno}, W. and {Osaki}, Y. and {Ando}, H. and {Shibahashi}, H.},
        title = "{Nonradial oscillations of stars}",
         year = 1979,
       adsurl = {https://ui.adsabs.harvard.edu/abs/1979nos..book.....U}
}

@ARTICLE{Kurtz:1982:roAp,
       author = {{Kurtz}, D.~W.},
        title = "{Rapidly oscillating AP stars.}",
      journal = {\mnras},
         year = 1982,
        month = sep,
       volume = {200},
        pages = {807-859},
          doi = {10.1093/mnras/200.3.807},
       adsurl = {https://ui.adsabs.harvard.edu/abs/1982MNRAS.200..807K}
}

@ARTICLE{Ledoux:1951:Criterion,
       author = {{Ledoux}, P.},
        title = "{The Nonradial Oscillations of Gaseous Stars and the Problem of Beta Canis Majoris.}",
      journal = {\apj},
         year = 1951,
        month = nov,
       volume = {114},
        pages = {373},
          doi = {10.1086/145477},
       adsurl = {https://ui.adsabs.harvard.edu/abs/1951ApJ...114..373L}
}

@ARTICLE{Cowling:1941:Approximation,
       author = {{Cowling}, T.~G.},
        title = "{The non-radial oscillations of polytropic stars}",
      journal = {\mnras},
         year = 1941,
        month = jan,
       volume = {101},
        pages = {367},
          doi = {10.1093/mnras/101.8.367},
       adsurl = {https://ui.adsabs.harvard.edu/abs/1941MNRAS.101..367C}
}

@ARTICLE{Born:1928:AdiabaticTheorem,
       author = {{Born}, M. and {Fock}, V.},
        title = "{Beweis des Adiabatensatzes}",
      journal = {Zeitschrift fur Physik},
         year = 1928,
        month = mar,
       volume = {51},
       number = {3-4},
        pages = {165-180},
          doi = {10.1007/BF01343193},
       adsurl = {https://ui.adsabs.harvard.edu/abs/1928ZPhy...51..165B}
}

@ARTICLE{Deheuvels:2026:NearCritical,
       author = {{Deheuvels}, S. and {Ballot}, J. and {Ligni{\`e}res}, F. and {Li}, G. and {Villate}, M.},
        title = "{Near-critical magnetic fields in Kepler red giants}",
      journal = {arXiv e-prints},
         year = 2026,
        month = apr,
          eid = {arXiv:2604.09901},
        pages = {arXiv:2604.09901},
          doi = {10.48550/arXiv.2604.09901},
archivePrefix = {arXiv},
       eprint = {2604.09901},
 primaryClass = {astro-ph.SR},
       adsurl = {https://ui.adsabs.harvard.edu/abs/2026arXiv260409901D}
}

@article{Wilczek:1984:GaugeStructure,
  title={Appearance of gauge structure in simple dynamical systems},
  author={Wilczek, Frank and Zee, Anthony},
  journal={Physical Review Letters},
  volume={52},
  number={24},
  pages={2111},
  year={1984},
  publisher={APS}
}

@book{Shapere:1989:GeometricPhases,
  title={Geometric phases in physics},
  author={Shapere, Alfred and Wilczek, Frank},
  volume={5},
  year={1989},
  publisher={World scientific},
  series={}
}

@misc{Ihallaine:2026:MagneticGammaDor,
      title={Seismic signature of a magnetic field in the $\gamma$ Doradus star KIC 2309579}, 
      author={S. Ihallaine and J. Ballot and F. Lignières and L. Ferrié and S. Charpinet and M. Galoy and G. Li},
      year={2026},
      eprint={2605.22533},
      archivePrefix={arXiv},
      primaryClass={astro-ph.SR},
      url={https://arxiv.org/abs/2605.22533}, 
}

@ARTICLE{Gough:1990:MagneticPertTheory,
       author = {{Gough}, D.~O. and {Thompson}, M.~J.},
        title = "{The effect of rotation and a buried magnetic field on stellar oscillations}",
      journal = {\mnras},
         year = 1990,
        month = jan,
       volume = {242},
        pages = {25-55},
          doi = {10.1093/mnras/242.1.25},
       adsurl = {https://ui.adsabs.harvard.edu/abs/1990MNRAS.242...25G}
}

@ARTICLE{Lignieres:2024:MagnetoGravitoInertial,
       author = {{Ligni{\`e}res}, F. and {Ballot}, J. and {Deheuvels}, S. and {Galoy}, M.},
        title = "{Perturbative analysis of the effect of a magnetic field on gravito-inertial modes}",
      journal = {\aap},
         year = 2024,
        month = mar,
       volume = {683},
          eid = {A2},
        pages = {A2},
          doi = {10.1051/0004-6361/202348243},
archivePrefix = {arXiv},
       eprint = {2311.13296},
 primaryClass = {astro-ph.SR},
       adsurl = {https://ui.adsabs.harvard.edu/abs/2024A&A...683A...2L}
}
